# Shadow Theory: Data Model Design for Data Integration

Jason T. Liu*

(Draft. Date:9/12/2012)

## ABSTRACT
In information ecosystems, semantic heterogeneity is known as the root issue for the difficulties of data integration, and the Relational Model is not designed for addressing such challenges, i.e. to re-use data that is modeled from other sources in the local data model design. Although researchers have proposed many different approaches, and software vendors have designed tools to help the data integration task, it remains an art relying on human labors. Due to lacks of a comprehensive theory to guide the overall modeling process, the quality of the integrated data heavily depends on the data integrators' experiences.

Based on observations of practical issues for enterprise customer data integration, we believe that the needed solution is a new data model. This new data model needs to manage the difficulties of semantic heterogeneity: different data collected from different perspectives about the same subject matter can naturally be inconsistencies or even in conflicts. In other words, existing data models are designed to support single version of the truth, and naturally they will have difficulties during data integration when data collected from different sources are based on different perspectives or at different levels of abstraction.

Therefore, we propose Shadow Theory to serve as the philosophical foundation in order to design a new data model. The kernel of the theory is based on the notion of shadow, which can be traced back to Plato's Allegory of the Cave over 2000 years ago. The basic idea is that whatever we can observe and store into databases about the subject matter are just shadows. Meanings of shadows are mental entities that exist only in the viewers' cognitive structures. Such mental entities are constrained by the viewers' internal model about the reality, especially the implicitly or explicitly chosen perspective(s) or ontology, if formally represented.

In this paper, we propose six basic principles to guide the overall data model design. Further, we also propose algebra with a set of basic operators to support data operations by their meanings, not by their logical structure. The representation is based on point-free geometry such that any meaning is represented as an area in semantic space, which can be decomposed or aggregated in different ways concurrently. We use W(hat)-tags to attach on shadows for their meanings, and E(quivalence)-tags to recognize what meanings can be treated as the same. We use enterprise customer data integration as an example to illustrate the data model design and operation principles.

## Categories and Subject Descriptors
H.2.5 [Heterogeneous Databases]

* The author can be reached at jasonthliu@acm.org

## Keywords
Data integration, information ecosystems, shadows, meanings, mental entity, semantic heterogeneity, tags,

## 1. Introduction

Data integration is essential for most enterprise systems that rely on data from multiple sources. Lenzerini defined data integration as to combine data residing at different sources in order to provide users with a unified view of these data [1]. Specifically, the goal of Enterprise Information Integration (EII) is to provide a uniform access to multiple data sources without having to load the data into a central place [2]. Consumes about 40% of IT budget, EII is cited as the biggest and most expensive challenge in [3], where Bernstein and Haas provide an overview for related difficulties and technologies.

In this paper, our main concern is in the conceptual modeling level, especially for data integration in large-scale information ecosystems. The context is the design activities for data modeling where a local (downstream) database needs to re-use existing data provided by upstream systems, and we focus on the specific issue that there are multiple sources for the same subject matter (e.g. customer). These data sources are designed for satisfying different business requirements based on different understanding, different perspectives or even different ontology implicitly.

Therefore, we use the term *data integration* in a more generalized sense in this paper, for *how we can design data models used in the context of information ecosystems to fulfill its local business functionality through data provided by multiple sources.* For example, ordering systems in an enterprise need to access customer data that exist in billing, marketing, or sales databases. However, these different sources have their own perspectives about the same subject matter. To develop an isolated model of customer data for ordering only is not an acceptable solution since it will not improve the overall performance for the enterprise, and isolated customer data will create more chances of inconsistencies or errors.

By data model design, there are two levels of meanings here:

(1) The generic principles to design the data models for specific applications, for example, to model enterprise customer data in a marketing database, or used in a data warehouse for business intelligence.

(2) The design of a generic data model like the Relational Model with algebra to serve as the foundation for data operations.

In general, the difficulties of data integration are due to heterogeneity at physical level, logical (structural) level, and semantics level. Since semantic heterogeneity is usually hidden behind structural heterogeneity, it is the most difficult one to identify, and the resolution by mapping between semantic heterogeneous data elements usually requires design decisions made by human data integrators (one of the reasons for why data integration still relies on human labors since the beginning of database era). Without a commonly accepted comprehensive theory to guide such design decisions, data integrators rely on their own experiences and personal preferences like performing an art in current practice.

The trends of related research are more in making such manual process automatic or semi-automatic (e.g. see the survey for matching schemas automatically in [4, 5]), in order to reduce the dependence of human work. Researchers usually start developing and testing their solutions from relative small number of data sources, and push for scale up later. Significant progresses were made in design software tools at application level to review schema and identify potential mapping candidates.

However, for large-scale information ecosystems, data integration faces different kinds of challenges. For example, it is not uncommon for an enterprise to have 500 or more different data sources that provide similar, overlapped, inconsistent or even conflicting customer data. First challenge is due to the number of schema, which indicates the scale of complexities for mapping between different logical representations. Second challenge is due to the dynamic nature for subject matters themselves, that millions of enterprise customers are in endless M&A activities. Third challenge is due to the different business requirements for each individual data models, that each one has its procedures to decide when to update their data due to M&A results. When there is a need to integrate individual information ecosystems due to merger of the service providers, each merged department like the integrated billing or ordering usually has to perform their own integration, since enterprise-wide enterprise customer data integration for all departments is slow or even not possible to happen.

Under these surface issues, the difficulties of semantic heterogeneity can be summarized as the two basic characteristics:

(i) There exist <u>different representations</u> for the <u>same meaning</u>,

(ii) There exist <u>different meanings</u> for the <u>same representation</u>.

Since Relational Databases are the dominant backbone for enterprise systems, we will focus on Relational Model to make comparison with in this paper. Different representations can be interpreted as different schema, which is explicitly represented in relational database systems. But semantics is not; it is something out of the scope of Relational Model. Although explicitness was one of the original objectives of the Relational Model (as summarized by Date: "*the meaning of the data should be as obvious and explicit as possible*" see p.68 in [6]), there is no mechanism to enforce meanings to be explicitly represented. What is even worse is that there is no mechanism to prevent multiple/heterogeneous meanings be represented by the same data values or by the same schema.

Hence, people think Relational Model as unintepreted, for example, Neven described for Relational Model that "*A database is an uninterpreted finite relational structure*" [7]. When compared with the design of ontology, researchers further explained, "*Database schemas often do not provide explicit semantics for their data. Semantics is usually specified explicitly at design-time, and frequently is not becoming a part of a database specification, therefore it is not available*" [5, 8].

Therefore, we believe the current difficulties encountered during data integration are the natural results due to this weakness. Besides, it is well know that data integration methodologies are constrained by the data models they can support [9]. We believe

**Content**

the first step for the solution to overcome difficulties of semantic heterogeneity is to design a new data model that can explicit represent meanings (not necessarily in a formal or complete way), and explicitly identify what meanings are different versus what meanings are the same based on subjective criteria of similarity, commonality, or supporting evidences.

One level deeper to the logic foundation, the challenge is about multiple versions of the truth. The traditional data models Relational Model has the implicit assumption to support single version of the truth (i.e. functional dependency), and its logical foundation is based on predicate logic, which relies on the semantics of the languages to defines the truth of each sentence. When each perspective is supported by experts and its user communities who establish their business operations on top of their views, data integration in information systems unavoidably involves with inconsistencies or even conflicts due to different perspectives. Bring them into an integrated data model requires the data model to support multiple versions of the truth, which simply makes the underlying logic foundation no longer valid. News reports about the same subject matter on internet can be an extra example. First, there exist numerous sources, and second, the contents are not always consistent due to different ways to collect data, and third, the perspectives to interpret meanings behind data chosen by each author can be very different or in conflicts.

Under such context, our main concern is how we should design the data model for the local database. Traditional data model design approaches do not include these considerations, nor do them consider the characteristics of information ecosystems that data flow from source (upstream) systems to downstream systems.

Here, we use the term *source systems* to indicate the origin of the data (i.e. the original system that models the subjects and provides such data to other systems to use), *downstream systems* as the systems that receive, use, and extend data directly or indirectly from source systems for their own application purposes. The term *upstream systems* indicate the relative data provider role for the systems that share their data, no matter they may or may not be the source systems of the data.

If inconsistencies or conflicts exist in reality and cannot be resolved, what should the data integrator do? If a uniform access mechanism is established to access all of these data, can we say these data are really integrated together? In other words, we have to ask the question that resolving structural heterogeneity cannot make the data really integrated; on the opposite, the uniform logical representation and access mechanism make the semantic heterogeneity problem hidden behind the surface. As a result, semantic heterogeneous data stored in the same table[1] or the same column such that through by table names, column names, or domain values alone we cannot uniquely recognize the meaning of the data. When a table or a column is like a data container holding data with uncertain meanings, the power of Relational Algebra is lost as complex criteria is required in order to perform data operations efficiently for the desired semantics.

In this paper, we propose a different direction: we should bring such inconsistencies or conflicts into the scope of the data model with the goal to help users recognize them, and to manage them for reaching their functional objectives for business operations. We use data space to model different logical representations, and semantic space to model different versus the same meanings. Further, we need a set of operators that can directly perform in semantic space, not just in data space with complex interpretations.

In this paper we propose a **Shadow Theory** to serve as the philosophical foundation to address the needs to model semantic heterogeneity. The concept can be traced back in Plato's Allegory of the Cave over 2000 years ago, and the basic idea is that whatever data we can observe and record into database are just shadows. Shadows are generated as the results of projection process from the subject matter(s) in the real world to wall-like surfaces of the database system requirements. The properties of shadows can be classified as the following three categories:

C1. Properties due to the characteristics of the **subject matter**.

C2. Properties due to the characteristics of **wall-like surface**.

C3. Properties due to the **projection process**.

Since a database is designed to satisfy specific business needs, only limited shadow properties can be included in the scope. Therefore, the design process of data modeling is actually like the projection process that subjectively filter and transform observable of the subject matter into shadows. The chosen perspective, level of abstraction, and ontology (implicitly) play the key roles in such projection process. Therefore, properties of shadows are not only due to characteristics of the subject matter, but rather a combination of all these three categories.

Based on Shadow Theory, the basic design principles can be summarized as the following six principles:

1. What we can observe and store in database are only shadows.

2. The meanings of shadows exist as mental entities only in the viewers' cognitive structures. We can use W(hat)-tags, short as W-tags, to uniquely anchor with mental entities with explicit representation for their perspectives. A shadow can be attached with multiple W-tags, and each instance of the W-tags indicates the specific meaning of the shadow based on the associated perspective.

3. Semantic Heterogeneity is due to differences among these mental entities, as well as how shadows are projected from the subject matters onto wall-like surface for system requirements.

4. No matter how different shadows may be, if they represent the same meaning, we can treat them as the same. E(quivalence)-tag, short as E-tags, are designed for such purpose, it can be attached to pairs of W-tags to indicate the two meanings as mental entities are semantically equivalent.

5. Meaningful data integration should be performed by semantic equivalence, not constrained by their different logical representations. However, it is a subjective decision for data integrators to decide the criteria for which W-tags can be treated as the same with supporting evidences. The data model should provide basic mechanism to perform operations based on semantic equivalence.

---

[1] Although there exist logical differences between a table and a relation in Relation Model (that duplicate rows are allowed in tables, but not in a relation), we use the two terms for the same meaning in this paper since our focus is in their generic sense of the logical structure. To avoid confusions about the term relation under different context, we will use the term table to indicate we mean the logical structure.

|  | Representation for _different meanings_ | Representation for _the same meaning_ |
|---|---|---|
| **Shadow Theory** | Only one standard mechanism: different **W-tag** instances represent different meanings as mental entities existing in viewers' cognitive structures. | Only one standard mechanism: an **E-tag** between two instances of W-tag indicates that the two meanings as mental entities can be treated as the same semantically, no matter how different the shadows can be in data values or logical structures. |
| **Relational Model** | Any part in two data models that are different may or may not represent different meanings, e.g. different data values, domain names, table names, attribute names, or different logical structure. | To represent the same meaning, and to be recognized as the same by the operators in Relational Algebra, it must be (i) **same data values** in the **same domain** (as extended data type), or (ii) the **same set of attributes** with the **same data values and domains** (for equivalence of two tuples). |

**Figure 1** Comparison for representation mechanism for the same or different meanings.

6. In addition, usage of the integrated data requires users to specify their desired perspective first, and evaluate W-tags with existing E-tags (may be established by others) to decide whether such E-tags can function as a bridge to cross the boundaries of individual perspectives in order to achieve specific functional objectives.

In this paper, we will use a scenario of enterprise customer data integration to illustrate related issues and how our proposal can better support data operations under semantic heterogeneity. Specifically, the objective of our proposed algebra is to support operations by meanings without constraints due to logical representations.

Compared with Relational Model, we made a very different tradeoff for representing what are the same versus what are different, as illustrated in Figure 1.

(1) **Representation of different meaning**:

Shadow Theory: The only standard mechanism for representing different meanings is by different W-tags. That is, different W-tag instances represent different meanings as mental entities existing in viewers' cognitive structures.

Relational Model: There is no standard mechanism for representing different meanings. Therefore, any part in two data models that are different may or may not indicate whether they represent different meanings. For example, different data values, domain names, table names, attribute names, or schema structures in different data models may or may not have different meanings.

(2) **Representation of the same meaning**:

Shadow Theory: An E-tag between a pair of W-tags is the only standard mechanism. That is, an E-tag between two instances of W-tag indicates that the two meanings as mental entities can be treated as the same semantically, no matter how different their shadows can be in data values or logical structures.

Relational Model: To represent the same meaning, and to be recognized as the same by the operators in Relational Algebra, it must be either (i) the same data values in the same domain (as extended data type)[2], or (ii) the same set of attributes with the same data values and domains (for equivalence of two tuples). That is, if different users want to use Relational Model to represent the same meanings, and expect their representations to be treated as the same by the Relational Algebra, the only native solution is to use the same data values, the same domain, and the same logical structure for each tuple. However, it is still not guaranteed that the same attribute values in the same table must represent the same meanings for all tuples within the same table (especially when overloaded with semantic heterogeneous data). Further, the notion of unique keys as attribute values to uniquely identify tuples can only represent the same meanings for semantic homogeneous data. In other words, the same meaning represented by different tuples cannot be treated as equivalent in Relational Algebra, as a result of semantic heterogeneity.

Shadow Theory is just the first step in the overall solution to manage semantic heterogeneity and to guide overall data integration process. On top of the new proposed data model, we can bring existing data integration algorithms and techniques into the overall picture. Our efforts echo the big direction Hass described in [11] that "*Experience with a variety of integration projects suggests that we need a broader framework, perhaps even a **theory**, which explicitly takes into account requirements on the result of the integration, and considers the entire end-to-end integration process*".

A mathematical theory is abstract and acquires meanings only through interpretation, and a mathematical model is a mathematical theory endowed with interpretation; further, a theory may have many models and be interpreted in many ways [12, 13]. As for a theory in physics, Stephen Hawking described it in a simple way as "*… that a theory is just a model of the universe, or a restricted parts of it, and a set of rules that relate quantities in the model to observations that we make it. It exists only in our minds and does not have any other reality (whatever that might mean).*" (see p.10 in [14])

---

[2] We understand that researchers have disagreements on whether there should have strong domain (type) checking when compare data values in Relational Model. For example, Date proposed that "*two values can be compared for equality in the relational model only if they come from the same domain*" (see p26 in [6]). Here we refer to Codd's weaker definition (p.47, [10]) that DBMS only need to check whether the basic types (e.g. INTEGER, BOOLEAN, CHAR) are the same when perform comparing values.

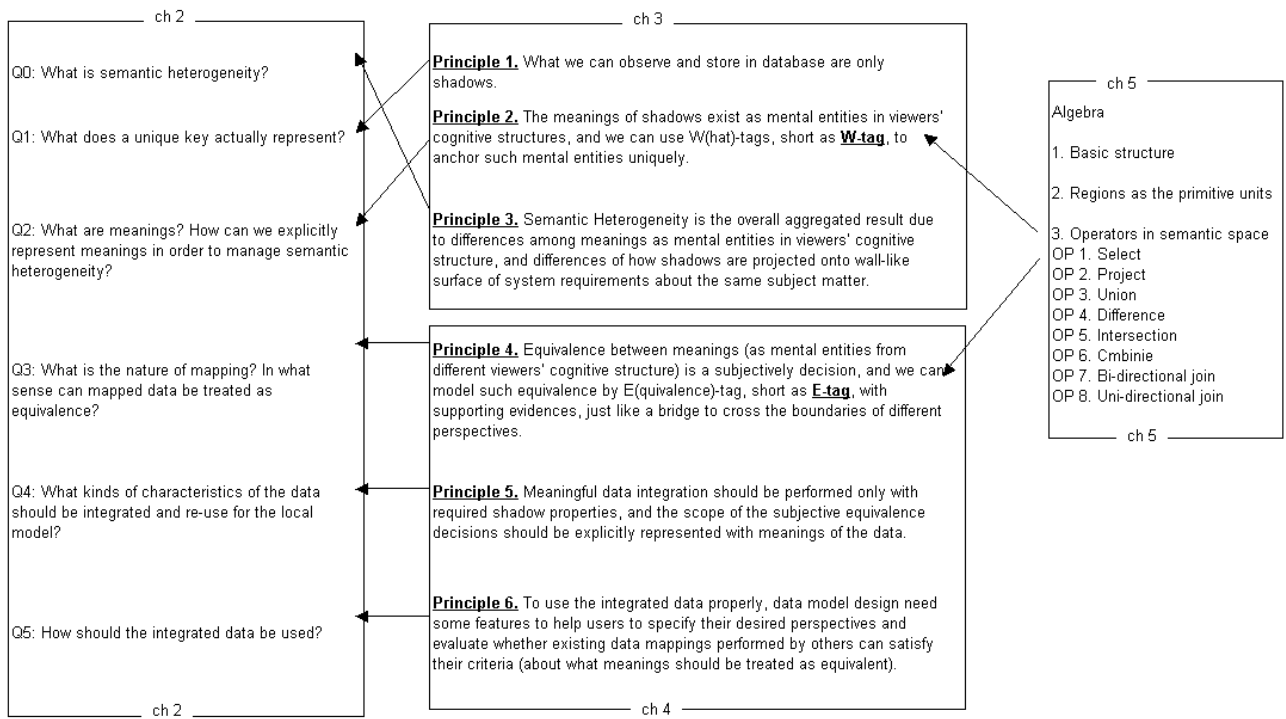

**Figure 2 Organization of this paper.**

Shadow Theory is not a mathematical theory as we need to include meanings themselves as the subject matter, and it is also not a theory for physics since the reality we want to investigate must include interpretations subjectively made by specific viewers from certain chosen perspective(s) and ontology. The subject matters that Shadow Theory must include are

(1) **Data**, treated as shadows, stored in some physical ways, generated by some methods automatically or through human to provide. This also includes logical models implemented in database systems, like schema in Relational Model, which govern how data is stored or accessed.

(2) **Semantics**, meanings as mental entities existing only in specific viewers' cognitive structures. This also includes the conceptual models that motivate the design of logical models implemented in database systems.

(3) **Reality**, existing in human business operations and in the environments of information ecosystems where, in addition to inconsistencies or conflicts may happen between different business operations, different people may have different ways to represent the same meaning and different people may recognize different meanings for the same representation.

Since all of these are results due to human activities, the center of Shadow Theory is really try to model human behaviors for how they perform business operations based on meanings of chosen perspectives/ontologies, subjectively collect and transform data, design logical models to manage data, and want to integrate their data plus associated meanings together.

Figure 2 illustrates the organization structure of this paper. In section 2, we will raise a sequence of questions encountered during data integration in information ecosystems, including:

Q0: What is semantic heterogeneity?

Q1: What meaning does a unique key actually represent?

Q2: What are meanings? How can we explicitly represent meanings in order to manage semantic heterogeneity?

Q3: What is the nature of mapping? In what sense can mapped data be treated as equivalence?

Q4: What kinds of characteristics of the data should be integrated and re-use for the local model?

Q5: How should the integrated data be used?

In section 3, we will propose Shadow Theory as the philosophical foundation, and discuss the first three principles to answer the first three questions. In section 4, we will focus on data integration and propose three principles to answer the rest questions. In section 5, we will discuss the basic structure and properties of point-free geometry that we need for supporting operations in semantic space. We will then discuss each of the basic operators for involved issues and comparisons with those operators in Relational Algebra. In section 6, we will discuss the logical foundation to support multiple version of the truth, followed by comparisons with Relational Model and existing data integration approaches. There are still many unsolved issues need future research, especially for data integration that involves many different fields. Section 7 concludes our proposal in this paper.

## 2. Enterprise Customer Data Integration

In general, the difficulties of data integration in information ecosystems are due to heterogeneity at physical level, logical (structural) level, and semantics level. Physical heterogeneity is obvious due to different hardware and software systems co-exist for various reasons. Structural heterogeneity is due to different approaches to model the logical structures of the subject matter, and it is usually observed as differences of schema (since Relational Databases are the dominate backbone in most enterprises). As for semantic heterogeneity, it is hard to be recognized since it is usually hidden behind schema, blended with structural heterogeneity. It is recognized as the most difficult to be resolved since the beginning of database era [9, 15, 11, 16].

In this paper, we will use the example of enterprise customer data integration to illustrate various issues due to semantic heterogeneity and due to the nature of information ecosystems. In general, the need to integrate enterprise customer data can be summarized as to combine customer data modeled by different service providers which are merged together as a single one (i.e. each kind of customer data once dominated his own information ecosystem before mergers). The objectives for such integration include supporting various enterprise business functionalities like billing, ordering, repair, and son on. The customers are enterprises that subscribe certain services from one of the service providers, and they themselves are constantly involved in M&A activities. We will use Scenario 1 to further illustrate the context of the requirements.

**Scenario 1.** Figure 3 shows three kinds of Enterprise Customer IDentifier (short as ECID): legal entity-based, location-based, and contract-based. Before merger, let's assume service provider B, G, and W have established consistent enterprise customer data within their own information ecosystems. Due to the different nature of their products, service provider B, G, and W choose to model enterprise customers from the perspectives of legal entities (for clear financial responsibility), locations (for precise location where service is provided at), and contracts (for explicit service types and the business terms). After merger, there is a need for the federated service provider to integrate these different ECID in order to support efficient business operations. In the following sections, we will discuss related issues encountered during data integration. The first one is about the notion of unique keys under semantic heterogeneity and the related data model design conventions.

## 2.1 The notion of unique keys and semantic heterogeneity

Before we start, we need discuss the definition of semantic heterogeneity in the context of data integration for information ecosystems.

**Question 0:** What is semantic heterogeneity?

Researchers for data integration have investigated semantic heterogeneity for several decades. In an easy and straightforward way, Halevy described semantic heterogeneity as the following "*When database schemas for the same domain are developed by independent parties, they will almost always be quite different from each other. These differences are referred to as semantic heterogeneity*" [15]. Similarly, semantic heterogeneity is described in [16] as "*… information on a common object …does

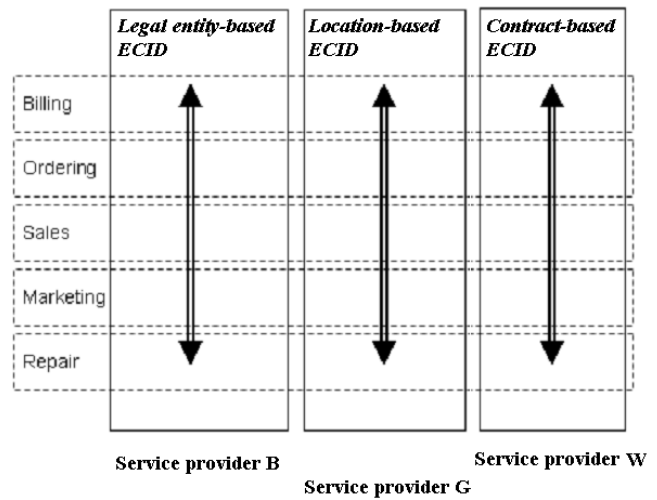

**Figure 3** A scenario that service provider B, G, and W merged together as a single one. Their existing models for Enterprise Customer IDentifier (short as ECID) used in individual departments need to be integrated across business organizations like billing, ordering, sales, marketing, and repair.

*not conform to a common representation … makes query processing a big challenge, because there is no structure on which to base indexing decisions and query execution strategies.*" Specific for relational Model, Brodie and Liu suggested the following criteria to test about semantic heterogeneous data: "*Relational data descriptions that are 1) verified by an authoritative expert to be the 2) same entity (considering set inclusion) and for which there is 3) no simple, complete entity mapping between the relational schemas for the data descriptions, e.g., that 4) cannot be defined in relational expressions and then the relational data descriptions are said to be semantically heterogeneous*" [17].

These definitions provide us a clear description about the nature of semantic heterogeneity. However, they do not provide a way to help us to model semantic heterogeneity. If we want to overcome related difficulties, we need a way to describe the characteristics of semantic heterogeneity such that we can include its behaviors in the model.

For information ecosystems, naturally there are many different data modelers perform data modeling for the same subject matter independently. Although these models may impact each other implicitly or explicitly, there is no single universal and centralized control for how modeling tasks should be performed. Hence, even there exist domain experts, the chances are that these domain experts may have different opinions about the same subject matter and they cannot agree with each other.

Under such context, we can describe semantic heterogeneity with the following two basic properties:

**Basic Properties of Semantic Heterogeneity**: There exist

(i) Different representations for the same meaning.

(ii) Different meanings for the same representation.

To further illustrate the characteristic that **there exist different representations** (data instance and schema) **for the same**

**meaning** in the ECID example, and unique keys are used in these different representations to model the same meaning, we will use Figure 4 to show the comparisons of different ECID. Location-based ECID relies on the uniqueness of locations where services are provided to identify enterprise customers (i.e. each unique location is assigned with a unique ECID). The advantage is that this approach can avoid the dynamic nature of enterprise organizations; in other words, as long as services are provided at the same physical locations, the unique ECID remain the same, only attributes need to be adjusted when M&A activities happen. For contract-based ECID, the advantage is due to the unique association with contracts with products, service terms, and customer information explicitly documented. However, since contracts are limited to specific products and service areas with specific refresh cycles, the same enterprise customer may hold multiple contracts, and old contracts may not be refreshed immediately when M&A activities happen. Legal entity-based ECID follows the uniqueness of legal entities registered with government. The advantages include clear financial responsibilities and third parties of business vendors can maintain and provide such information periodically. However, the dynamic nature of M&A is the major disadvantages such that the "same" enterprise customer may be assigned with different unique keys due to its legal status changes.

|  | Location-based | Legal entity-based | Contract-based |
|---|---|---|---|
| **Uniqueness:** | No change as long as location the same! | Change when there are M&A or legal activities! | No change as long as the contract remains |
| **Refresh rate:** | By request | Every 30 days | Between 1 to 5 yr, or by request |
| **Meaning differences:** | Implies the existence of services at the specific location | Implies the legal responsibilities of any services used | Implies the special rates for special products |
| **Application needs:** | Existing information ecosystems rely on the "location" meaning for service, invoice, repair, …etc | Existing information ecosystems rely on the "legal" meaning to calculate invoices, profits…etc. | Existing information ecosystems rely on the "contract" meaning to calculate price, invoices,…etc. |

**Figure 4** Characteristics of different kinds of ECID

Obviously, there may not exist 1:1 mappings between these different kinds of ECID. A legal entity may have multiple locations, and a single location may host multiple legal entities. So do their relations with contracts. In addition to these three perspectives, there many other ways to model enterprise customers, e.g. by accounts, by invoices, based on business process or sales team structures to divide into smaller parts, or by different functionalities [18] [19] [20] [21]. Obviously, there exist inconsistencies or even conflicts between these different modeling approaches. Even within the enterprises themselves, they may have different internal view for how they identify different sub-organizations. For example, from the perspective of billing department of an enterprise customer *X* to view itself, there are 5 different sub-organizations due to financial responsibilities, while from the perspective of repair department of *X*, there may be 20 different sub-organizations based on internal repair process, and from the perspective of ordering department of *X*, there may 3 different sub-organizations due to services or products required.

For subject matter like enterprise customers, there is no single modeling perspective that can satisfy any kind of requirements. Therefore, without the support of a single universal truth, any attempt to build a global representation with different kinds of ECID integrated together is a difficult challenge. The root issue can be traced to the foundation of data modeling about unique keys.

**Convention # 1** Relying on unique keys as data values stored in databases to model the subject matters.

Traditionally, we rely on the notion of unique keys as data values stored in databases to model subject matters. When the subject matters are fuzzy like enterprise customers, data modelers will try to identify them by some kind of unique characteristics based on a specific perspective. Such practices contribute to the difficulties of data integration.

Researchers have known the limitation of unique key since the beginning of database era. For example, Codd pointed out that the difficulties of using unique keys include (1) unique keys are subject to change, (2) different unique key may denote the same *thing*, (3) data of the *thing* may exist before the unique key exist, or after the unique key ceases (p409, [22]). Further, Kent pointed out that the required properties for applying unique key successfully are: (1) **Immutable**: a unique key should be immutable such that an entity is represented by the same key value through its life time; (2) **Singular**: a unique key should be singular that two keys should not represent the same objects [23].

In Scenario 1 of ECID example, we do know that ECID is not immutable due to the dynamic nature of enterprises, which are constantly evolving through M&A in the real world. And we also know that ECID are not singular as there are multiple ways to represent the same enterprises (and their sub-organizations). Therefore, using unique keys to model enterprise customers is a compromised data model decision, and it is not within the original assumptions for how Relational Model is designed for. Chances are, that during the continuous expansion for applying Relational Model in serving business needs, data modelers may not know it is out of its theoretic boundaries. Then, the question is, what is the boundary? If starting with the most basic one, when a unique key cannot uniquely represent the subject matter, then we have to raise the question:

**Question 1:** What does a unique key actually represent?

In the practice of data integration, it is common that all kinds of different unique keys about the same or similar subject matter are loaded into a single table with attributes to indicate their differences (sources, types…and so on). Under such practices, this question is even more interesting, since the same thing in reality may be represented by different (kinds of) unique keys in this single table. After violating the properties of singularity, our usage of unique key is in a grey area whose interpretation is subject to different users.

Of course, people can use a single table to store semantic heterogeneous data. The cost for doing so is at the complexities for identifying the meaning(s) for each row, i.e. by reading the values of certain attributes to decide which row represents what kind of meanings. In this way, a table, or more precisely, a relation in Relational Model, is like a data container that contains different meanings, and we do not know exactly what kind of meaning until we decode data values. In order to understand the meaning represented by a tuple, sometimes it may even need to check a combination of several attribute values within the same table or across different tables.

For example, if we collect the three different kinds of ECID in Scenario 1 and load them into a single table illustrated in Figure 5. The first attribute *ECID* is for the unique keys from different source systems, and the second attribute *Type* to indicate which kind of ECID it is. The value *L* is for location-based ECID, *G* is for legal entity-based ECID, and *C* is for contract-based ECID. To avoid potential collision among unique keys, we only set the combination of *ECID* and *Type* to be unique within this table. Here is how we can violate the **singularity** criteria that the same enterprise customer may have 5 unique keys of legal entity-based ECID, 20 unique keys for location-based ECID, and 7 contract-based EICD.

Further more, the exact meaning for the attribute *status* cannot be determined by its attribute name or domain value, which is the only mechanism for Relational Model to represent meanings of the data[3]. That is, the same value *A* can represent different meanings due to different lifecycles of the customer data required in the original data sources:

(i) For location-based ECID (*Type = L*), it means active service at the specific location for this customer.

(ii) For legal entity-based ECID (*Type = G*), it means active legal status, but it does not indicate where the services for this customer is active or not.

(iii) For contract-based ECID (*Type = C*), it means active contract status. The customer information is what was recorded in the contract, and it may not reflect current reality due to organizational changes or M&A activities.

This example illustrates the second characteristics of semantic heterogeneity that **there exist different meanings for the same representation** (i.e. the same data value in the same attribute for the overloaded schema). The root of these different meanings is due to the different lifecycles for the different kinds of ECID in their original business semantics. For example, legal entity-based ECID can have two different statuses: *A* is for active legal entities and *I* implies it is no longer active. However, for contract-based ECID, there are three statuses for its lifecycle: *A* stands for active contract status, *P* stands for pending status, and *I* stands for inactive contract status. For location-based ECID, there are four cycles: *A* represents active services at the specific location, *D* represents that the services are disconnected and maybe reconnected later, *I* represents the services are connected but not in use, and *F* represents that the customer was at final status like a logical deletion.

---

[3] For attribute names, Codd explained that the reason to have distinct column name other than domain names is that "… *such a name is intended to convey to users some aspect of the intended meaning of the column* …" (see p.3 in [10]). As for different domains, so called extended data type, are "…*intended to capture some of the meaning of the data*" such that the system can tract the difference when the same basic data types (e.g. INTEGER, CHARACTER, BOOLEAN) used to represent semantically distinguishable types of real-world objects or properties (see p.43 in [10]).

| ECID | Type | Customer name | Status |
|------|------|---------------|--------|
| 765125 | L | XYZ Inc | A |
| 0065120 | G | XYZ | A |
| XZ87065 | C | XYZ Group | A |

**Figure 5.** Semantic heterogeneity may be hidden behind a common logical structure. The same domain value 'A' for '*Status*' attribute can represent different meanings: For legal-based ECID (Type = L), it means *active service status* (at the specific location); for legal entity-based ECID (Type = G), it means *active legal status* (but it does not indicate with active service); for contract-based ECID (Type = C), it means *active contract status* (but it does not indicate if the service of specific accounts are still active).

For information ecosystems, there are extra issues for unique keys when data flows from upstream systems into downstream systems by replication or ETL process. Not only the desired **immutable** property may be lost, the meanings represented by the unique keys also evolve with implicit semantic dependencies.

First, we will explain the **ghost problem**. In theory, the same unique keys should have the same behaviors in upstream systems as well as in downstream systems. In practice, it is often not in this case due to different local business requirements, and upstream systems are usually not designed for the purpose of downstream systems. For example, let's assume a marketing database provides its legal entity-based ECID to downstream systems like portal for external users or a data warehouse for internal business metrics for orders. The business requirements for the marketing database only ask to keep historical records for 2 years, while the downstream systems may be required to keep for 7 years. Hence, when a unique key is physically deleted in the upstream systems, downstream systems cannot perform the same. In addition to different business requirements, another reason that prevents downstream systems to delete such data is due to there the existence of other references, for example, ordering records which has to-be-deleted ECID as foreign keys, which cannot be deleted. These remained ECID are like ghosts which existing only in the downstream systems, and we have problems about the meanings they represent since their existence in the upstream systems is gone.

Similarly, there is **unique key re-use problem** that the same unique key may be re-used again for representing a different *thing* in the upstream systems. Again, in downstream systems, such unique keys may be locked due to being foreign keys for other purpose. Then we have the issue for the meanings of the unique key, it is difficult to recognize the semantic differences since it is just data values.

Another scenario may happen in information ecosystems is that upstream systems may try to "rollback" mistakes in which they accidentally delete unique keys. Since no transactions can be scale up to support large number of systems, the common solution is to re-insert of the same unique key. We call this problem as **rollback problem** since the headache is in the downstream systems; if the downstream systems cannot follow deletions at first place, then this kind of rollback by insertion behaves the same as unique key re-use problem. It is hard for downstream systems to understand if the same unique keys should or should not represent the same meaning as before.

These problems are due to the implicit semantic dependencies between upstream and downstream systems. One may think that we better to force upstream systems to manage their unique keys properly to avoid issues in downstream systems. Such ideas are in a direction opposite to traditional design approach that the data model design of downstream systems depends on the data models in upstream systems. We call such kind of dependence as **reverse system dependence**, since the needs to keep unique keys be interpreted properly in the whole information ecosystems require coordination between different system requirements, and upstream systems need have special policy for unique management.

However, this direction cannot go too far due to many reasons. First, upstream systems are usually not designed to support for the business interpretation in downstream systems. Schedule wise, upstream systems are usually designed before downstream systems, and hence it is hard to predict potential applications in non-existing downstream systems. Second, when the scale of the information ecosystems grows big, there is no way for upstream systems to consider business requirements for every one of their downstream systems. Third, the upstream systems may be out of the control of the enterprise; for example, ECID may be provided from a third party vendor 0r from government, who will not modify their design for the downstream systems. Fourth, there may be real reasons that the unique keys have to be changed. For example, when an enterprise customer split into two different legal entities, a new unique key must be generated and only one of the offspring can inherit the old unique key, or when two enterprises merge as a single legal entity, one of them may need to drop their unique key as the result.

With these observations, we hope to show readers about the problems of unique keys due to semantic heterogeneity and due to the nature of information ecosystems: unique keys may be neither singular nor immutable, they may be changed for reasons, their meanings are confusing. Applying traditional data modeling approaches under such context simply misuse the notion of unique keys, which can function properly only within a semantic homogeneous environment. The bottom of the issue is that the traditional data models like Relational Model are not designed to manage semantic heterogeneity. Semantic homogeneity as the hidden assumption for Relational Model can be observed from Codd's description in the context of explaining why duplicate rows are not allowed for the purpose of sharing among users: "*If, on the other hand, the data is shared or is likely to be shared sometime in the future, then all of the users of this data would have to agree on what it means for a row to be duplicated (perhaps many times over). In other words, the sharing of data requires the sharing of its meaning. In turn, **the sharing of meaning requires that there exist a single, simple, and explicit description of the meaning of every row in every relation** … *" (see p.6 in [10]) , also "*when hundreds, possibly thousands, of users share a common database, it is essential that they also share a common meaning for all of the data therein that they are authorized to access ...*" (see p.19 in [10]).

In the ECID integration example, the situation is that such data are shared among different groups of users and each group holds different perspective about the same subject matter (in order to satisfy their business operations). Relational Model has no mechanism to help managing inconsistencies or conflicts. The consequence of overloading semantic heterogeneity into relational schema is that the computing power of Relational Algebra is jeopardized, as reported in [17] that an extra criteria for Relational Database (in addition to the original 13 Codd's rules in [24]) should be "*The full power of relational technology applies to semantically homogeneous relational databases, and the more overloading with semantic heterogeneity, the less efficiency the models have.*"

On the other hand, we need to reconsider where the uniqueness comes from: should such uniqueness be constrained within a perspective or the underlying ontology? One major trend of research is to apply ontology to help data integration [4, 25, 26]. Indeed, if formally represented, the different perspectives for the data to-be-integrated can be viewed as different ontology supported by business semantics. For example, legal entity-based ECID can be viewed as an ontology system in which legal entity is the only primitive concept for enterprise customers.

However, ontology alignment itself is a big challenge with long history of debates (see [27] for recent survey). For example, the description in "GRDI 2020 Roadmap" has a good summary for the root issue of research across disciplines: "*...several difficulties can occur when sharing representations across different research communities of practice. They communicate using different 'jargon'. Much of this cannot be translated in a satisfactory way into terms used by other communities, since it reflects a different way of acting in the world (a different ontology and epistemology). There is the risk of interpreting representations in different ways caused by the loss of the interpretative context. This can lead to a phenomenon called 'ontological drift' as the intended meaning becomes distorted as the information object moves across semantic boundaries (semantic distortion)...*" (see p11 [28]).

In addition to the ontology alignment issue, another difficulty about ontology is due to the dynamics of evolving semantics when data flow from one system into another. The meanings of the unique keys are not static; rather they evolve slightly due to the subjective needs of the downstream systems. For example, the assumption of atomicity is a subjective local decision, not a universal consensus in information ecosystems. A unique key in upstream systems may be assumed to be atomic due to associated perspectives of business semantics, while it may need to be decomposed in sub-components with different meanings in downstream systems. Or, a unique key may need to be aggregated with extra information from different sources in order to establish meaningful uniqueness in downstream systems. Take the case of a telephone number: it may be an atomic data element in a upstream system to represent a customer, but it is decomposed into three elements as NPA-NXX-XXXX[4] in downstream systems where NPA and NXX are foreign keys referring to area code and central office. The same telephone may also be required to combine with a special customer identifier in order to distinguish different owners of the same telephone number during different periods of time.

---

[4] North American Numbering Plan, NPA is area code, NXX is Exchange code, and XXXX is station code.

Such dynamics in information ecosystems creates significant difficulties for applying formal ontology into data integration, especially for the approach that expects a global representation that can satisfy the needs for every system. Interested readers may check the details reported by Hepp about the difficulties for applying ontology in [29]. The issues can be summarized as (i) Ontology engineering lag versus conceptual dynamics, (ii) Resource consumption, (iii) Communication between creators and users, (iv) Incentive conflicts and network externalities, (v) Intellectual property rights.

Readers may notice that the involved issues for data integration in information ecosystem are actually beyond the scope of ontology, more into **epistemology** about the limitation of what different users know about the subject matter. Ontology focuses on recognizing the *thing* and relations of the *things*, while the dynamically evolved meanings of the same unique keys indicate the understanding about the subject matters is changing, due to different users' business needs as well as due to the constraints of their business knowledge. On the positive side, it is during data integration that we have the chance to recognize the similar *things* and asking the question like: What do the users in source system know about the *thing*? What are the differences of the knowledge about the same *thing* in data sources one compared with data source two? Should we take the advantage to include such information and represent explicitly in the integrated data model?

## 2.2 Schema and related issues

Next, we will look into schema related issues. Here we refer the term schema in generic sense[5] as a set of formulas describing the data structures in databases. The problem of schema integration is defined as the activity of integrating schemas of existing (or proposed database) into a global unified schema, and database integration is defined as to produce the global schema of a collection of database based on their local schema [9]. Further, the research for global schema can be classified as classified as (1) GAV, global-as-view, the global schema is expressed as a view in terms of the local schema from different sources, and (2) LAV, local-as-view, requires the global schema to be specified independently from the sources, and the relationships between the global schema and the sources are established by defining every source as a view by the global schema [1]. On the other side, researchers also report that the approach of establishing standards or global schema for integration problems has limited success only in domains where the incentives to agree on standards are very strong [15, 31].

Further, researchers have investigated the definability and computational complexity in data exchange, using source-to-target tuple-generating dependencies to specify data exchange between a relational source and a relational target [32, 33], or even for nested source-to-target dependencies [34]. They proposed to use second-order tuple-generating dependencies as the "right" language for composing schema mappings, since the source-to-target complexity of mapping composition may not be expressible (not closed or un-decidable) in First Order languages [32] [35-37].

With these observations, we can observe the following convention:

**Convention # 2** Schema as the logical structure of data is the center of data model design, and naturally people try to manipulate schema design for overcoming the difficulties encountered during data integration.

However, we observed several issues within information ecosystems. The fundamental one is that the schema of the upstream systems may not be available for data integration in downstream systems. Often, data integrators in the downstream systems can only access some kind of interface definition for data provided, and hence the local data model is designed based on such loose semantic dependency (between data in upstream and downstream systems, not between schemas). Readers may wonder why schema is not available for data integrators, the reasons could be:

(1) **Business or legal constraints**: Non-technical reasons may prevent source systems from sharing the details of their schema design to downstream systems.

(2) **Heterogeneous technologies**: Source systems may be implemented by different technologies that schema may exist or too difficult for data integrators to understand. For example, [38] reports that "*Relational Data Universe is less than 15% of the Digital Universe*" such that many data sources do not have the notion of schema.

(3) **To reduce impacts to downstream systems due to schema evolution in upstream systems**: When data flows to multiple downstream systems in a hierarchical way, if every system share its schema for downstream systems, then schema change in upstream systems may impact many downstream systems, even the whole information ecosystems. Hence, it is a practice that owners of the upstream systems prefer to hide their schema as internal design in order to reserve their freedom for the need of schema evolution later.

(4) **To avoid bottleneck of scale up for the whole information ecosystems:** In addition, if dependencies among schema are reserved and propagated to every downstream system, the whole information ecosystem has difficulties to scale up. That is, project managers and system architects of the information ecosystems prefer to only allow data flowing around without dependencies of schema, such that they can reduce the overall complexity and cost of each individual project.

Without schema, the uniqueness of unique keys is not protected. As a result, the unique keys provided from upstream systems are just data values with assumed uniqueness and there is only weak semantic dependency (depending on the interpretations determined by the data integrators of downstream systems). This makes the meanings of the data even easier to evolve, since it is up to the conceptual understandings of the data integrators in downstream systems to determine what their schema should be. Under such conditions, Figure 6 illustrates the dependency between upstream and downstream systems that only data flow with weak semantic dependency.

Even if all details of every schema in every upstream system are available for downstream systems to perform data integration,

---

[5] This includes a database schema as the overall design of the database, a relation schema, or "*a set of formulas in the database language that specify integrity constraints imposed on the database*" [30].

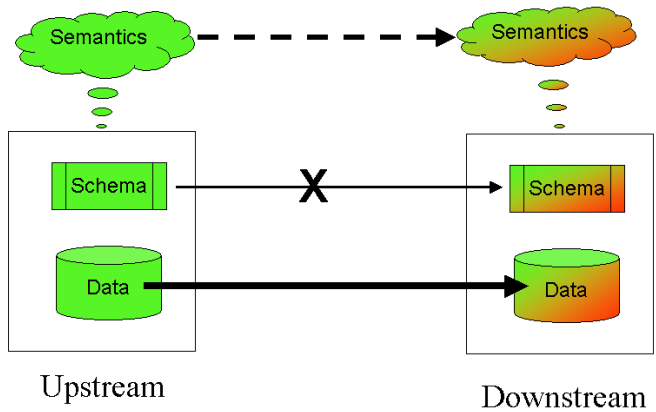

**Figure 6** When schema upstream systems is not available for downstream systems to perform data integration, there only exist weak semantic dependency (represented as dashed line ) between them.

there are two major issues. The first one is about how downstream systems can design their local schema to resolve structural heterogeneity for the inconsistent or even conflicting schema in upstream systems for data about the same subject matter.

Generalization by building IS-A relations is the general direction to modeling different kinds representations for the same subject matter. The more generic ones are in the higher level of the abstraction, while less generic ones are at lower levels of abstraction. To implement such relations by relational schema, similar to implement tabular representation of generalization for entity in ER model (see section 2.9.5 [39]), there approaches can be summarized as:

(1) Every kind of entity is implemented as a table with their own attributes, and every instance of the entity is stored in the table of its the lowest level with a foreign key referring to its associated unique key (representing this instance) in the table at higher-level generalization.

(2) If the generalization if disjoint and complete, that is, for every entity in the generalization, it can only be classified as one kind at any level, and every one in higher (or lower) levels must also exist in the lower (or higher) levels, then the implementation by schema can be done by only tables for every kind in the lowest level, with common columns due to attributes in the higher level(s) of generalization.

(3) A single table for the top-level kind of entity with attributes collected from every kind of the entity, and with unique key(s) formatted into a common form in order to identify their different representations (e.g. to identify their different sources or types, combined with the unique keys in their original models). Another mapping table is required to establish the IS-A relations through referring to pairs of the unique keys in the common form. In the ECID example in Scenario 1, Figure 5 illustrated this approach for the schema of the single table. The mapping table (not displayed) needs to be flexible in order to support different logical structure like flat, hierarchy (only one parent for each child), or graph (multiple parents for a child, and multiple child for a parent).

To scale up with more kinds of data collected for the same subject matter, the first and second approaches will end up with many different tables with different structures, and the cost of scale up is at the expense of structure of heterogeneity. For the third approach, it will end up with many different kinds of data stored in a single table, and the cost of scale up is at the expense of semantic heterogeneity. Due to the relative easier access and query programming, the third approaches are popular in data warehouse for large-scale data integration, in the form of star schema or snowflake schema with big fact tables with a set of dimension tables [40].

Although it is popular, resolving structural heterogeneity at the expense of semantic heterogeneity does trigger serious issue. That is, there exist **different representation** (as data instances) **for the same meaning**, one of the basic properties for semantic heterogeneity as we described in previous section. Dimension tables can help to filter and transform data in fact tables like to query at different levels of abstraction for semantic homogeneous data. However, for semantic heterogeneous data, the uniqueness of fact records within the same fact table is just for logical representation, not for their meanings in desired perspectives.

The root issue is that meanings are not explicitly represented by schema or data instances in Relational Model. Codd expects that "*…there exist a single, simple, and explicit description of the meaning of every row in every relation*." (see p6 in [10]) in order to share among users. Semantic features are mainly on domains, primary keys, and foreign keys, but Codd's noticed "*Domains, primary keys, and foreign keys are based on the meaning of the data. These features are quite inexpensive to implement properly, do not adversely affect performance, and are extremely important for users. However, most DBMS vendors have failed to support them, and many lecturers and consultants in relational database management have failed to see their importance.*" (p.vii) [10].

Note that explicit representation of meaning may not need to be in formal with complete information. It was one of the original objectives (not formal or complete representation) of the Relational Model: "*the meaning of the data should be as obvious and explicit as possible*" (see p.68 in [6]), and Date summarized Codd's basic underlying criteria as information principle: "*The entire information content of the database at any given time is represented in one and only one way: namely, as explicit values in attribute positions in tuples in relations.*" (p.295 in [6]).

However, this is just an expectation; there is no mechanism to force meanings to be explicitly represented in the model, and no mechanism to prevent multiple meanings from being represented by a single row in a relation. Rather, it is up to the users who can use Relational Model for representing *explicit* meanings if they design the model properly. Hence, people think Relational Model as unintepreted, for example, Neven described for Relational Model that "*A database is an uninterpreted finite relational structure*" [7]. When compared with the design of ontology, researchers further explained, "*Database schemas often do not provide explicit semantics for their data. Semantics is usually specified explicitly at design-time, and frequently is not becoming a part of a database specification, therefore it is not available*" [5, 8].

It is very easy (or even recommended) for users to overload semantic heterogeneity into domains, keys, and data values. For example, to resolve the known difficulties about schema evolution, overloading schema with different meanings is in the proposed solutions classified as in [41]: (1) reduce schema changes by adding the abilities into conceptual and data models the scope to accommodate modest changes to definition, (2) reuse current schema definition by changes of the application/wrapper for multiple extensional data, (3) accommodate schema changes seamlessly as much as possible.

Now, let's switch to a different angle to think about how meanings represented in Relational Model. Set Theory is the foundation for operations, and a relation is a set with tuples as its elements, with there is no order constraint among these tuples. Since a set is a collection of **distinct *things*** considered as a whole, the issue is **whether the notion of a set can be used to represent semantic heterogeneity**.

(1) Within semantic heterogeneous environments, different ontological primitive *units* are used due to their chosen perspectives or desired levels of abstraction. Hence, the same *things* in real world can be represented differently without easy mapping between these different representations. For the ECID example in Scenario 1, the same enterprise customer can be modeled as a set of legal entity-based ECID, a set of location-based ECID, or a set of contract-based ECID, but the problem is that there is no simple mapping among the atomic elements in these different sets.

(2) Further, even two meanings represent exact the same *thing*, there are unlimited different ways to decompose their meanings such that the relations among the atomic elements within the set cannot be represented by the notion of a set. For ECID example, such decomposition may have different kinds of logical structures, flat, hierarchy (i.e. every child can have only one parent), or graph (any child can have multiple parent, and a parent can have multiple children). Hence, if we use a set (of tuples) to represents an enterprise customer, we cannot explain how a legal entity-based ECID partially overlaps with a contract-based ECID in terms of what they represent.

(3) What is even harder is that the definition of being the same *thing* is a subjective judgment according to individual viewer's cognitive structure, and we can observe a spectrum for different criteria of similarity that people can use to make such decisions. However, a set that equals to another set must hold exactly the same elements within the set. Therefore, we cannot use the notion of a set to represent the property of semantic heterogeneity that there exist different representations for the same *thing* (since the different representations as sets cannot be equal if they have different kind of number of elements). In next section, we will further explore the issues for the notion of equivalence, its underlying subjective criteria of similarity, and performing mappings among different domains.

The result is, that a relation (in terms of Relational Model) is a set of tuples, but semantically we have issues to decide whether the meanings represented by different tuples are for **distinct *things*** or not under semantic heterogeneous environments. Readers can imagine that, if we load 500 different kinds of ECID into a single table, it is actually very difficult to calculate the answers for simple question example like how many enterprise customers exist in this table. It is in semantic homogeneous environment that Set Theory can efficient help to model the data instance in this table, since there is a common agreement on the ontological primitive units like each tuple (or a group of tuples) represent a unique enterprise customer.

Therefore, with all these issues we discussed in this section about schema, we may say that schema design, as the kernel for Relational Model, even it is available for data integration in downstream systems, is not designed to explicit represent meanings with prevention mechanism for users to overload with semantic heterogeneity, and the underlying logical foundation of Set Theory cannot be used to model semantic heterogeneity.

To overcome semantic heterogeneity, it will be very helpful if we can first represent meanings explicitly. But, we need to ask the following:

**Question 2: What are meanings? How can we explicitly represent meanings in order to manage semantic heterogeneity?**

This second question is one step deeper than the question we asked in section 2.1 about what meanings unique keys actually represent in the context of information ecosystems and semantic heterogeneity. Such questions are usually not included in the field of data model or data integration, as we usually assume we have common agreement about what meanings are.

In section 3.4 we will summarize different approaches to model meanings, and chose one that can satisfy our needs. On top of this choice, we will propose to use point-free geometry as our logical foundation to model operations for meanings. That is, we will treat a meaning as an area in semantic space, not as an atomic point. Before discussing the details of our proposal, we will continue digging into the requirements for data integrations, in next section, especially about the notion of equivalence for the purpose of mapping.

## 2.3   Equivalence, Similarity, and Data Integration

The notion of equivalence that is supported by Relational Algebra is only by the same data values from a common domain: "*The fundamental principle in the relational model is that all inter-relating is achieved by means of comparisons of values, whether these values identify objects in the real world or whether they indicate properties of those objects. **A pair of values may be meaningfully compared**, however, **if and only if these values are drawn from a common domain**.*" P.8 [10].

Therefore, Relational Model does not support semantic equivalence for different representations about the same subject matter in its basic operations. Since it is the nature of semantic heterogeneity for information ecosystems to have different models about the same *thing* in reality, data integrators are forced to rely on relations between attributes to model mappings. For the ECID example in Scenario 1, since these different kinds of ECID can not be compared semantically under Relational Model to determine if they are equivalent or not, mappings between different kinds of ECID are done by treating the unique keys of each kind of ECID as a domain, and different kinds of ECID unique keys are in different attributes for the mapping tables.

Let's now use Scenario 2 to explore further details about mappings in information ecosystems.

**Scenario 2. Figure 7** shows a potential sequence of mapping events performed by different departments along time axis. Due to legal constraints or local business requirements, each department needs to do their mappings under different criteria. For example, when a M&A plan between two enterprise customers is announced, marketing and sales departments may have the needs to immediately perform mapping for related different kinds of ECID in order to predict potential business opportunities across all products in all service providers.

However, commercial vendors who provide legal entity-based ECID data cannot perform such operations until M&A plans are legally valid. Once M&A is legally true, business vendors provide merged legal entity-based ECID such that existing unique keys may be revised[6].

Billing, service, and repair departments still cannot follow the unique keys changes to remap involved different kinds of ECID until official requests or authorization received from customers directly. As a result, mapping between ECID within the same perspective (old unique key, versus new unique key), and mappings between different kinds of ECID due to unique keys changes in their original models, are not performed at the same time across all different departments.

Unlike a single database that all users can be guaranteed to see consistent data through the help of transactions, inconsistencies or even conflicts in information ecosystems, e.g. inconsistent mappings between ECID, may takes months, years, or forever to be resolved.

Further, frequent M&A of the service providers themselves makes the ECID integration task even more complicated. Different kinds of ECID used within each service provider, like legal entity-based, location-based, or contract-based, creates difficulties for each involved departments that need to perform their own integration for their kernel business operations, e.g. integrated billing or integrated ordering across different products from service provider B, G, and W. With such dynamic environments, individual departments usually have to perform their own integration without waiting for the enterprise wide customer data integration, which may takes a long time or forever to happen.

Combining these factors, we can use Figure 8 to shows a 3D diagram to illustrate the complexities for ECID data integration. The three axis include:

---

[6] One example for the reason that unique identifier for ECID may need to change is that when two ECID merged as a single enterprise, this new merged legal entity may keep one of its existing unique identifier, or get a totally new one unique identifier (discard both existing ones). It is up to the vendor's data model policy.

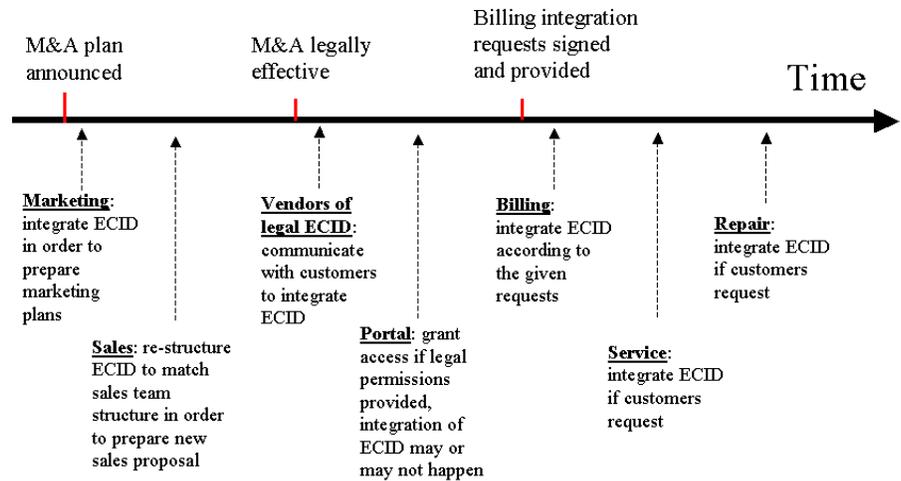

**Figure 7** Different departments have different legal and business requirements for how and when they can map different kinds of ECID due to unique keys changes triggered by M&A events.

(1) Data integration across different kinds of ECID perspectives holding by individual service providers that were used to be a single information ecosystem before merger,

(2) Data integration across business departments like billing, ordering, and repairs, whose modeling and mapping about ECID are based on local business requirements or legal constraints,

(3) Data integration due to M&A activities among enterprise customers.

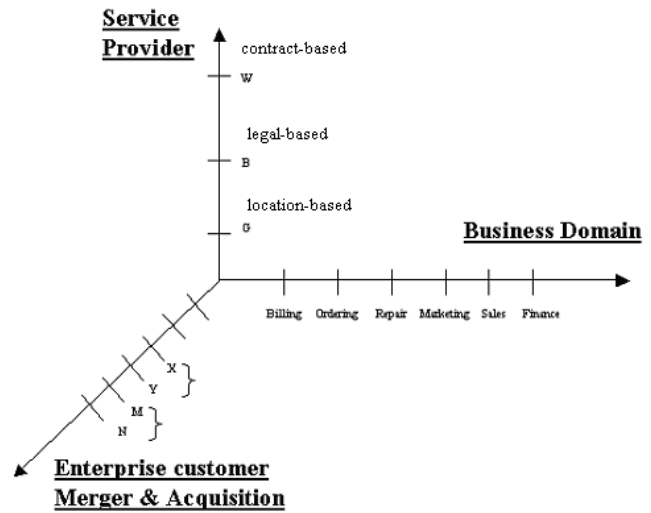

**Figure 8** Three dimensions for enterprise customer data integration: (1) across different kinds of ECID perspectives hold by service providers that were that were used to be a single information ecosystem before merger, (2) across different departments (e.g. billing, ordering, or repairs) whose modeling and mapping about ECID are based on local business requirements or legal constraints, (3) M&A activities among enterprise customers.

There are many issues involved in this scenario for mapping between different ECID as different domains. First, **single version of truth** (for the correctness of the mapping) **is not valid when each perspective holds only limited knowledge about the reality**. The idea of a global schema is really a confusing wish: if there exist a single version of the truth, then the job is to identify it. If people cannot identify what is the truth, then as long as people can coordinate their perspectives to a common one, then global schema can still work. Unfortunately, in many fields involved people have difficulties to reach consensus, or it is too slow and too costly to reach agreement due to scale of the information ecosystems.

To further explain this point, Figure 9 illustrates a Venn diagram about a specific enterprise customer *X*. The dashed circle represents the reality about the enterprise customer *X* in any possible perspectives. The three solid line circles represent what can be observed from the perspective of legal entity-based ECID (observed through the products of service provider B), location-based ECID (observed through the products of service provider G), or contract-based ECID (through the products of service provider W). Hence, mapping between any two kinds of ECID is only limited to what can be observed in both perspectives.

That is, the limited knowledge held by different perspectives may not bring consistent information about the subject matter, in addition to their different chosen ontology. We will not be able to have single version of the truth for mapping under such conditions, and single version of the truth can happen only when all of the different perspectives cover the same area in such Venn diagram. That is why we mentioned in section 2.1 that we observe issues about epistemology, in addition to ontology alignment problems.

Second, the traditional design principle based on **functional dependency is not valid for mapping between semantic heterogeneous data**. The reason is that functional dependency is a generalization of the notion of keys, and in section 2.1 we have explained that there are many issues about keys under semantic heterogeneity. Specifically, the issue happens when certain functional dependency that is valid within a perspective (i.e. semantic homogeneity) encounters another set of functional dependency that is valid under a different perspective: inconsistencies or conflicts between the two perspectives may not have resolution, and it indicates that we can not use the functional dependency across their boundaries. For the ECID example in Scenario 1, mappings from legal entity-based ECID to location-based ECID are difficult since each perspective has only limited knowledge about the subject matter, we simply cannot use either one kind of the keys to uniquely determine the mapping as illustrated in Figure 9.

We can use Figure 10 to further describe the issues due to different decomposition structures combined with different ontological primitive units, even if each perspective can have full truth about the enterprise customers. Assuming mappings between legal entity-based ECID and location-based ECID is *m*:*n*, i.e. one legal entity with *n* locations, and one location with *m* legal entities. However, such *m*:*n* mapping cannot really represent how their meanings overlapped together. Since each perspective has different ontological primitive units, and these "atomic" units may actually overlap partially about what they really represent. However, neither perspective has the capabilities to describe such partial overlapping due to lack of finer ontological units.

One impact of the above two issues is that the composition of mapping is not valid, i.e. the composition of individual mappings from *A* to *B* and from *B* to *C* may not equal to mapping from *A* to *C*. That is, in the ECID example, if we have mapping from legal entity-based ECID to location-based ECID, and mapping from location-based ECID to contract-based ECID, we still do not know the mapping from legal entity-based ECID to contract-based ECID.

With these difficulties, we need to raise the fundamental question about mapping:

Question 3: **What is the nature of mapping?** In what sense can mapped data be treated as equivalence? Does such equivalence uni-directional or bi-directional?

When data integrators map a legal entity-based ECID to a location-based ECID, do they mean a legal entity is equivalent to a location? Or they mean these ECID represent the same subject matter? What is the notion of "the same"? Can we subjectively treat different representations as the same? How about similarity? Are such mappings directional such that mapping from legal entity-based ECID to location-based ECID is different than the reverse direction?

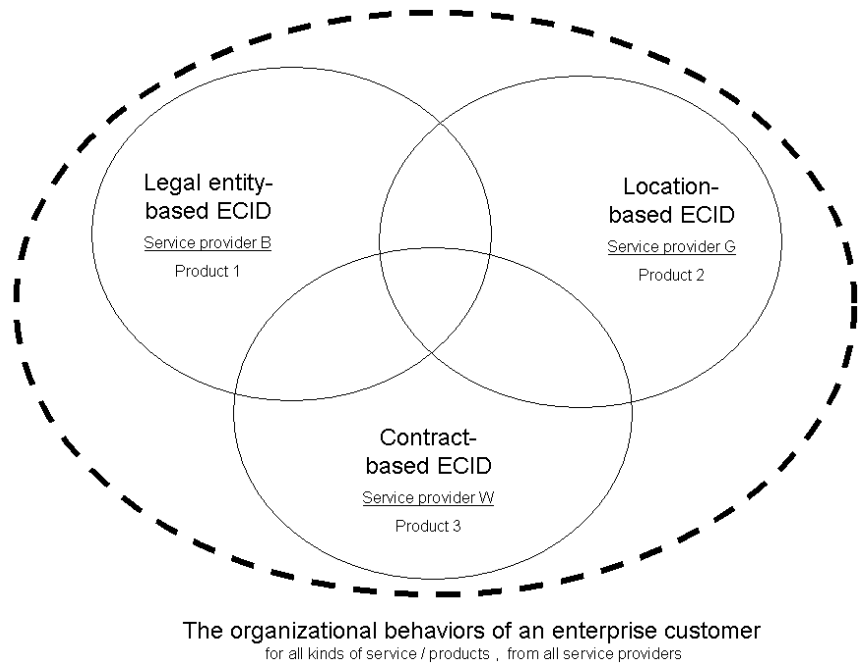

Figure 9. A Venn diagram illustrates about enterprise customer *X*. The dashed circle represents the reality about the enterprise customer X in any possible perspectives. The three solid line circles represent what can be observed from the perspective of legal entity-based ECID (observed through the products of service provider B), location-based ECID (observed through the products of service provider G), or contract-based ECID (through the products of service provider W).

Although in Relational Model that a pair of values may be meaningfully compared if and only if these values are drawn from a common domain, it is common in our daily lives that we make comparisons of different representations based on their meanings without worrying about whether they are same domain or not. Further, when we focus on specific criteria, we can ignore their difference (in their representations or in their meanings) and treat them as the same. The factor of different representations does not prevent us from making such comparisons or decisions.

If we consider the general relations between data elements from different sources, Batini proposed the following classification [9]: (i) identical, (ii) equivalent, (iii) compatible, (iv) incompatible, (v) different abstraction levels, and (vi) not direct linked. For equivalence, it is defined as for 1:1 correspondence, and Batini further classified equivalence among data elements can be based on (i) behavioral: X1 is equivalent to X2 if for every instantiation of data element X1, a corresponding instantiation of X2 exists that has the same set of answers to any given query and vice versa, (ii) mapping: X1 and X2 are equivalent if their instances can be put in a one-to-one correspondence, (iii) transformational, X1 is equivalent to XR2 if X2 can be obtained from X1 by applying a set of atomic transformations that by definition preserve equivalence.

In this approach, mapping is only one kind of equivalence, and there are some hidden assumptions for the notion of equivalence that are not valid within semantic heterogeneous environment. First, the criteria are based on logical representation, instead of semantics with explicit scope or limitations. That is, even there is 1:1 correspondence, it is possible that the two different representations may have different meanings such that their 1:1 correspondence can be applied only within certain context. For example, different kinds of ECID may have similar attribute values for their company names such that people treat them as the same meaning. However, they may not represent the same enterprise customer from every aspect in involved applications; even they do represent the same enterprise customer, the semantic heterogeneity between different kinds of ECID (that we have described in section 2.1) may lead to issues like users may use the wrong kinds of ECID to make orders. In the ECID example of Figure 3, a user may chose a legal entity-based ECID in service provider B, (instead of its equivalence location-based ECID) to make orders in product of service provider G who can only support location-based ECID. The 1:1 correspondence between legal entity-based ECID and location-based ECID does not indicate the equivalence of a legal entity and a location.

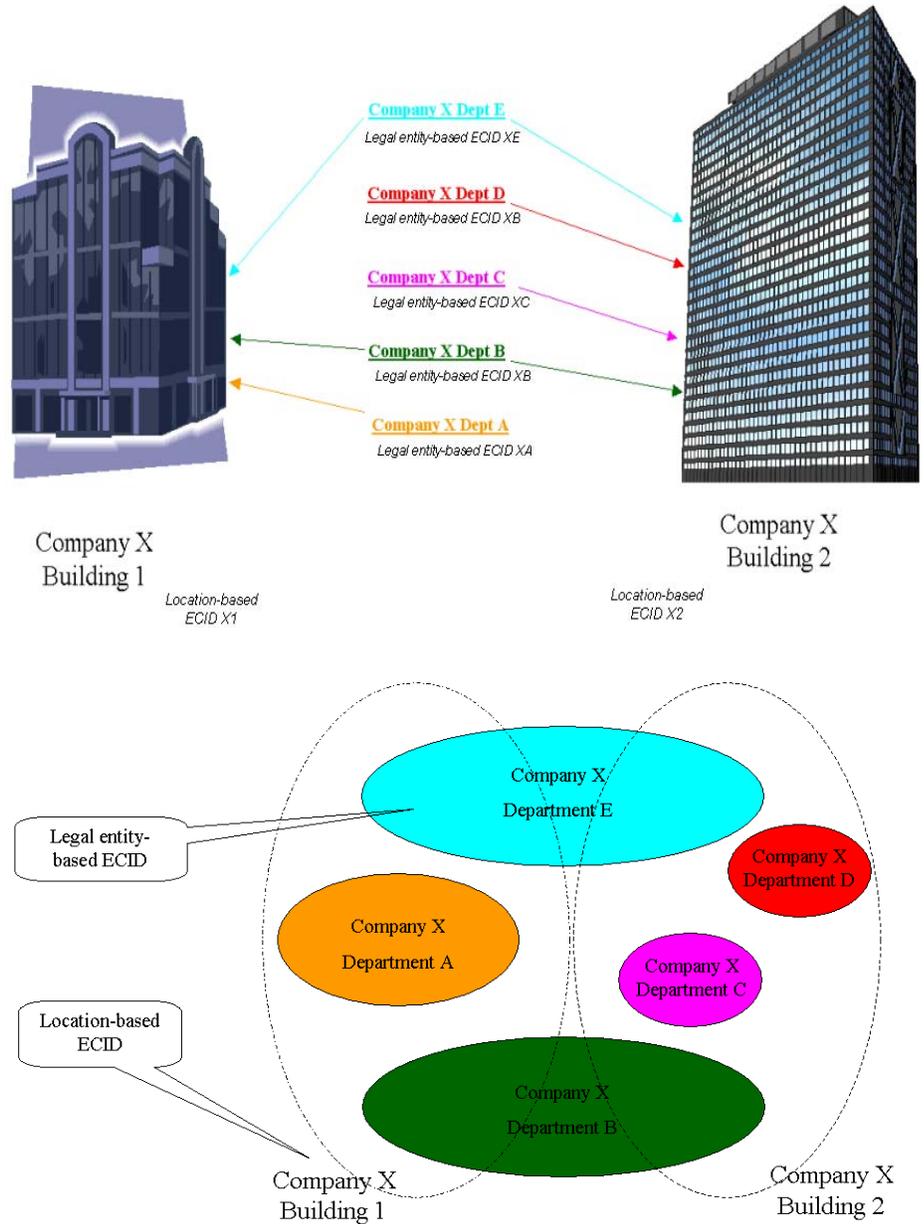

**Figure 10** Mapping between legal entity-based ECID and location-based ECID can be $m:n$, i.e. one legal entity with $n$ locations, and one location with $m$ legal entities. However, such $m:n$ mapping cannot really represent how their meanings overlapped together, as illustrated here.

The same issue also is in the theory of attribute equivalence proposed in [42], since equivalence is defined characteristics of attributes based Entity-Relationship model. Such characteristics include uniqueness, cardinality, domain, static and dynamic integrity constraints, allowable operators, and scale. Different types of equivalence are defined for different situations, and on tope of these varieties, equivalence between objects as well as equivalence between relationships are also defined. However, the fundamental issue is due to the limitation of underlying logical representation, i.e. the subjective criteria of entity versus relationship, the foundation of Entity-Relationship model. For example, an ECID may be modeled as an entity, an attribute of an entity, or a relationship. If data integrators need to integrate 10 ECID modeled as entities, 7 ECID modeled as attributes, and 4 ECID modeled as relationships, the various kinds of definition for equivalence simply make the notion of equivalence very complex; and such extra complexities have little help to manage or reduce the complexities of semantic heterogeneity.

The second assumption is the single version of the truth, and each different model must have full knowledge of the reality. In the example of Figure 8, we have explained the issue that single version of truth is not valid when each perspective holds only limited knowledge about the reality. In other words, if we want to define equivalence based on 1:1 correspondence in the ECID example, then, we need that (a) all different sources of ECID must cover 100% of all organizational behaviors for all enterprise customers, (b) all different perspectives of ECID must have exactly the same primitive ontological units for the definition of an enterprise customer, (c) for all products, every business department must perform mapping of different kinds of ECID with the exactly the same criteria, and (d) all systems must have the same policy for managing unique keys to prevent data replication related problems in information ecosystems like ghost, key reuse, and rollback problems (see discussion in section 2.1).

Looking for modeling equivalence in a different way, many research efforts are performed in mapping based on semantics techniques. In general, the idea is that the relation between data elements from different sources is a set of formulae that provide certain kind of semantic relationships [43]. One direction is to view equivalence or mapping to be based on similarity or by certain criteria; that is, the more commonality the different data elements share, the more similar they are, or from the opposite direction, the more differences they have, the less similar they are [44]. Researchers also proposed measurable ways for mapping, for example confidence levels, distance measurements, or probabilistic-based model are proposed in [5].

Applying similarity concept for mapping seems very natural for people as it can be observed from our daily lives. However, to apply in data integration for information ecosystems, many fundamental issues cannot be fully addressed. For the ECID example, one issue is that two different kinds of ECID may share no commonality but differences in every sense. Readers can imagine that different kinds of ECID are like different characteristics of an elephants; from any perspectives, an elephant head is not similar to an elephant tail. However, when some one have evidences to prove both the elephant head and elephant tail represent the same elephant, we can treat the different representations as (semantically) equivalent since they represent the same meaning of a specific elephant.

Therefore, semantic similarity or commonality may help to identify potential candidates for mapping, but it cannot serve as the decisive criteria for mapping. That is, the mapping between data elements from different sources is a subjective design decision in the level of logical representation, so does the notion of similarity in the level of conceptual representation, which is based on what the different users can observe about the subject matters, plus their background knowledge and chosen perspectives. However, these two can be correlated, but not necessary be consistent with cause-effect relations.

This observation leads to an interesting question between these two levels. Can the mechanism of a relation in Relational Model properly model similarity and distinguish with the mapping design decision? If we use the generic notion of equivalence to include both similarity and mapping, then the question can be described as the following: in the approach A or B that we have classified in the beginning of this section, overload domains with semantic heterogeneous data, or use relations with different attributes to represent semantic heterogeneous data, can they properly represent the subjective notion of equivalence we need for data integration?

Now let's review the rich varieties about similarity and mapping based on what researchers have pointed out. In the survey of [4], Rahm and Bernstein summarized the concept of similarity and treated a mapping as a similarity relation, which can be directional or no directional over scalars ($=$, $<$), functions (addition), semantic relations (is-a, part-of), or set oriented operations. The work in [45] classified related mapping as three major categories: equivalence, set theory, and generic (semantic) relation. Sheth introduced the concept of semantic similarity with the following four levels, semantic equivalences, semantic relationship, semantic relevance, and semantic resemblance, to be used with abstraction mechanism like aggregation, generalization, mapping (1:1, n:1), and functional dependency in related context representation for describing relations between objects [46] [47].

Most approaches focus on developing application level solutions for data integration difficulties. Hence, mappings of heterogeneous data are usually modeled by different attributes within the same tuple, since the operators of Relational Algebra cannot support semantic equivalence naturally. As a result, there are extra complexities introduced to deal with constraints due to the chosen data models, which not only make the design complicated, but also limits how semantic mapping can be applied to resolve challenges for data integration.

In contrast, we believe the root issue is the weakness of existing data models to represent similarity and mapping in an explicit way. In general, the difficulties for data integration are the natural results due to such weakness: existing data models are not designed for this purpose. The data model is supposed to help us, not to prevent us (data integrators) to make semantic comparisons like simple mathematical operations.

Specific for Relation Model, the question is that why we cannot have a standard way to model the varieties of similarity by extending the notion of equivalence in basic operations. In Codd's own words: "*The fundamental principle in the relational model is that all inter-relating is achieved by means of comparisons of values, whether these values identify objects in the real world or whether they indicate properties of those objects. A pair of values may be meaningfully compared, however, if and only if these values are drawn from a common domain.*" (see P.8 in [10]). That is, for semantic homogeneous environment, for example, there is only one kind of ECID universally accepted by every data model in the information ecosystems, and only one universal schema as the logical structure, then we can compare the unique key values of ECID to determine whether they represent the same enterprise customers or not.

For semantic heterogeneous environments, we need a way to make simple and meaningful comparison of the representations to determine whether they have the same (or similar) meanings or not. For the ECID example, that is to be able to compare ECID without worrying their structural heterogeneity or semantic heterogeneity. The challenge is that how a data model can represent semantic similarity among data elements that are represented with different logical structures, and how the limitations of such equivalence can be represented for users to design mapping correctly.

Of course, the desired data model should also consider the practical requirements in information ecosystems, and does not assume single version of the truth such that data integrators must have full knowledge of the reality in order to make their data model design decisions. In next section, we will take a close look of the expectations and practical needs.

## 2.4 Expectations and Practical Needs of Data Integration

In the context of information ecosystems, it is common that the data received by downstream systems is not just second hand, but third hand or fourth hand, i.e. the upstream streams that provide data is not the original source systems; instead, the upstream systems receive the data from somewhere else. The reason upstream systems receive such data is not for the purpose to pass to downstream systems, but for using the data for local purpose. Therefore, they may filter data, revise based on their judgments, or mix data with extra information before providing to downstream systems. As a result, downstream systems need to perform the same in information ecosystems since they may receive the same or similar data from multiple sources.

Differences, inconsistencies, or conflicts are blended such data streams flowing from sources to systems in the chain for reasons, including errors made by data integrators who may misunderstand the original meanings. However, based on what we observe from practices, the biggest factors are different business semantics for systems in information ecosystems to support different kinds of business operations. The different local business requirements force each system in information ecosystems to view the data from their specific perspectives.

If we accept such phenomenon is the norm in information ecosystems, then there are some fundamental questions need to be considered before any one perform data integration:

**Question 4:** <u>What kinds of characteristics of the data should be integrated and re-use in the local model</u>? How data integrators should perform their design activities such the integrated data can be meaningful for users and re-useable again later for different needs? And how we should handle the inconsistencies or conflicts?

These questions are usually not the focus of traditional data model design, which usually make simple assumption that the designers know everything about the reality, about the business needs, and about the solutions to resolve inconsistencies or conflicts. Considering the fundamental advantages of database, the traditional answers include the following areas that database can perform efficiently than files can: data redundancy and inconsistency, difficulty in accessing data, data isolation, data integrity problems, atomicity of updates, concurrent access by multiple users, and security problems [39].

Now consider database in information ecosystems, these advantages do not exist. One observation is that current data models are not designed for information ecosystems, since the basic underlying mechanism to manage inconsistencies and redundancies is based on the assumption of single version of the truth. We have explained why single version of the truth is not valid in the context of information ecosystems in previous sections, we now can summarized this as a convention as the following.

**Convention # 3:** <u>Data model should hold consistent data for all users; therefore, it is easier to design data models by supporting only single version of the truth.</u>

Following the same principle, this convention can be extended in data integration as the assumption that there exists a unified schema that is consistent globally; hence, the challenge for data integration is to search for such unified schema. For example, Lenzerini defined data integration as the following: "*Data integration involves combining data residing in different sources and providing users with a unified view of these data*" [1].

Further, assumed consistent under single version of the truth, mappings among data can be applied sequentially such that if A is mapped to B and B is mapped to C independently, then the wish is we can build mapping from A to C by combining the individual mappings together. We realize such assumption is an indication for the wish that inconsistencies or conflicts in real world business semantics can magically be resolved in the model world, and data integrators should be the magicians to make it happened without the need to change the real world.

However, for data integration in large-scale information ecosystems, is it reasonable to expect inconsistencies or conflicts between different perspectives can be resolved simply due to data modeling activities? How about the extreme case of information ecosystems like Internet in which data flow from one source to another frequently and constantly with subjective human interpretations?

Another consequence can be observed in the criteria people proposed to evaluate data integration project. We will discuss the details of three common ones: completeness, correctness, and understandability with clear semantics.

### 2.4.1 Completeness

One definition of completeness is defined as the following: every property or characteristics carried by existing schema/model must also exist in the integrated schema/model [9] [48]. In information ecosystems when there are needs for inconsistent or conflicting data to co-exist in the integrated database, then we violate the fundamental objective of database design to avoid redundancy and inconsistency. As a result, the criteria of completeness for data integration itself lead to conflicts with the criteria of consistency for the underlying data model design where data integration is implemented.

One implicit assumption for the completeness of data integration is that the completeness is evaluated from the perspectives of upstream systems, not from the perspectives of the downstream systems. In other words, it assumes all of the required information for the downstream local systems is available in the data sources such that if data integration is performed properly, then the downstream systems should be able to fulfill all of its local needs. This assumption is valid if the integrated database is to replace the existing data source without new functionality included, but not valid when new requirements or new information is required to be mixed into the data, or both the data sources and the integrated database need to co-exist but satisfy different business requirements.

For the ECID integration example, following the traditional view, completeness can be interpreted that all existing enterprise customer data from every available data sources are integrated into a single database (physically or virtually). It has nothing to say about how complete such integration can cover the enterprise customers in reality. However, the practical business wishes are that complete ECID integration should provide any information for enterprise customers, even if they are future customers that never have business with the service providers before. Therefore, even one can integrate all existing ECID from upstream systems, he still fails to satisfy the needs of the local downstream systems.

Another issue about completeness is due to that data integration may be performed before applications are designed or even specified. Therefore, there is no way that designed decisions make by data integration can be complete (in the sense of all existing data are integrated, or in the sense to have all information about the subject matter in reality). The concept of CDI (Customer Data Integration) that is now promoted in industry is one such example. Bt definition in [49], CDI is "*the combination of the technology, processes and services needed to create and maintain an accurate, timely and complete view of the customer across multiple channels, business lines, and potentially enterprises, where there are multiple sources of customer data in multiple application systems and databases*". In other words, the idea is simply that an enterprise service provider should integrate all of their customer data such that applications can have a standard way to use such information. The rationale of such expectation is based on the commonsense example of a library: different kinds of publications are collected and systematically stored without asking users how they want to use the information. Why customer data cannot be collected and systematically stored in the same way without asking how applications or end users want to use them? Or, data is treated like water; just like water can be "integrated" in reservoir, why can we "integrate" customer data such that any one can use it later for any kind of purpose.

Such business expectations do not consider many fundamental issues. In addition to the fact different perspectives can make data integrations with different results (as we have explained before), but also the chicken-and-egg dependences among different data integration projects. That is, since the integrated billing, ordering, repairs and so on all rely on the integrated customer data, it is often decided to perform CDI first and hopefully to design integrated billing, ordering, repairs later on top of the CDI results. However, without the detailed requirements for the applications (i.e. how integrated billing, repair, or ordering need to use the CDI), data integrators of CDI can only try to predict potential needs and make their design decisions.

Under such conditions, applying the criteria of completeness into such chicken-and-egg issue is a funny concept. If it is only evaluated from upstream system, completeness is not possible due to the inconsistencies or conflicts of the original model perspectives. If it is only evaluated from downstream systems local requirements, completeness may not be possible as future applications are not designed or even specified yet.

### 2.4.2 Correctness

One way to define correctness of data integration is that source data and the integrated data are mapped precisely such that it can be evaluated based on query against the original sources and integrated data model with exactly the same answer. Such query answerability is proposed as one of the important criteria for data mapping in [43], and further used to evaluate model management composition operator [35]. However, for data integration in information ecosystems, we have multiple data sources provide similar or inconsistent data about the same subject matter. There is little chance to run the query in the integrated data model with results back exactly the same as run the same query in every one of the original data sources, since subjective filtering or data selection criteria is performed during the integration process.

A different approach is proposed in [32] such that the focus switches from queries to data instance space by definition in terms of the schema mappings alone, without the need of reference to a set of queries. However, it suffers the same issues as query answerability based approach for data integration in information ecosystems. In addition, there could exist different design decisions about mappings between data instances due to data integrators' subjective criteria, and inconsistent or conflicting mapping decisions may have individual supporting evidences and user communities to justify as the correct answers.

Both of the above approaches have some implied assumptions. First, as discussed in the criteria of completeness, they simply assume that all information in data sources should be integrated (such that the query results or data instances can be compared between integrated database or the original data sources). However, we observe that the needs of downstream systems often focus on a portion of the meanings that is useful to satisfy local business requirements. As a result, the criteria of correctness should not be based on the original meanings. The other portions of original meanings that are not in the focus are subjectively filtered out or revised in the downstream systems.

Second, when there exist multiple versions of the truth in information ecosystems, both of the two approaches have the issue as correctness may be based on a specific version of the truth. Since no one has full knowledge about the reality, inconsistencies or even conflicts due to different perspectives chosen the original data sources simply make the criteria of correctness not helpful in evaluating data integration.

The third one is another assumption in traditional data modeling that schema design is done with full understanding of the data it needs to represent.

Convention # 4: <u>The design of a data model is performed with full understanding of the data to be represented</u>.

For example, the functional dependency assumes data modelers can understand the meanings of data and capture their relations in the data model. In practice, the dynamic, fast-changing nature of business environments, and the rich variation of business semantics brings the simple fact that there exist infinite ways to describe the same *thing* in reality, and the same data may represent multiple meanings. Data integrators have to perform data model design based on their current understanding without full knowledge. When more meanings or different aspects are learned, data integrators have to revise the data model to revise design decisions made earlier.

This triggers another known difficulty about schema evolution [16, 50]. Specifically, schema in Relational Model is not really for satisfying such try-an-error approach; it can be observed in Codd's own terms that schema is "*an irregular part consisting of predicate logic formulas that are relatively stable over time*" [22].

Researchers for this issue have proposed different solutions, the directions include the the following three categories as classified as in [41]: (1) <u>reduce schema changes</u> by adding the abilities into conceptual and data models the scope to accommodate modest changes to definition, (2) <u>reuse current schema definition</u> by changes of the application/wrapper for multiple extensional data, (3) <u>accommodate schema changes</u> seamlessly as much as possible. With our explanation in section 2.1, readers can see that approaches (1) and (2) try to resolve logical structure issues by creating more semantic heterogeneity.

Another push from business is the tradeoff between correctness (i.e. with better understanding of the meanings) and speed of design process for faster project deliver. It is due to the popularity of agile software development that expects incremental results in very short period of time. Interested readers can find an overview of the principles in [51]. For the ECID integration example, following the traditional data model design approach, we better to have full understanding the meanings of different kinds of ECID before we perform schema design. When the scale is to integrate ECID from 500 different data sources (and there is no global standard representation among them), the time and efforts required to capture the meanings and semantic heterogeneity for one-shot design is tremendous such that business simply cannot afford to wait. Incremental data integration becomes the only acceptable choice: any data integration design decision made today, may need to be revised tomorrow. Relatively, correctness with full understanding of the meanings is something that can be satisfied in such context.

### 2.4.3 *Understandability with clear semantics*

One expectation of the integrated data model design is to be understandable by users with clear semantics to support application [9, 43]. However, there are reasons why this simple and basic request is highlighted, and the difficulties are especially amplified for data integration in information ecosystems. First, the scale of complexity, there is no single person who can have the full knowledge of the reality, and understand the full meanings of existing data in every data source. This constrains the data integrators as well as the users of the integrated: if data integrators cannot have full understanding, users can have even less. And the most important part is that every one has bios about how they view the data due to his chosen perspectives, preferred levels of abstraction, or hidden ontology used in the business operations. Under such context that there are multiple versions of the truth, understandability with clear semantics is subjective for evaluating data integration projects.

For ECID data integration example, the specific design decisions to support integrated revenue calculation may not be understandable for business users whose operations are in the integrated ordering area. One issue is due to that users with a different perspective has real challenges to understand data integrations performed from another perspective, just like people in different cultures or in different religions have real difficulties to understand the behaviors of others. Another issue is due to the users understanding (or expectation) of the data integration. In practice, it is common for users with very naive expectation without understanding the actual limitation due to the data sources, available technologies, and the data integration methodology itself. Just like the issues we have reported previously, ghost problems, key re-use problems, rollback problems, or semantic heterogeneity are common issues in information ecosystems and the "current" resolutions requires cooperation among systems which is difficult for large scale information ecosystems. Without a consensus or a global standard, it is not a single data integrator can resolve in his local integrated database. The situation is just like students with different background to understand textbooks designed for college students versus designed for elementary school students.

In addition, the current data models are not designed to help different users to understand the design. That is, the schema structure is a rigid logical representation and it users' responsibilities to understand the design spirit, the data model is not designed to help users to understand the meaning of the data by providing interactive explanations. Currently, the majority of systems in information ecosystems are based on Relational Model, which we know from experience that there is no easy way to understand complex schema design without the help of the designer or knowledgeable users. As a result, when users have questions about the meanings of data they see, they cannot get answers the database directly. They have to ask people to find out the answer, and the answers depend on the knowledge level of the people being asked.

How about documentation? As the current way to assist users to understand the data model, it is actually the only alternative other than relying on people to explain meanings of data. Traditionally, documents for system requirements, analysis, or schema design should include explanation about the meanings. Such documents live outside database in a passive way, and only if perfectly managed, such documents can carry the latest meanings that match with the data sit inside databases.

There are also practical difficulties hard to overcome for documentation, even if one tries very hard. First, documentation are usually for specific purpose or business perspective, it is difficult to capture all perspectives possible in information ecosystems. Due to the nature of project-oriented management, documents for requirement, analysis and design are usually for specific project. Once the project is delivered, meanings of the data described in the documents start to getting outdated as other related projects may make changes to the shared data model. Second, documents cannot evolve automatically when data model evolves. Remember, data model changes are for catching up with the evolutions of business semantics and desired business operations. As a result, documentation outside database simply creates another gap from the business semantics. Third, the efforts of documents simply cannot catch up with the scale of information ecosystems in which many changes concurrently happen with or without dependencies among them. Documentations cannot capture the common phenomenon that the meanings of data in original data sources may evolve when the data flow to downstream systems.

We can describe the convention as the following:

**Convention # 5.** <u>Explicit meanings of the data and explanations to help users understand the data model design are usually not in the scope for data model itself. Users need to go to other sources like human or documentation to understand.</u>

This convention is the root problem for understandability with clear semantics, just like what we have quoted in section 2.2, "*Database schemas often do not provide explicit semantics for their data. Semantics is usually specified explicitly at design-time, and frequently is not becoming a part of a database specification, therefore it is not available*" [5, 8].

Therefore, the expectation for understandability with clear semantics is actually at the data model level, not at data integration level. This actually leads us to think the need of a new data model that can provide explicit meanings of data in order to help users to understand the data model, hence, data integration can be built on top of such new data model. The challenge is that we extend the scope of a data model: Not only it should provide mechanism to model data and to manage data, but it also need to capture meanings of data and help users (with various backgrounds) to understand the data in order to the data properly.

Therefore, we need to raise our next question as the following:

**Question 5: How should data models help users to understand the meanings of data?** For data integration, can data models help users to recognize the problems due to semantic heterogeneity such that users can manage inconsistencies or even resolve conflicts in their business operations first?

If we view information ecosystems like jungles, what users need the most is something like GPS that can guide them to where they need to go. Specifically, it will be very helpful for users if the integrated databases can interact with users to explain the meanings of data, where the original data are from, how the integrated data are mapped, what events happen, which process perform the change, and why the processes do so (i.e. following which business logic rules), and even detect inconsistencies or conflicts from different perspectives.

To proper model business logic rules is the next challenge behind this question. In information ecosystems, different perspectives that trigger the difficulties of semantic heterogeneity also trigger inconsistencies or conflicts among different business logic rules. In traditional data modeling approaches, such needs are usually addressed by

(i) Referential integrity in schema level in a declarative ways, or

(ii) Procedure-oriented implementation in programs inside or outside database.

The special need of information ecosystem is that the business logic rules involve with data flowing from one system into another. Even if the original model can enforce referential integrity locally, there is no generic ways to enforce the same business logic outside of the original data model, especially when there are different perspectives out there.

As a result, different business logics rules that happen in information ecosystems are spread everywhere. If a user want to understand what happened to certain data (e.g. why expected data failed or changed), not only he needs to check the implementation at schema level about related business rules, he also needs to investigate all related store procedures, application programs, or even the workflow engines.

Further, overloading schema with semantic heterogeneous data reduces the capabilities of referential integrity that can be performed at schema level. To an extreme, the majority of business logic must reside in application programs. We can use the ECID integration example to further explain the situation.

**Scenario 3.** Assuming that there is two kinds of data need to be integrated from various databases: ECID and service accounts. Different ECID represents different kinds of representation about enterprise customers (as illustrated in Figure 4), while different kinds of service accounts represent the various conceptual entities holding by enterprise customers to indicate the specific services or products provided by the service providers. For example, for billing systems, service accounts are the entities with invoices charged to enterprise customers (e.g. billing account); for service and repair organizations, service accounts are the entities that represent the service used by enterprise customers (e.g. a phone number); for ordering organizations, service accounts include those entities associated with order details; for network monitoring systems, service accounts are the entities with statistics about service usage by enterprise customers (e.g. circuit).

For business logic rules, service provider B expects that every service account must have exactly one legal entity-based ECID to be the owner for financial responsibilities. This is implemented in the traditional schema design as foreign key such that a service account cannot exist without a proper foreign key to the legal entity-based ECID.

A different business logic rule is in service provider G due to the nature of different products (e.g. network): a service account may have zero, one, or multiple location-based ECID to be the owner. Zero indicates this service account is for internal use, one or more indicates that financial responsibilities of such services accounts are shared among involved enterprise customers. As a result, this rule is implemented in data models not as foreign keys but as mapping to support zero to multiple associations between ECID and service accounts.

Now we have problem during data integration to implement different business rules. If we overload schema in Relational Model with semantic heterogeneous data like we have discussed before, we can use a generic ECID table to hold both kinds of ECID with a flag to indicate legal entity-based versus location-based. In the same way, a generic service account table to hold the two kinds of service accounts in service provider B and G. What happen now are the following:

(i) We can no longer represent business logic expected by service provider B by referential integrity at schema level for service accounts. The issue is that the referential integrity of required foreign key of ECID is not expected for service accounts in both service B and G.

(ii) Actually, the generic service account table cannot have a foreign key to the generic ECID since multiple ECID may share the same service account in service provider G. Therefore, there is a need for a generic map table between generic service account and generic ECID. In this generic map table, we cannot enforce that every service account from service provider B must have a legal entity-based ECID at schema level.

(iii) Now, we can explain how the data exchange timing issues may damage the referential integrity modeled in service provider B. Let's assume the legal entity-based ECID arrives the downstream system from a marketing sources through daily file transfer, while the associated service accounts arrive to the same downstream from a billing database through near-real time messages. To satisfy the business expectation in service provider B that every service account must have one and only one legal entity-based ECID, data integrators can not insert the service accounts when they arrive; instead, a buffer is required to hold received service accounts until the legal entity-based ECID arrive. The alternative is to compromise the data integrity such that services account may be inserted without the associated legal entity-based ECID.

(iv) Since there is no guarantee about data delivery in information ecosystems, extra logic rules are required for handling the case that associated foreign key of legal entity-based ECID may never arrive, or arrive later than acceptable time delay. The downstream systems have to face the tradeoff between the availability of the service accounts (which is partially correct if the associated ECID are missing) versus the validity of the business logic originally implemented as referential integrity.

(v) The root problem is due to that the foreign keys and the records they referring to are not bounded together in information ecosystems. Any business logic rules that have the same needs have to face similar issues. For example, applications for ordering, billing, or repairs all need to record customer data with their internal activities in downstream systems, while the upstream systems who provide data model of ECID may revise (even delete) the records concurrently. This is how ghost problems, unique key re-user problem, or rollback problems we described in section 2.1.

The main point for this example is to explain that the capability of Relational Model to support business logic decreases when overloaded with semantic heterogeneity. Even for semantic homogeneous environments, it is already not sufficient to help users understand the cause-effect relations for what happened in database. For example, there are only two choices about the requirement of references like foreign keys: must, or optional. There is no mechanism to model the concept "should" between these two levels such that we can allow foreign keys be missing due to data exchange delay.

Although such support may exist in application level like workflow engine outside of database, we believe the real need is to have a way to model business rules in data model level. Especially for surviving in the jungles of information ecosystems, what users needed the most is some navigation guidance among different semantics to help them interpret data, changes happened to data, and the business logics behind the changes. Documents may be also helpful for this purpose, but they can only serve as the secondary backup due to their maintenance difficulties to catch up with the changes in information ecosystems. Since data are managed in data model with the ability to continually evolving, why the meanings of data can not be maintained in the same way? That is, the responsibility of a data model also includes automatically adjust the explanation when design changes as well as trace what happened and explain the logic rules behind the scene.

## 2.5    Summary

To explain the motivation what we need to propose a new data model for the purpose of data integration, we raised a sequence of questions in this chapter based on a practical ECID integration example. The questions are:

Question 0: What is semantic heterogeneity?

Question 1: What does a unique key actually represent?

Question 2: What are meanings? How can we explicitly represent meanings in order to manage semantic heterogeneity?

Question 3: What is the nature of mapping? In what sense can mapped data be treated as equivalence? Does such equivalence uni-directional or bi-directional?

Question 4: What kinds of characteristics of the data should be integrated and re-use in the local model? How data integrators should perform their design activities such the integrated data can be meaningful for users and re-useable again later for different needs?    And how we should handle the inconsistencies or conflicts?

Question 5: How should data models help users to understand the meanings of data? For data integration, can data models help users to recognize the problems due to semantic heterogeneity such that users can manage inconsistencies or even resolve conflicts in their business operations first?

We have illustrated how the root issues about semantic heterogeneity, structural heterogeneity, and the practical needs in information ecosystems can dangle together. Specifically, we explained why and how Relational Model cannot efficiently model semantic heterogeneity due to its original design purpose. Since data integration methodologies are limited by the data models they can support [9], and data integration itself must be implemented by a data model, we believe the solution is not to develop data integration techniques on top of existing data models, but to design a new data models for the purpose of data integration. With such a new foundation like a platform, existing data integration approaches or new algorithms can be applied to help users to manage such challenge.

In next two chapters, we will propose a Shadow Theory with six basic principles to answer these six questions we raised in this chapter. In chapter 5, we will further develop algebra to extend Relational Algebra to support operations needed in Shadow Theory. In chapter 6, we will compare our proposal with Relational Model and discuss related issues.

## 3. Shadow Theory

Based on the analysis described in section 2, we believe that the difficulties encountered by data integrations are the natural results due to limitations of existing data models, especially Relational Model, which is not designed for managing semantic heterogeneity but overloaded with semantic heterogeneous data for use in information ecosystems. The need is to address the fundamental issues about semantic heterogeneity at data model level, and based on the fundamental data operation utilities to establish solutions for data integration difficulties.

We understand that inconsistencies and conflicts in real world cannot be simply be resolved in model world, therefore, our objective is to help users (who may be from different background with inconsistent or even conflicting understanding about the same subject matter) to be able to see the data explicitly from different perspectives or at different levels of abstraction, such that they can manage data with semantic heterogeneity or even resolve conflicts in real world first.

Specifically, we focus on designing a data model that can manage dynamic meanings of the data without constrained by their (existing) logical structures, but still can be backward compatible with Relational Model for practical reason. In addition to support multiple versions of the truth, we also need to support the business practices about agile development process such that the data model can easily adjust to quick evolving business requirements. In other words, we need the data model and integration can be designed mainly at semantics level that can follow business semantics directly.

In this section, we will first introduce the fundamental concepts in Shadow Theory (section 3.1), then the six principles for applying such philosophy into data integration for information ecosystems (section 3.2). In section 3.3, we will provide precise definitions of shadows. Following the sequence of first three principles, we will answer the first three questions raised in section 2:

Question 1: What does a unique key actually represent?

Question 2: Where are meanings represented in a data model? In data values? In logical structures? In both of them? Or somewhere else?

Question 0: What is semantic heterogeneity?

The rest questions will be address in section 4 when we discuss the principles for data integration.

### 3.1 Shadow Theory

**Shadow Theory** is proposed as a starting point to answer the needs to model semantic heterogeneity. The basic philosophy is that whatever data can be observed and recorded into database about the subject matter are only shadows. Shadows are generated as the results of certain projection process from the subject matter(s) in the real world to wall-like surfaces of the database systems. Hence, we can classify properties of shadows as the following three categories:

C1. Properties due to the characteristics of the **subject matter**.

C2. Properties due to the characteristics of **wall-like surface**.

C3. Properties due to the **projection process**.

Since a database is designed to satisfy specific business needs, practically, only limited properties can be stored and model to satisfy its system requirements. Therefore, the design process of data modeling is actually like the projection process that determines (subjectively transform and filter) observable of the subject matter into shadows (i.e. data records). The chosen perspective(s), level(s) of abstraction, or formally the ontology used by the users play the key roles in such projection process. Physically, the projection process also includes the execution of data collection (directly or indirectly to interact the subject matters), data exchange (data flow from upstream systems to downstream systems), data filtering, transformation, and aggregation (to combine partial information into a bigger view of the subject matter). Shadows, as the result of such process, are not simply due to characteristics of the subject matter, but a combination of these three property categories. In addition, the concept of epistemology can be introduced here when we collect different shadows from different data sources in order to integrate them together; what the users know about the subject matter also limits how their design process of their data models.

Figure 11 shows an abstract diagram to illustrate shadows and their properties. In the ECID problem described in section 2, enterprise customer data collected and stored in database are all shadows, and we can classify their properties as the following:

(C1) <u>Properties due to characteristics of the subject matter</u>, e.g. customer names for how they identify themselves and customer contact information are characteristics due to the customers (may be provided by the customers directly).

(C2) <u>Properties due to characteristics of the systems</u>, e.g. a local unique key of ECID that represents a specific customer can be a property only for the system, not due to the real world since the customer may not have knowledge of this unique key.

(C3) <u>Properties due to characteristics of projection process</u>, e.g. the chosen model perspective like location-based ECID is due to the projection process, which determines the meaning of the uniqueness for the ECID; if Relational Database is used for storing data, then the characteristics of Relational Model also contribute to the properties of the shadows.

Obviously, shadows of the same subject matter(s) may look very different and confusing! Why? We know any differences in walls

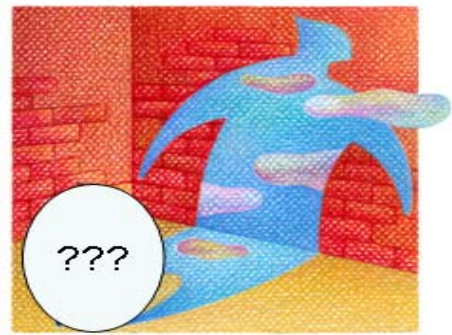

**Figure 11 Shadow Theory.** What we can observe and store in database are only shadows of the subject matter(s). The properties of shadows can be classified as: C1. Characteristics of the subject matter(s), C2. Characteristics of the wall-like surface (i.e. system requirements), and C3. Characteristics of the projection processes that subjectively filter and transform observables into shadows.

(system) will make the shadows different, so does differences in the projection processes (model perspectives) as they subjectively filter and transform shadows. The subject matters themselves also contribute the differences by their inconsistent behaviors (which then observed by different model perspectives).

In terms of Shadow Theory, we can think the task of data integration is like a game to identify which shadows from different walls corresponding to the same subject matters, and under what contexts we can treat them as the same. For the example of ECID integration, Figure 12 illustrates an overall diagram for such challenge. Imagine the enterprise customers are like elephants: legal-based ECID, location-based ECID, and contract-based ECID are the observed shadows of these elephants from different perspectives. Since each individual service provider can only collect limited information, none of the individual model can have the full truth of everything. The perspective of legal entity-based ECID is like applying a subjective filter (i.e. the ontology of legal entities they use, represented as a hierarchy in the diagram) to look at elephants with the results of unique tails identified. The perspective of location-based ECID is like applying a different subjective filter (i.e. the ontology of location, represented as an hierarchy) to look at elephants with the results of unique heads identified. The perspective of contract-based ECID is like applying the filter of account for contract and records the side view of elephants. The worst issue is that, these different perspectives do not have a common understanding for what an elephant is. In addition, the enterprise customers may not show consistent behaviors across different perspectives (e.g. to avoid tax issues, to inherit marketing brands, to confuse their customers), just like two elephants may seem have a common tail or two heads associated with six legs. Further, the complexities are significantly increased by frequent business M&A activities, it is just like moving legs of one elephant to another but function as tails, or merging two heads as a single one. The answers for interpreting what happened really depend on the viewers' understanding of the subject matter, or subjective judgments about the cause and effects based on chosen perspectives, and are constrained by the limitation of the viewers' knowledge.

With such pictures to understand the tasks of data integration in information ecosystems, we will propose six common sense principles as the overall guidelines in the next section. Then, we will discuss the details for how to answer the first three questions we raised in section 2 by applying these guidelines.

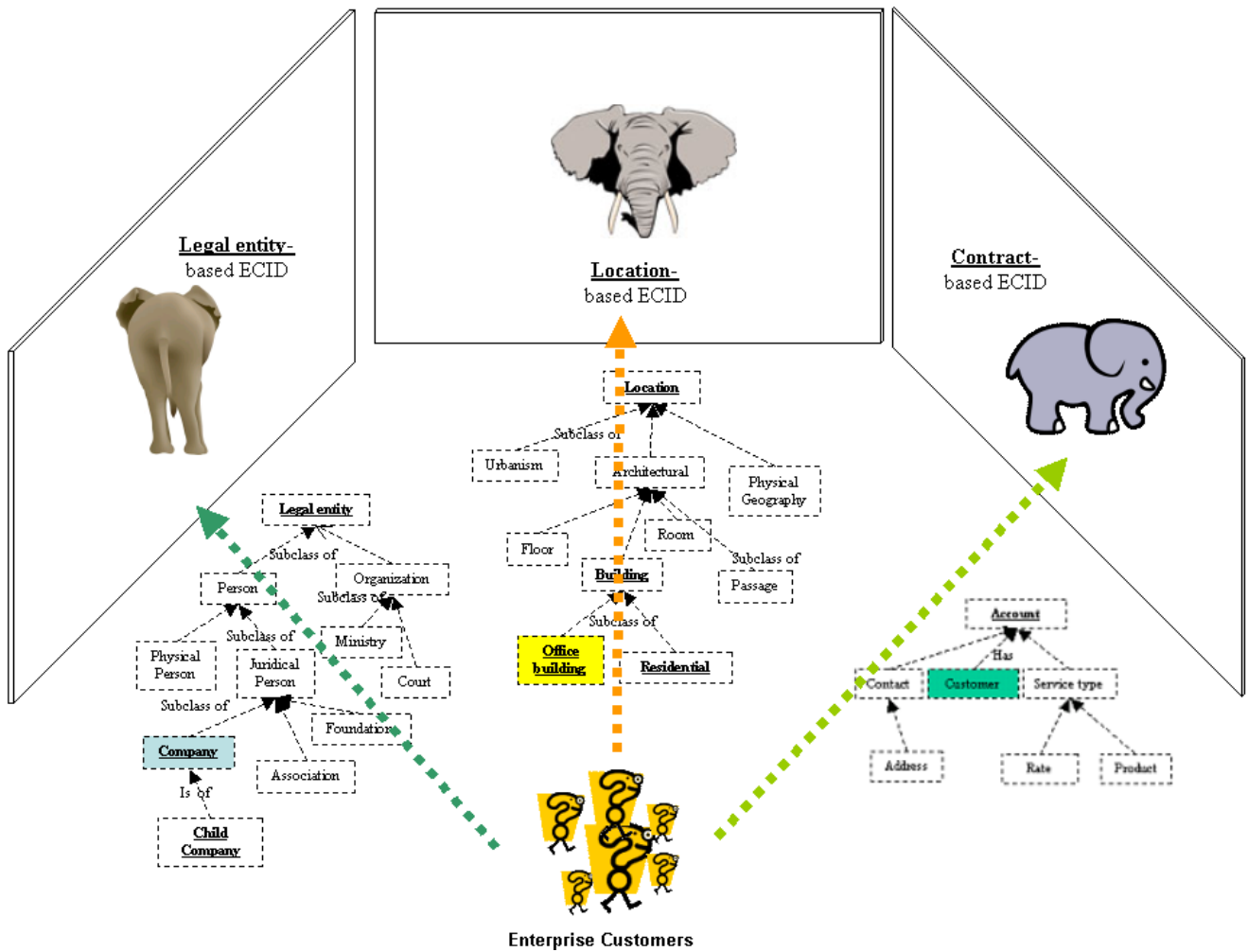

**Figure 12** Various kinds of enterprise customer data collected and stored in database are like shadows. The different model perspectives, or ontology if it is formally modeled, play the key roles for filter and transform shadows into specific characteristics and structures; for example, the uniqueness of legal entities, locations, and contracts are just like different perspectives and representations of elephants as illustrated here.

## 3.2 Six Principles for Data Integration

The six principles for data integration can be summarized as the following:

**Principle 1. What we can observe and store in database are only shadows**.

Any data element that is used to describe a specific subject matter is a shadow. Such shadows are the results of certain projection process from the subject matter (the *thing*) to wall-like surfaces of the database systems. Since the subject matter can be represented in different ways, there exist different shadows that represent the same *thing*. The correspondences among shadows that represent the same thing may not be able to be represented as simple 1:1 mapping, since the notion of "the same" (or similar) is a subjective judgment made by the viewers, and the primitive units for the viewers to view the subject matter depends on their chosen perspective or ontology .

**Principle 2. The meanings of shadows exist as mental entities in viewers' cognitive structures, and we can use W(hat)-tags, short as W-tag, to anchor such mental entities uniquely.**

Shadows themselves do not carry meanings; it is the viewer who decides what meaning(s) to associate to the particular shadow. That is, meanings of shadows are like mental entities that live in viewers' cognitive structures, named as semantic space. Such mental entities are constrained by the viewers' chosen perspectives, levels of abstractions, and overall understanding about the subject matter in reality. However, they are not constrained by the logical structure of the shadows in data space.

**Principle 3. Semantic Heterogeneity is the overall aggregated result due to differences among meanings as mental entities in viewers' cognitive structure, and differences of how shadows are projected onto wall-like surface of system requirements about the same subject matter.**

In addition to that different shadows can be projected from the same subject matter in reality to different wall-like surfaces following different projection criteria, there may also exist different interpretations for the same shadow collected and stored in database. Hence, semantic heterogeneity can be viewed as the overall results due to differences among meanings (of shadows) as mental entities from different viewers' cognitive structures, and different shadows projected from the same subject matter onto different all-like surfaces.

**Principle 4. Equivalence between meanings** (as mental entities from different viewers' cognitive structure) **is a subjectively decision, and we can model such equivalence by E(quivalence)-tag, short as E-tag, with supporting evidences, just like a bridge to cross the boundaries of different perspectives.**

The most important feature for Shadow Theory is that, no matter how big differences the logical structures or physical representations can be, different shadows may be viewed as the same by association to the same meaning. That is, meanings as mental entities in specific semantic space can be treated as equivalent under subjective criteria, perspective, and with supporting evidences. It is like to establish bridges among mental entities to cross the boundaries among viewers' cognitive structures. Such bridges can be combined together to associate different kinds of mental entities (as long as the criteria for each bridge can be satisfied under the specific context), or be revoked when evidences become invalid.

**Principle 5. Meaningful data integration should be performed only with required shadow properties, and the scope of the subjective equivalence decisions should be explicitly represented with meanings of the data.**

Since not all the three categories of shadow properties (i.e. due to the subject matter, the system requirements, or the projection process) are always required to be integrated for each project, data integrators have to clarify the objectives of data integration. After the objectives are specified, meaningful data integration needs to explicitly represent not only shadows and associated logical structure, but also the meanings of the data from viewers with the criteria why such data can be treated as equivalent. Scalable (in the sense of complexity)data integration requires explicit representation of the scope and limitation of such criteria with supporting evidences, such that in the future data integrators can be reused the representations again for other integration purpose. With the goal to help users manage semantic heterogeneity, consistency is only maintained within individual perspectives, and inconsistencies or conflicts should be included in the scope of integrated data model such that users can see the issues (due to realities of different perspectives) and hopefully identify resolutions in physical world first.

**Principle 6. To helps users to understand and use integrated data properly, data models need some features to explain the meanings of data, including modeling perspectives, business logic rules, and the criteria for decision decisions made for semantic equivalence.**

Following the above principles, users who want to use the meaningfully integrated data requires to select available model perspective(s) and level(s) of abstraction in order to properly formulate semantics to be queried. The system should be able to expand query results with bridges whose supporting evidences meet the semantics criteria. Further, the data model design does not assume complete knowledge is available; instead, it needs to allow easy adjustment in order to support data integrators with incremental understanding about the subject matter, as well as agile development process due to dynamic business environments.

In the following sections, we will discuss the details for each principle. First, section 3.3 will trace the philosophy foundation of shadows back to Plato's cave. Second, we will review different approaches for modelling "meanings" in section 3.4, and how we can explicitly represent meanings in order to model semantic heterogeneity. With the philosophy foundation and approach for modelling meanings, we can propose a representation to manage semantic heterogeneity in section 3.5

Discussion about representation of equivalence and similarity is introduced in section 4.1, which is one of the core challenges that we cannot avoid for data integration. In section 4.2, we will further discuss what shadow properties should be integrated and for what kinds of data integration objectives, as well as the criteria to evaluate success of integration. The features to support users to use integrated data is discussed in section 0.

## 3.3 Shadows in Plato's Cave

Philosophically, Shadow Theory only assumes the existence of shadows, but not the existences of the *things* that shadows are projected from. The reason is that we do now have the full knowledge about the *things* and we need to support multiple versions of the truth.

It is contrary to the common assumptions of most conceptual and ontological modeling approaches (e.g. [52] [53]) that it is usually assumed explicitly or implicitly about the existence of the subject matter in real world. However, the philosophy foundation can be traced back in Plato's Allegory of the Cave. Over 2000 years ago Plato observed the inherent nature of multiple views of reality, and the rich philosophical development can be summarized as the following. A group of "prisoners" live chained inside a cave, and what they can see are only shadows cast on the wall, which are due to *things* passing in front of a fire. What each prisoner sees is different to each other as it depends on their perspectives, experiences, and knowledge of the *things*.

In an information ecosystem, the data stored in each database are like shadows projected to wall-like surface of the system requirements. Users for each database are like prisoners who can only see shadows, and how to interpret the meanings behind shadows about the subject matter depends on the users' understanding about the subject mater, the chosen perspectives, and desired levels of abstraction. The full truth is not available since any database can hold only limited descriptions, and there exist unlimited ways to describe the same subject matter in reality. Usually, the objective of a database is to satisfy its local requirements based on specific business semantics. Hence, data flow from upstream systems into downstream systems is like copying shadows collected from one perspective and store in a different place, and such copied shadows also be viewed from a different angle without the original context. The universal truth about the subject matter is not the most important concerns for individual users; it is up to the downstream systems to decide what data can satisfy their local needs: if they do not like the shared shadows, they can try to reach other sources to find what they need.

With such philosophy foundation, we now can review questions raised in section 2.1 and propose a philosophical answer.

Question 1: **What does a unique key actually represent**?

As we have explained, in information ecosystems, there are reasons why unique keys may be neither singular (two different keys should not represent the same object) nor immutable (an entity is represented by the same key value through its life time). This is the basic challenge for semantic heterogeneity as we lose the foundation for the convention to use unique key to build data models.

Based on the first principle of Shadow Theory,

**Principle 1. What we can observe and store in database are only shadows.**

and we know the fact that properties of shadow can be classified as due to the subject matter, due to the wall-like surface, or due to projection process, we can say that the real meaning of a unique key is to uniquely **identify a shadow,** not the *thing* in reality**.** When viewers think a unique key representing the *thing* uniquely, it is a subjective data model design decision based the chosen perspective and selected level of abstraction. In the ECID example, a unique location-based ECID and a unique legal entity-based ECID can represent the same enterprise customer; therefore, the uniqueness of such ECID can be interpreted as uniqueness of the shadows of this enterprise customer that projected to a system with the model perspectives of location-based or legal-entity-based.

To further proceed, we need a precise definition for a shadow. By dictionary[7], a shadow is a partial darkness or obscurity with a part of space (from which rays from source light are cut off by an interposed opaque body), and therefore may reflect the image of the interposed body. For the purpose to model data in information ecosystems, we define a shadow as the following:

**Definition 1. Shadows of entities.**

> Defined from a specific perspective and at certain level of abstract, a shadow is a data element used for representing some *thing* conceptually. We use the term subject matter to indicate the *thing* in the model. Each shadow can be assigned with a unique key, and the unique key represents the shadow whose properties are combinations of characteristics due to the subject matter, the wall-like surface (i.e. system requirements), or the projection process.

Intuitively, the above definition seems enough to describe the notion of shadow we need for a single database. It can cover the relation between data in a model and the subject matter in realty. However, it cannot express the characteristic of information ecosystem about data flowing from upstream systems into downstream systems. There are different kinds of roles for individual data models in such environments. Some are like producers to generate data for others to use; some are like consumers that rely data from others to fulfill its local functionalities or business operations. The chance of a single database that uses only self-generated data is very low; most databases will need some second-hand data provided from somewhere else. That is, the systems may not be able to have interactions with the subject matter directly.

There are some impacts due to such data flow from upstream systems to downstream systems. For second hand or even third hand data, not only the meanings of data may evolve during data exchange process, but also that extra issues like ghost problems, unique key reuse problems, or rollback problems may happen to pollute the meanings. To capture such loose semantics dependency, we can further revise the definition by extending the meaning of projection process as the following:

**Definition 2. Shadows of entities flowing among databases in information ecosystems.**

> Defined from a specific perspective and at certain level of abstraction, a shadow is a data element describing some *thing* conceptually. We use the term subject matter to indicate the *thing* in the model. Each shadow can be assigned with a unique key, and the unique key represents the shadow whose properties are combinations of characteristics due to the subject matter, the wall-like surface (i.e. system requirements), or the projection process. *The projection process can be from the subject matter in reality to shadows on wall-like surface, or from shadows on one wall-like surface to another different wall-like surface (i.e. projection from the meanings modeled in*

---

[7] Merriam-Webster dictionary published by Merriam-Webster, Inc., Springfield, MA, USA.

*one database to the meanings modeled in another database) , or a mix of both kinds.*

We may use the term second level projection (or third level, fourth level, … and *n*th level) to highlight the impacts due to systems involved in the middle of data flowing, from original data source to a specific local system. Note that for shadows to go though every project process, logical structure or data formats may change, and the meanings may also evolve through subjective criteria mediated by systems in the middle.

Readers may wonder whether the subject matter that shadow represents must be an entity, like the notion of entity in Entity-Relationship Model [54]. However, there is a problem about **semantic relativism** [55] which concerns about the ability to view and manipulate data in the way most appropriate for the viewers semantically, not forced by the chosen data model. Take the example to model a marriage, it is the viewer to decide how to view such information: as an entity, as an attribute, or as a relation. If a data model forces such distinction by choosing to model a marriage as an entity, the model itself creates difficulties in data integration for downstream systems, which need to manage the conflicts that another data source may model a marriage as a relation.

If shadows are defined to represent only entities, there seems existing a hidden assumption about separate representations for entities and other types like relationships or attributes of entities. Such classification is unnecessary since we do not have full knowledge of the reality about what the shadows represent, or we can have a consensus from every perspective about what the shadows represent. For example, if we use the terms of Entity-Relationship model to describe the ECID example, some ECID are like representing entities with unique keys, some ECID are like representing attributes of something else (e.g. attributes of accounts) without any uniqueness, and some ECID are like representing a relationship (e.g. an ECID does not exist by itself, but only exist to serve cross reference between accounts and other records).

Since the objective for Shadow Theory is to help model and manage semantic heterogeneity, we want to avoid such trap that make data integration even harder. Therefore, what we need is a generic data type to cover all kinds of data types in existing data models, including entity, relationship, and attribute in Entity-Relationship Model, or even object in Object-Oriented database. This generic data type should represent generic meaning such that we can reduce the unnecessary classification to minimum.

Another reason we need to avoid the notion of entity is that, Shadow Theory only assumes the existences of shadows, not the existence of the *thing* as an entity. The existence of the *thing* is actually a concept in the viewers' cognitive structure, such that the viewers can subjectively choose to view the thing as an entity, a relationship, or as an attribute of an entity (or of a relationship). We will use the term **semantic space** for the meanings existing in viewers' cognitive structures, and use the term **data space** to describe where data stored in database. That is, any piece of data in data space is a shadow in generic sense, and whether viewers want to treat the meanings represented by shadows as entities or relationship should be in semantic space only.

In this way, we need to revise the definition of shadows as a generic notion for any kind of data in data space, which is a result of projection process. The generalized definition can be described as the following:

**Definition 3. Generic Shadows.**

1. A shadow is a piece of data in any kind of formats or structures, physically or conceptually.

2. The meanings of shadows are due to the mental entities in viewers' cognitive structure, such that the viewers associate their mental entities to the shadows.

3. The viewers' cognitive structures are the most important parts for the meaningful existence of shadows; shadows can be associated to meanings only if the viewers have ways to conceptually recognize (interpret) the shadows.

4. The projection process can be mix of either one of the following:

    - From the subject matter in reality to mental entities in viewers' cognitive structures conceptually, with or without to shadows stored in storage physically.

    - From shadows in one storage to shadows in another storage physically (i.e. data flow from upstream systems into downstream systems in information ecosystems).

    - From mental entities in one viewer' cognitive structure to mental entities in another viewer's cognitive structure conceptually (i.e. projection from meanings as mental entities recognized by one viewer to the meanings as mental entities recognized by another).

5. Properties are shadows are a combination of characteristics due to the subject matter, the wall-like surface (i.e. system requirements), or the projection process (i.e. the chosen perspective or level of abstraction, and the modeling tools).

Here we can perform some comparison with Relational Model to clarify the concept. In Relational Model, data can be classified as two types: atomic and compound, and Codd explained that the reason there is only one type of compound data is that "*any additional types of compound data add complexity without adding power.*" (p.6 in [10] ). To certain degree, we are actually pushing the principle further by removing the atomic data type. In section 2.1, we have reported that the assumption of atomicity held in one data model may not be valid in another data model (due to the difference of the underlying ontology), and the consequence is that we can no longer apply Set Theory since we do not have a common primitive "unit" to view the same distinct *things* (see section 2.2).

That is, in terms of Relational Model, a shadow can be in the form of an attribute value, a portion of an attribute value, a tuple, a portion of a tuple, or a collection of tuples with the same or different attributes organized in some way. Our objective is to explicitly model semantics without constraints of logical structure in data space, hence our approach does not eliminate logical structure, but to creates more flexibility for how shadows can be decomposed or aggregated in different perspectives or at different levels of abstraction. Before we discuss how we can represent such logical structure in section 3.4.3, we need to review the question about representation of meanings in next section.

## 3.4    W-tags: Meanings of Shadows

Before we can discuss how to represent meanings in a data model, we need to ask the question: what are meanings? It is actually a long-standing philosophical dispute concerning the meaning of "meaning." In the survey performed by Gärdenfors (ch5 in [56]), different approaches of the answer are classified according to whether semantics is referential or not, and whether the meanings of expressions are some kinds of objects or not. The categories are:

(1) **Meanings exist in the communicative function**. This approach does not think meanings are referential to something out there in the world, instead; in this functionalist tradition, meanings of expressions are determined in the context of their use. The key is that meanings need to be understood through communication; if no one can understand, then where is the meaning?

(2) **Meaning is something out there in the world**. This realist semantics approach can be further classified as extensional versus intensional:

   (2a) **Extensional**: The constituents of the language are mapped onto a "world": names are mapped onto objects, predicates are mapped onto sets of objects or relations between objects, and sentences are mapped onto truth values. Frege's semantics and Tarski's theory of truth can be classified as in this category.

   (2b) **Intensional**: The set of linguistic expressions is mapped onto a set of "possible worlds". The meaning of a sentence is taken to be a proposition that is identified with a set of possible worlds, i.e. the sentence is true in the set of worlds. Situation semantics is an alternative in which situations are partial descriptions of the world. It maps meanings to a "polarity value" that expresses whether the fact holds in the situation or not.

(3) **Meanings are mental entities in persons' conceptual structure**: In cognitive approach, meanings are described as mappings from the expressions to mental entities in persons' conceptual structures, and such conceptual structures can be seen as the persons' internal representation of the world. In addition, meanings that can be communicated among people are due to the common mental entities held in these people's conceptual structures.

To make comparison easier, Figure 13 lists the illustrative diagrams Gärdenfors made to describe (2a), (2b), and (3) (see Figure 5.1, 5.2, 5.3, and 5.4 in [56]).

In Plato's cave, the shadows are the same to all prisons, and the reason each prisoner capture different meanings is due to differences in their perspectives, experiences, and knowledge of the world. Therefore, the cognitive approach is the closest one that can satisfy the needs to explain meanings for Shadow Theory. Since different prisons have different cognitive structures of the world, and mental entities they choose as the meanings of shadows are based on personal subjective decisions.

That is, meanings of shadows do not refer to something out there in the world; instead, **meanings of shadows refer to mental entities in viewers' cognitive structures**. Such cognitive structures are the viewers' internal representation of the world, and the mental entities reflect the subjectively filtered modeling of the *things* in reality. Hence, cognitive approach can explain the two key features of semantic heterogeneity that

- different representations (i.e. different shadows) for the same meaning (i.e. the same mental entity in one viewer's cognitive structure), and

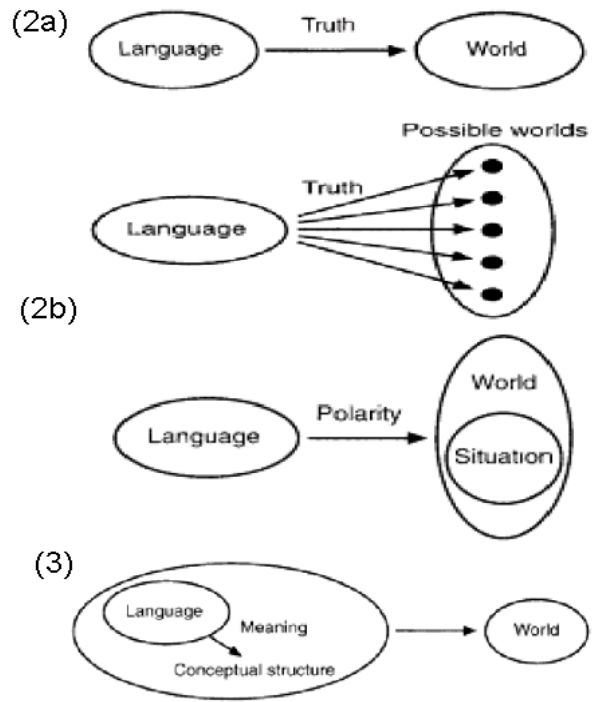

**Figure 13.** Gärdenfors illustrated the differences among different approaches concerning the meaning of "meaning" with these diagrams, original as Figure 5.1, 5.2, 5.3, 5.4 in [56]. (2a) represents extensional realistic semantics that meanings are something out there in the world. (2b) represents the intensional realistic semantics that meanings are true values of sentences to possible worlds (top) or situations (bottom). (3) represents cognitive approach that meanings are mental entities in people's cognitive structures, and communications are performed through common mental entities.

- different meanings (i.e. different mental entities in different cognitive structures) for the same representation (the same shadow).

Based on this approach, we review the question 2 raised in section 2.2:

Question 2: What are meanings? How can we explicitly represent meanings in order to manage semantic heterogeneity?

We can propose the second principle for Shadow Theory to answer the question about what meanings are:

**Principle 2. The meanings of shadows exist as mental entities in viewers' cognitive structures,** and we can use W(hat)-tags, short as W-tag, to anchor such mental entities uniquely.

With the first principle, we treat all data stored in database as shadows in data space. Now with the second principle, we think meanings of shadows are mental entities in viewers' cognitive structure (i.e. semantic space). Now we can proceed to the question how we can represent meanings.

*3.4.1 Meanings represented as Areas in Semantic Space*

In the ECID example described in section 2.1, we have explained that different kinds of ECID can be designed to model the same enterprise customers with different decomposition structure, and the assumption of atomicity is a local decision made within the specific perspective or ontology. Since viewers with different cognitive structures may agree that their different mental entities actually represent the same *thing* semantically, we need a representation that can support concurrent different decomposition structures of the same *thing*, and there is no common agreement on the primitive units how this *thing* can be decomposed into, as well as there is that there is no limitation for when decomposition must stop.

Compared with Relational Model whose foundation is based on First Order Logic, atomic data elements like points in geometry are the required assumption that cannot be avoid. Obviously we cannot apply such concept in Shadow Theory, the notion of an area in geometry as the primitive unit can better serve our needs: an area can be decomposed into smaller areas without limitations, and there can exist multiple ways to decompose the same area concurrently.

The **point-free geometry** initiated by Whitehead in [57, 58] is based on such notion to use regions as the primitive ontological units. It is further developed by other researchers into axiomatic systems grounded in mereology and mereotopology [59]. We choose this as the mathematical foundation to design algebra for supporting data operation based on Shadow Theory. In section 5.2 we will provide readers a summary for the properties we need from point-free geometry, and further how we can implement the representations we will discuss next.

In short, we propose to model meanings (as mental entities) in semantic space with areas to be the primitive ontological units. Any area can be further decomposed without limitation, and there can exist different decomposition for the same area concurrently. Therefore, two regions that fully overlap can be viewed as the same, representing the same meanings. Even if not fully overlapped, they may still be treated as the same (approximately) based on subjective criteria that define the rules of similarity, commonality, or other logic conditions like one region is included in the other. In section 4.1.1 we will further discuss the mathematical foundation for model equivalence.

*3.4.2 W-tags: to Explicitly Anchor Unique Meanings*

Next, the question is how can we explicitly represent meanings in order to model semantic heterogeneity? Since meanings are mental entities in viewers' cognitive structures, we can use some tag-like mechanism to anchor each different mental entity explicitly, such that viewers can see a tag and recognize its associated meaning without ambiguity. If one viewer never has this mental entity in his cognitive structure, we need the tag carries certain descriptions to help the viewer to establish one.

Our first need is to define the kind of tags to anchor meanings as mental entities that are used to identify what the *things* are, i.e. the results of classification based on the viewers' ontology, implicitly or explicitly. We call this as **W(hat)-tag**, short as W-tag. The following are some rules for W-tags:

**W-tag Rule # 1.** <u>Each W-tag should anchor with a unique meaning as a mental entity existing in a viewer's cognitive structure named semantic space.</u> In order to identify the differences of the perspectives in different viewers' cognitive structures, the implementation of W-tag should always carry an identifier for such purpose. For example, *legal entity-based ECID* is a W-tag to anchor the shadows with the meaning that the ECID is modeled from legal entity-based perspective.

So far, we have used the term perspective in common sense.. To further proceed, we need a precise definition here. By dictionary[8], the relative explanations include (1) the appearance to the eye of objects in respect to their relative distance and positions, and (2) the interrelation in which a subject or its parts are mentally viewed. We use the first meaning in the descriptions for metaphor of shadows, and we use the second meaning when we describe how we model meanings as mental entities in semantic space.

Later when we discuss algebra and calculus, we will continue use the second meaning for the term perspective with a symbol for representation. Specifically, we assume that mental entities in a single perspective are consistent or at least not in conflicts. Although human can hold inconsistent mental entities in their cognitive structures, we need to restrict our representations for simplicity reason such that inconsistencies or conflicts must not exist within a single perspective in our model.

Next, the following two rules define how we can use W-tags to attach to shadows in data space. The reason we request each shadow must have at least one W-tag (even just a place holder like *something unknown*) is simply to have at least one way to access the shadow in data space from semantic space.

**W-tag Rule # 2.** <u>A W-tag can be attached to shadows, any piece of data, and there is no constraint of what kind of logical structure the shadows should have.</u> Unlike a relation in Relation Model, which has fixed rigid logical structure for attributes to fit in, W-tags can be attached to any kind of shadows without worrying about the logical structures of the data.

**W-tag Rule # 3** <u>Any shadow must be attached with at least one W-tag in order to be managed in semantic space</u>. If a shadow has multiple meanings, then it should be attached with multiple W-tags.

Philosophically, Shadow Theory can only assume the existence of shadows (due to some *things* through projection process), but not the existence of the *things* themselves. The reason is that we cannot know what the *things* are without choosing a perspective and ontology. However, if we choose one, then the knowledge about the *things* will be constrained by the choice. Since our objective is to resolve semantic heterogeneity by modeling such different choices made by different people, we do not want to establish the model with potential bios in the foundation.

Therefore, there is standard about what W-tags should be attached to what shadows; that is, we do not assume **single version of the truth**. Since different viewers can have different cognitive structures to model the reality, they may have different W-tags representing "the same" *thing*. Furthermore, the criteria for what is "the same" *thing* is a subjective decision made by data integrators; in other words, in the cognitive structure of the data integrators, the mental entities anchored by W-tags and imported (learned) from different viewers due tot heir perspectives, there exist enough similarity or other characteristics such that data

---

[8] Merriam-Webster dictionary published by Merriam-Webster, Inc., Springfield, MA, USA.

integrators believe the different mental entities actually represent "the same" *thing* in reality.

We need a dictionary-like mechanism to explicitly provide descriptions for W-tags in human readable way. The purpose of such dictionary is not to provide absolute definitions like a perfect representation for the meanings represented, but rather a representation that the viewers who already have such mental entities can recognize the unique association between W-tags and the mental entities. For other people who do not have such mental entities in their cognitive structures, such descriptions can serve for them to learn the associated concepts in order to establish mental entities in their own cognitive structures. In this way, different users compile their own dictionary about their W-tags, and data integrators can collect them into where integrated data exist, and make needed revision or add extra information in order to suit for the purpose of data integrators. For application domains where consensus can be reached within specific groups of people, a domain expert can be the one to collect common mental entities with descriptions accepted by people in the group to recognize the common mental entities unambiguously.

### I. W-tags are extended notion of names

Here we need to make some comparisons with Relation Model in order to better explain the basic concept of W-tags. A W-tag is not like a pointer to data; it is more like an **extended notion of names** (of attributes, tables, or domains) to overcome the issue of semantic heterogeneity by

(1) Representing explicit meanings to data instance level (instead of schema level) to mental entities.

(2) Recognizing uniqueness explicitly due to the mental entities in specific viewers' cognitive structure, not necessarily due to the property of the subject matter in reality.

In Relational Model, combined with domain values, names of attributes, or name of tables are the mechanism to represent the meanings of data. At schema level, Codd explained that the purpose of a column name includes: "*1. such a name is intended to convey to users some aspect of the intended meaning of the column; 2. it enables users to avoid remembering positions of columns, as well as which component of a tuple is next to which in any sense of "nextness;" 3. it provides a simple means of distinguishing each column from its underlying domain. A column is, in fact, a particular use of a domain.*" (p.3 [10]). At data instance level, he explained that different domains (as extended data type) are "…*intended to capture some of the meaning of the data*" such that the system can tract the difference when the same basic data types (e.g. INTEGER, CHARACTER, BOOLEAN) used to represent semantically distinguishable types of real-world objects or properties (see p.43 in [10]).

However, as we have illustrated by the ECID example in section 2.1 that the meaning(s) of an attribute value may not be able to be determined uniquely by its attribute name or domain values when the table is overloaded with semantic heterogeneity. The purpose of W-tags is to overcome such issue by explicitly anchoring with the unique meanings in viewers' cognitive structure, and also bypass the limitation of logical structure for the data. The advantages of using W-tags over names (of attribute, table, domain) include:

(1) To provide users descriptions about the represented meanings at data instance level, such that the viewers who already have the mental entities in their cognitive structures can uniquely identify the meanings.

(2) To enable users to avoid remembering the logical structures or the formats of the data, such that data query can be done by W-tags without detailed descriptions for the associated logical structures like what is required today in SQL. We will provide details in the ECID example in section 5.3.1.

(3) Semantic relations among W-tags can prove the associations among these mental entities, and we will discuss the details and proposed usage rules in next section).

(4) To provide a mechanism to identify different meanings for the same representation. For similar but different meanings, different W-tags will explicitly represent their differences, and the semantic relations among W-tags can represent their similarity.

For semantic homogeneous environment, W-tags can be simplified as table names or attribute names in Relational Model. In fact, we can simulate the functionality of relational schema by templates of W-tags. We will discuss how to perform so in section 5.4.

### II. WID in Semantic Space: the Alternative to Unique Keys in Data Space

Here we need to pay attention for the notion of uniqueness. The uniqueness of W-tag instances is due to the uniqueness of the mental entities in the viewers' cognitive structures. For a group of people with common mental entities in their cognitive structures, the uniqueness is due to the commonly accepted mental entities in the shared semantic space to represent *things*; however, it is not necessarily due to the uniqueness of the *things* themselves in reality.

We believe this is the critical point for resolving semantic heterogeneity, and the reason is simply due to the fact that we cannot know what the *things* are without choosing a perspective implicitly or explicitly, with or without an ontology formally rperesented. For Shadow Theory and our purpose to design data models for data integration purpose, we need to carefully model such differences explicitly to avoid injection of potential bios to prevent users from recognizing where the issues of semantic heterogeneity are.

Therefore, we need adjust the notion of unique keys and propose the concept of unique **W-tag Instance Identifier** (WID) as the following rule:

**W-tag Rule # 4.** Any instance of W-tag carries a unique W-tag Instance Identifier (WID). When a new instance of W-tag is created, the database system should assign a unique WID to attach on the shadow, and the WID represents a unique meaning as mental entity in viewers' cognitive structure. Note that the uniqueness of WID is due to the cognitive structure of the specific viewer, not due to the *things* in reality.

The main difference between WID and surrogate keys described in [22, 60, 61] is that, surrogate keys are unique due to the logical representation in data space, not due to the unique existence of mental entities in viewers' cognitive structures: the semantic space about how viewers understand the reality from their specific perspectives. Therefore, we can only treat surrogate keys as shadows captured by database systems, and their uniqueness may not be able to be associated to a meaning.

For the practical data integration, WID and unique keys can work together in the integrated data model in order to support data operations in semantic space. As we described in section 2.1 that relying on unique keys is a convention common in data model design, but when shadows flow out of their original source systems into downstream systems, the uniqueness is no longer valid due to semantic and structural heterogeneity that the original data model cannot control.

Specifically, we need to overcome the issues of ghost problem in information ecosystems (as discussed in section 2.1) that when upstream systems delete the unique keys, downstream systems cannot perform the same due to business requirements (e.g. upstream systems hold historical data for 2 years, while downstream systems may be required to hold for 7 years). We can move the unique keys into historical archive of the W-tag instances, but keep the WID and their connections with other W-tags. Therefore, downstream still can recognize the meaning of the shadows that no longer exist in their source systems. For example, when upstream system physically delete a unique key for a legal entity-based ECID, downstream systems which are required to hold their associated ordering records or billing invoices can continue rely on WID to keep all of the W-tags still connected together.

This also can help to resolves the unique key reuse problem that when the same unique key is used for representing something else in the upstream systems, the downstream systems can continue hold the same WID for the W-tag instance but without the old unique keys values. We also resolve the rollback problems, as all we need to do is just move shadow values from archive to where they were.

For referential integrity, in section 4.1 we will introduce E-tag and semantic equivalence that we will use as an alternative solution for the relation between primary key and foreign key. Such referential integrity is established on top of WID, the existence of W-tag instance for the unique existence of mental entities, not by the unique key values. One may wonder when we should really delete the W-tags and their semantic relations in the integrated database of downstream systems. It depends on the business requirements or legal constraints, just as it is now in practical applications. Compared with current practice to design extra historical tables without referential integrity, our approach provides a simpler solution at data model level to support the gap between systems requirements.

### 3.4.3 *Decomposition of Meanings in Semantic Space*

Next, we will discuss decomposition mechanism in semantic space and their associations in data space. In data space, shadows can be decomposed into sub-shadows with some principles to organize the overall structures. For Relational Model, the decomposition happens during schema design process with a set of principles (i.e. Normal Forms) to evaluate the results. That is, the representations of the subject matters are decomposed to a set of tables that are decomposed to individual columns, and these tables and columns should interact with each other following the basic mechanism supported by Relational Algebra.

Naturally, such decomposition concept is semantic in nature, however, the implementation focus on logical representation, i.e. the objective of the design theory focuses on whether (i) each particular table/relation is in "good" form, (ii) the decomposition is a lossless-join decomposition, such that Relational Model can support correctly and efficiently. For example, chapter 7 in [39] summarizes the comprehensive design theory based on functional dependencies and multi-valued dependencies for such purpose. Since we have realized that the notion of key is not universal applicable in information ecosystems under the context of semantic heterogeneity, and function dependency is essentially a generalization of the notion of key, the traditional schema design principle is not helpful when we have to include inconsistencies or even conflicts from different perspectives.

### I. *Generic Decomposition Mechanism in Semantic Space*

Therefore, we have a need to re-think about how we can and should decompose shadows in data space, and mostly importantly, how we can perform such decomposition in semantic space. Semantically, decomposition of a shadow corresponds to the decomposition of the mental entities (the meaning) of the *thing*, in other words, decomposition of the meanings in semantic space. Here we will first visit two decomposition scenarios in the ECID example before we propose a solution based on Shadow Theory.

Graphically, Figure 14 illustrates two decomposition examples in Venn diagrams. (1a) illustrates that the W-tag of an *Enterprise Customer* can be decomposed to two sub-W-tags of the same type *Enterprise Customer*; that is, within the same perspective like legal entity-based, an enterprise customer includes two sub-components and each one is represented as a legal entity-based ECID again. We can also use a hierarchy to represent their decomposition relations like (1b), which is closer to our common views about organizational structures. We can call such decomposition as **homogeneous decomposition** since it is the same kind of W-tag used for the sub-components.

Figure 14 (2a) shows a different scenario for decomposition that the W-tag of an *Ownership* contains two required sub-components which are under different W-tags: *Enterprise Customer*, and *Service Account*. That is, an *Ownership* is due to an *Enterprise Customer* owns a *Service Account*. In terms of Entity-Relationship Model, this is like a relation between *Enterprise Customer* and *Service Account*. To avoid semantic relativism issue (see section 3.3) to make data integration harder, we treat it as a W-tag to represent the meaning without separate representation of entities versus relations. We can also use a hierarchy to represent their decomposition relations like (2b). We call such decomposition as **heterogeneous decomposition** since the sub-components are different kinds of W-tag than parent's W-tags.

Since each shadow and any of its sub-shadows is required to have at least one W-tag, the decomposition of shadows can be parallel with a decomposition of W-tags in semantic space. However, it is not required in the reverse direction, i.e. not every W-tag in a decomposition structure in semantic space must have associated shadows in data space. The reasons include:

(1) Databases are not required to collect all of the data corresponding to every mental entity in viewers' cognitive structures; it is neither practical nor useful since W-tags in semantic space can be applied as generic template for logic operations even without materialized for each instance.

(2) For information ecosystems, we need keep W-tag even after the data are physically deleted in data space in order to resolve issues like ghost problems, unique key reuse, or rollback. The idea can be explained through Shadow Theory that the disappearance of a shadow is not equal to disappearance of the *thing* in reality, and even the *thing* actually no longer exist in the real world, the associated mental entity may still live in viewers' cognitive structures as historical records.

For example, a shadow is attached with W-tag *car*, and the decomposition structure of *car* in semantic space includes W-tags of *wheel*, *engine*, *seat*, *windows*, and son on. It is not required for the database to have shadows attached with *wheel* in the database: we know the existence of wheel(s) semantically, it is just the database may not collect such shadows and store them physically. Therefore, we have the following rule:

**W-tag Rule # 5.** <u>Parallel structure is not required between the decomposition of W-tags in semantic space and the decomposition of shadows in data space</u>.

One interesting and confusing characteristics about meaning is that it can be modeled as different level levels of abstraction. It triggers our special attention since it is one of the major issues encountered in data integration, and researchers reported that incompatibility among levels of abstraction is one of the root issues [9]. The reason is that traditional data modeling usually only focus on specific level(s) implicitly, and there is no explicit representation for the notion of different levels of abstraction. Hence, even there is flexible mechanism to support meanings propagated from one level to another, it is difficult for users to recognize the special effects due to meanings propagated through levels of abstractions. However, such level shifting is common in human communication, and we need to explicit represent involved concepts in order to support the notion of equivalence or similarity. Now we will define levels of abstraction by decomposition structure. In section 4.1.3 we will further discuss level shifting and semantic equivalence such that different meanings at different levels of abstraction may refer to the same subject matter and being treated as the same.

**W-tag Rule # 6.** <u>Different levels of abstraction</u>**.** For heterogeneous decomposition, we call the W-tags in the lower level of a decomposition structure as W-tags at lower level of abstraction, compared with W-tags at higher level in the decomposition. For homogeneous decomposition, since W-tags of parent and child are the same, we call the W-tag instances are at different levels of abstraction.

For example, like what is illustrated in Figure 14 (1a), a legal entity-based ECID which represents the international headquarter of a bank, may be decomposed to several legal entity-based ECID which represent its headquarter branches in different countries, and be further decomposed to legal entity-based ECID which represent its branches in different states or providences. We can say there are three levels of abstraction in the decomposition structure: international level, country level, and state / providence level.

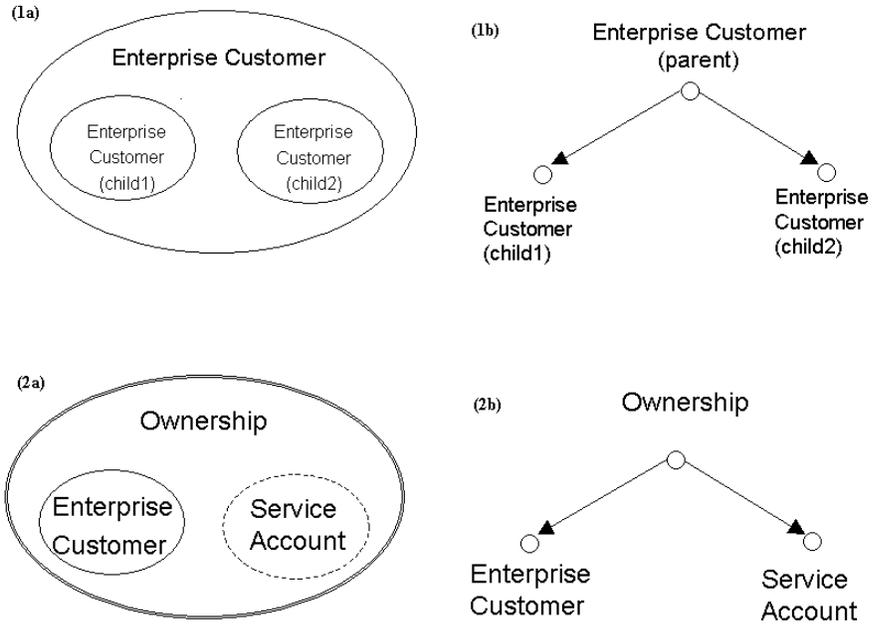

**Figure 14. Decomposition of meanings in semantic space.**

**Homogeneous decomposition**: (1a) illustrates that the W-tag of an *Enterprise Customer* can be decomposed to two sub-W-tags of the same type *Enterprise Customer*. We can also use a hierarchy to represent their decomposition relations like (1b), which is closer to our common views about organizational structures.

**Heterogeneous decomposition**: (2a) shows that the W-tag of an *Ownership* contains two required sub-components which are under different W-tags: *Enterprise Customer*, and *Service Account*. We can also use a hierarchy to represent their decomposition relations like (2b).

*II. Semantic Inheritance*

With the above rules and explanations, readers may sense that our objective is to use decomposition mechanism as the single generic representation for supporting semantic relations among W-tags. One challenge is that we need a mechanism to represent the propagation of meanings with an indication of the direction. Before we present our proposal, we will again use a scenario in ECID example to illustrate the phenomenon.

Figure 15 illustrates an example with two kinds of W-tags: *Enterprise Customer* and *Ownership*. To illustrate their interaction graphically, we adjust our graphical representation for *Ownership* from a hierarchy to an arrow to indicate the sub-W-tag on left is an *Enterprise Customer*, and the one on right is an *Service Account* (short as *Acct*), as displayed in the top two diagrams. The difference between solid double line arrow versus dashed line arrow is that, solid line arrow represents that the W-tag is explicit specified as a mental entity in the viewers' cognitive structure, and the dashed line arrow represents that the existence of this W-tag as a mental entity is due to meaning propagation through different levels of abstraction.

Note that they represent the same meanings (the Venn diagram in Figure 14 2a, 2b, and the arrow here), and we do not indicate *Ownership* is a relationship type like in Entity-Relationship model, as we have explained in previous sections to avoid semantic relativism issue. The only purpose for different kinds of graphical representations is for readers easier to understand the meaning propagation phenomenon. It is just another application of our first principle that the different graphical representations are only shadows, and we use these different shadows to represent the same meaning to service different **functional objectives** (i.e. Venn diagram one is to show the nature for area-based model in semantic space, the hierarchy one is to highlight the decomposition direction, and the arrow one is to use Relationship-like notion for readers to understand the complexities easily).

Figure 15(a), the hierarchy on the left represents a homogeneous decomposition structure for shadow *EC root*, which is tagged as *Enterprise Customer*. The sub-shadows *EC1*, *EC1.1*, *EC1.1.1*, and so on with the same kind of W-tag represent their semantic relations within the decomposition structure. That is, the decomposition structure can be used to represent the organizational structure for this specific enterprise customer.

The shadow *Acct1.1* on the right side is tagged as *Service Account*. The solid line arrow (1) represents the shadow tagged as *Ownership*(1), which can be decomposed to sub-shadows *EC1.2* and *Acct1.1* with their individual W-tags *Enterprise Customer* and *Service Account*. That is, this solid line arrow represents the meaning that the specific enterprise customer represented by *EC1.2* owns the specific service account *Acct1.1*, denoted as *EC1.2→Acct1.1* for short.

The dashed line arrow (2) represents the shadow tagged as *Ownership*(2) that can be decomposed to sub-shadows *EC1* and *Acct1.1* with W-tags *Enterprise Customer* and *Service Account* individually, short as *EC1→Acct1.1*. The existence of *Ownership*(2) is due to the existence of *Ownership*(1) and the semantic relations between *EC1* and *EC1.2*. That is, since *EC1.2* is a sub-organization of *EC1*, *EC1* also owns the *Acct1.1* that *EC1.2* owns.

Similarly, we can infer the existence of *Ownership*(3) since *EC root* is parent of *EC1* in the decomposition structure. We can describe this phenomenon as that the same meanings can propagate to different levels of abstraction in **upward direction** along certain decomposition structure. In other words, the notion that we think a parent company also owns the account its sub-company owns, can be viewed as meanings (as mental entities) in which their sub-components (also as mental entities) can shift their levels of abstractions towards higher levels in the decomposition structure.

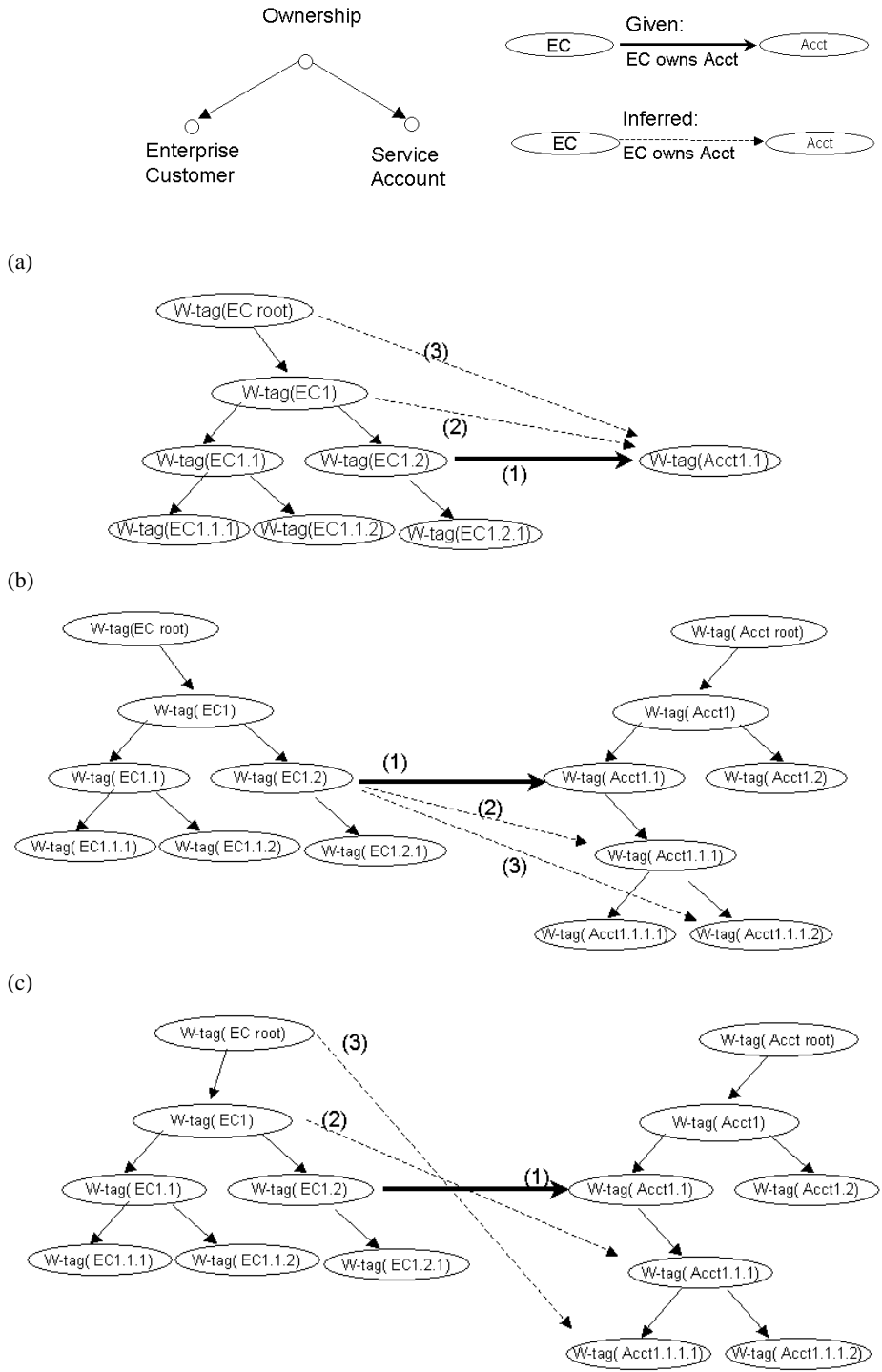

(a)

(b)

(c)

Figure 15 Meanings can shift to different levels of abstraction along the semantic relations in decomposition structures. Top graph: to help readers easier to understand the complexities, we alter the graphical representations of W-tag *Ownership* from hierarchy to arrows between its sub-components: from *Enterprise Customer* to *Service Account*. Solid line arrows are given, and dashed line arrows are the results due to level shifting. **Upward direction** is illustrated in (a) such that parent companies also own the accounts of its child companies. **Downward direction** illustrated in (b) such that a company owns the account as well as it sub-accounts. Graph in (c) shows that the two direction of level shifting can happen at the same time for the sub-components in W-tag *Ownership*.

Next, Figure 15 (b) shows similar level shifting but in **downward direction**. The hierarchy on the right represents a decomposition structure for service account that higher level accounts includes lower level accounts. The solid line arrow (1) represents the shadow tagged as *Ownership*(1), which can be decomposed to sub-shadows *EC1.2* and *Acct1.1* with their individual W-tags *Enterprise Customer* and *Service Account*. That is, this solid line arrow represents the meaning that the specific enterprise customer represented by *EC1.2* owns the specific service account *Acct1.1*, denoted as *EC1.2*→*Acct1.1* for short.

The dashed line arrow (2) represents the shadow tagged as *Ownership*(2) that can be decomposed to sub-shadows *EC1.2* and *Acct1.1.1* with W-tags *Enterprise Customer* and *Service Account* individually, short as *EC1*→*Acct1.1.1*. The existence of *Ownership*(2) is due to the existence of *Ownership*(1) and the semantic relations between *Acct1.1* and *Acct1.1.1*. That is, since *Acct1.1.1* is a sub-account of *Acct1.1*, *EC1.2* that owns the *Acct1.1* also owns *Acct1.1.1*.

Similarly, we can infer the existence of *Ownership*(3) since *Acct1.1.1* is parent of *Acct1.1.1.2* in the decomposition structure. We can describe this phenomenon as that the same meanings can propagate to different levels of abstraction in downward direction along certain decomposition structure. In other words, the notion that we think a company owns the account as well as its sub-accounts, can be viewed as meanings (as mental entities) in which their parent-components (also as mental entities) can shift their levels of abstractions towards lower levels in the decomposition structure.

Further, the two different direction of level shifting can work together as illustrated in Figure 15(c). We can think the meaning of shadow (2) *EC1*→*Acct1.1.1* inherits from the meaning of shadow (1) *EC1.2*→*Acct1.1* through (i) upward level shifting along the hierarchy (decomposition structure) of *Enterprise Customer*, and (ii) downward level shifting along the hierarchy of *Service Account*.

To model such phenomenon and especially the direction to which level meanings can shift, we can use the IS-A and HAS-A semantic relations proposed in Semantic Data Models [62] [55, 63] [64]. That is, if the semantic relations between parent and child W-tags in a decomposition structure are marked with IS-A, then the meanings can shift downward since any child W-tags still hold the IS-A relation. If the semantic relations between parent and child W-tags in a decomposition structure are marked with HAS-A, then the meanings can shift upward since the parent W-tags still hold the HAS-A relation.

Since we propose to use areas to represent meanings in semantic space, we can interpret such semantic relations as the following. In terms of Venn diagram, the child W-tags are within the areas of parent W-tags such that if the child W-tags have IS-A relations with parent W-tags, then meaning can shift downward the hierarchy. For example, if parent W-tags represent RED areas, child W-tags also represent RED area. If the parent W-tags have HAS-A relations with parent W-tags, then meaning can shift upward the hierarchy. For example, if child W-tags represent have the something X, then parent W-tags also have the something X.

We can call this property as **semantic inheritance**. The notion of IS-A and HAS-A can be specified to control the direction for how meanings can propagated/inferred across levels of abstraction. Hence, we propose the following rule:

W-tag Rule # 7. **Semantic inheritance among W-tags**. Meanings as mental entities that are anchored by W-tags can be propagated to different levels of abstraction along a decomposition structure. Specifically, when a specific W-tag $X$ is referred by $Y$ (i.e. to be included as a sub-component of W-tag $Y$ in $Y$'s decomposition structure $D^Y$), $X$ can shift to different levels of abstraction (from $X$ to $X1$ at higher or lower levels) along $X$'s decomposition structure $D^X$ and infer the existences of different $Y$ W-tags due to the semantic inheritance process.

(i) If the inheritance is in upward direction (i.e. $X1$ is at higher level than $X$ is), then the semantic relation between $X$ and $X1$ is marked as HAS-A.

(ii) If the inheritance is in downward direction (i.e. $X1$ is at lower level than $X$ is), then the semantic relation between $X$ and $X1$ is marked as IS-A.

Note that different shadows attached with different kinds of W-tags that exist in the same decomposition structure (but at different levels of abstraction) may be used by different systems to represent the same *thing* based on certain criteria subjectively. Another situation is that different shadows attached with different kinds of W-tags that exist in different decomposition structure may be used by different systems to represent the same *thing*. We use the term **semantic equivalence** to describe either case as the meanings of different shadows anchored by different W-tags can be treated as the same. It does not mean these shadows are the same shadow; it indicates that there exist overlapping areas in semantic space for their associated W-tags. If we need to distinguish the two situations, we can call the former as **vertical semantic equivalence** since the equivalence cross different levels of abstraction vertically, and the later as **horizontal semantic equivalence** since the equivalence happens between different decomposition structures (designed from on different perspectives). We will further discuss semantic equivalence in section 4.1.1 and the mathematical foundation.

Another major feature to model semantic inheritance is to model the situation that the original $Y$ W-tag and the inferred $Y$ W-tags may be treated as the same by data integrators to reach their functional objectives. In other words, although the sub-component $X$ and $X1$ are at different levels of abstraction, the original $Y$ W-tag and the inferred $Y$ W-tags may be used by different viewers to represent the same or overlapped meanings. Explicit model of this mechanism in semantic space gives us a tool to explain about the rich varieties of mapping based on similarity concept, as we can use decomposition structures to model similarity.

### III. Different Decompositions for the Same Meaning

W-tag Rule # 3 requests to have multiple instances of W-tags attached to a shadow if there are different meanings as mental entities in viewers' cognitive structure associated with the same shadow concurrently. These different W-tags can have different semantic relations with other, and hence have different decompositions in semantic space.

What we need to discuss here is about how to manage the situation that different decompositions may happen to the same meaning as a mental entity within a single perspective. Although human viewers can naturally model different decompositions for the same meaning, we need to explicitly distinguish different decomposition structures. One choice is to have another kind of identifiers for decomposition structure; that is, we use W-tags to uniquely anchor meanings as mental entities and we need an extra identifiers to recognize the different decompositions the W-tags can have. Another choice is to avoid this extra identifier by requesting that each W-tag can only allow one kind of decomposition; if there are different kinds of decompositions, then different W-tags are required.

Back to the basic principles of Shadow Theory, when there are different ways for decomposing the same meaning as a single mental entity in viewers' cognitive structure, it indicates differences in terms of their behaviors in semantic space. When the viewers associate different behaviors to the same mental entity, they naturally performed some kind of semantic equivalence operations, i.e. no matter how different the decompositions are, the meaning is the same, representing the same *thing* in the world.

Therefore, we choose the approach to force a W-tag can have only single decomposition structure, and we will need a mechanism to treat them as the same, which will be introduced in section 4.1 as Equivalence-tag, short as E-tag. With the help of E-tags, we can recognize which W-tags are associated to the same *thing* but different decomposition structure, no matter these W-tags are in the same or different perspectives. In this way, we further push for explicit representation such that different meanings as different mental entities and the same mental entities but with different behaviors (decompositions) in semantic space must be represented by different W-tags. The rule can be summarized as the following:

**W-tag Rule # 8.** Only one decomposition structure is allowed for a W-tag. If a meaning as a mental entity can be decomposed in different ways, then there should be different kinds of W-tags to identify each one of the different decompositions of this meaning.

After we introduce the notion of Equivalence-tag, short as E-tag, in section 4.1, we will use an example to explain how we can establish equivalence between these different W-tags in section 4.3.

## 3.5 Semantic Heterogeneity and Meaning Independence

Now we can review the very first question we asked:

Question 0: **What is semantic heterogeneity**?

In section 2.1 when we raised this question, we quoted some descriptions or definition from several researchers for readers to understanding the nature of the problem. Further, we also try to describe the two basic characteristics of semantic heterogeneity as:

(i) There exist different representations for the same meaning.

(ii) There exist different meanings for the same representation.

In traditional modeling approach, the difficulties to model these two properties are due to that we only include representations (e.g. schema) in the scope of the data model, but not explicitly represent meanings, or explicitly represent what are the same meanings, versus what are different meanings. What is even more difficult is how we can model the subjective decisions made by different viewers about what can be treated as the same meanings, based on various criteria like similarity, commonality, probability, and so on.

The efforts we spent in previous sections to establish Shadow Theory, and the reasons to choose Gärdenfors' approach that meanings are mental entities, are to overcome this difficulty such that we can have an explicit way to represent meanings. Note that explicit representation does not necessarily need to be in formal or complete ways like ontology. We can rely on W-tags to uniquely anchor with these mental entities that existing in different viewers' cognitive structures to fulfill the needs of explicit representations. Of course, users can capture all of their mental entities if they want to be formal and complete, but here we will just need the related ones that are required by the business needs to reach their functional objectives.

Now we can provide a definition based on Shadow Theory with the goal to support algebra we want to develop. We will first review the different approaches surveyed by Gärdenfors (that we have summarized in section 3.4) regarding the philosophical dispute about the meaning of "meaning" (see Figure 13), then we will discuss why mental entities can better explain semantic heterogeneity for our purpose.

If we choose functionalist tradition that meaning is in its communicative function, semantic heterogeneity can be viewed as differences among such communicative functions. However, for semantic heterogeneous data stored in various database within information ecosystems, it is difficult to explicitly represent the differences among communicative functions for the purpose of data integration.

If we choose the approach of realist semantics, semantic heterogeneity implies differences among *things* in the real world physically, or differences among *things* in possible worlds or situations. This is also difficult for us to model since we are trying to integrate data due to the same *thing* semantically, and truth values assigned to sentences offer little help as they can not explain the rich varieties of differences or similarity.

Now, for the cognitive approach in which meanings are mental entities in viewers' cognitive structures, semantic heterogeneity can be viewed as the differences among such mental entities. In this way, the factors of different perspectives and different levels of abstractions can be modeled as the characteristics of the cognitive structures (which reflect how different viewers model the reality), and we can also model the rich varieties for meanings and shadows. This can match very well what we proposed in section 3.4 for representing meanings through W-tags to anchor mental entities uniquely.

Hence, we can propose the following principle for define semantic heterogeneity for the purpose of data integration based on Shadow Theory:

Principle 3. Semantic Heterogeneity is the overall aggregated result due to differences among meanings as mental entities in viewers' cognitive structure, and differences of how shadows are projected onto wall-like surface of system requirements about the same subject matter.

Our objective is to represent semantic heterogeneity as explicit as possible. The basic idea is to use different W-tags to represent (i), and to use E(quivalence)-tag to represent (ii). Here we will further explore how different W-tags can explicit represent semantic heterogeneity, and E-tags will be introduced in next section.

Figure 16 illustrates three situations graphically for different meanings for the same representation. Case (1) is the simplest one, in which shadows cannot be further decomposed and they hold properties associated to the same subject matter. The result is that different viewers have different W-tags (i.e. W-tag1 and W-tag2) attached to the same shadow. In other words, the same subject matter is project to a shadow that is interpreted as different meanings corresponding to different mental entities in different viewers' cognitive structure. We can say that semantic heterogeneity is represented as different W-tags in this case.

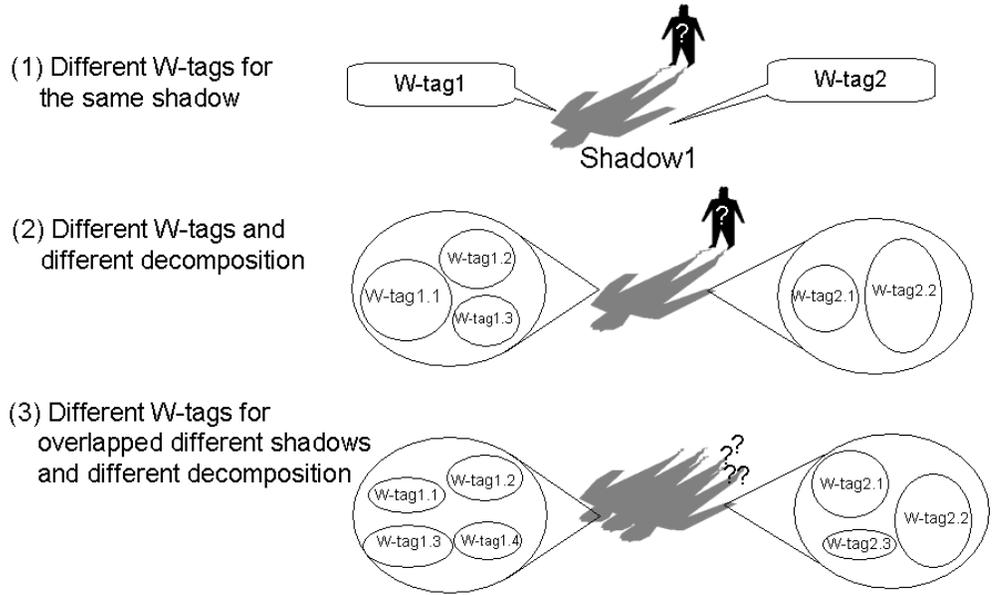

**Figure 16.** Three cases for semantic heterogeneity that different meanings for the same representation. Case (1) Different W-tags for the same shadow such that semantic heterogeneity can be represented by difference of W-tags. (2) Different sets of W-tags are attached to the same shadow due to different ways of decomposition. (3) Different sets of W-tags attached to overlapped shadows and the subject matters do not have universal accepted ontological units.

Case (2) includes the complexity due to different decomposition mechanism. Assuming the same shadows only hold properties associated to the same subject matter, the different decomposition mechanisms trigger different sets of W-tags attached to the same shadow. For example, W-tag1 is decomposed into W-tag1.1, W-tag1.2, and W-tag1.3, while W-tag2 is decomposed into W-tag2.1 and W-tag2.2. We can say that semantic heterogeneity is represented as different sets of W-tags, and there is no 1:1 correspondence among these W-tags.

Case (3) includes extra two complexities. The first one is that the shadows may be projected from *things* (represented by question marks in the illustration) that do not have a universally accepted ontological unit from different perspectives. The second is that the shadows may not be able to be cleared viewed as a single one, but more like overlapped shadows for the *things*. Combined with the factor of different decomposition mechanisms, we may end with W-tag1.1, W-tag1.2, W-tag1.3, and W-tag1.4 in the left side, while W-tag2.1, W-tag2.2, and W-tag2.3 on the right side. As a result, semantic heterogeneity is represented as different sets of W-tags, and the difference is due to the chosen ontology.

Case (4), not illustrated in Figure 16, introduces another level complexity that is one step closer to the practical issues we encountered in the ECID example: not only that we do not have a universal ontological unit for the *things* to project shadows, but also that different viewers may not be able to view the *things* with full knowledge about the reality (like viewing only a portion of an elephant). The challenge in this case is really about the definition what is the same and what is different in the sense of both logical representation and their meanings that viewers can capture in their cognitive structure. This is where the spectrum of similarity may happen, as viewers must make subjective decision through their chosen ontology. As a result, semantic heterogeneity is represented by differences among W-tags or sets of W-tags, and the perspectives where W-tags exist include scope or limitation for what the viewers can see about the subject matter.

In the ECID example we have discussed in section 2.1, there is no universal way to model an enterprise customer without a subjective decision about the ontological primitive units. Further, the scopes of shadows that service provider G, B and W can model about their enterprise customers are limited to their specific service areas and interactions with enterprise customers for specific products. In other words, we do not know what is the truth; especially, we cannot assume a single version of the truth that the organizational behaviors of all enterprise customers must behave the same (e.g. M&A activities).

These four cases may illustrate some basic situations for semantic heterogeneity that different meanings for the same representation. To fully explain the issues in ECID examples, we need to add the case (5), that different data models do not have the same logical representations for the overlapped shadows. This includes different data values, formats, domains, and logical structures chosen to represent specific shadow properties. As a result, semantic heterogeneity is represented as differences among W-tags or sets of W-tags that are attached to different shadows collected from various sources.

Even more, we need case (6) that the properties of shadows are not only due to the subject matters. There are many other factors that contribute to the properties of shadows recorded in databases. This is the place we need to use the three categories of shadow characteristics introduced in section 3:

(C1) Properties due to characteristics of the subject matter.

(C2) Properties due to characteristics of the systems.

(C3) Properties due to characteristics of projection process.

For case (1) to (5), we mainly consider shadow properties due to the subject matter. Here we need to think the impacts due to the system requirements and the projection process itself. For system requirements, they include specific business semantics, operations logics, required logical data formats, or chosen perspectives about how to model the subject matter. For the projection processes, they include the methodology used to build data models, like to represent data as an entity, an object, an attribute, or as a relation, in addition to their algebra or process methods.

We can further classify the differences, inconsistencies, or even conflicts into these three categories. It is a helpful exercise for data integrators to examine the nature of issue and clarify the objective of data integration projects. For example, should the downstream systems integrate all three kinds of properties from their upstream systems? Or, is the objective only about subject matter such that downstream systems can ignore those properties of (C2) and (C3) introduced by upstream systems? Could the issue be mainly due to different perspectives chosen by different upstream systems such that the best potential solution is for business to coordinate their perspective first?

However, (C1), (C2), and (C3) are often mixed together and difficult to separately represented explicitly, we cannot to represent semantic heterogeneity they triggered along this classification. Instead, we propose to represent them along the separation of semantic space and data space, which we have discussed in section 3.3. That is, semantic heterogeneity due to different meanings for the same representation, of for representation of the same (or similar) subject matter, can be represented as differences among W-tags or sets of W-tags in semantic space. We need another mechanism to fully capture the differences in data space, including any differences in logical representations that may be even triggered by semantic heterogeneity in semantic space.

Therefore, we proposed to use **P(rojection)-tag**, short as **P-tag**, as the mechanism to denote the properties of logical representations. It is used for data space only for the characteristics of shadows being projected onto wall-like surface, including data types, formats, logical structure, (logical) uniqueness constraints, or other factors required by chosen data models (e.g. the shadow is treated as an entity, an attribute, or a relation). A set of rules for proper using P-tags includes:

**P-tag Rule #1**. P-tags are attached to shadows to represent properties due to logic models or to satisfy system requirements.

**P-tags Rule #2**. P-tags are optional (i.e. in certain default logical representations), not required for every shadow, and multiple P-tags can be attached to the same shadow if they are consistent.

**P-tags Rule #3.** If therefore is a need to enforce data formats for shadows such that those shadows fail to satisfy the criteria should be prevented from being loaded into database, we can specify the P-tag as required. In such case, it must be associated with specific kinds of W-tags such that the enforcement is performed when shadows are attached with the W-tags[9].

**P-tags Rule #4.** If there is a need to convert shadows between different logical formats (e.g. different time formats), users of the databases can specify P-tags in their query such that database should retrieve shadows and convert accordingly.

### 3.5.1  *Meaning Independence versus Data Independence*

One advantage for separate mechanism of W-tags and P-tags is that, data integrators can have the opportunities to consider what should be integrated along the separation of semantic space and data space. Since P-tags may not directly due to subject matter, the downstream system may not need to follow the same logical representations.

Since shadows do not hold meanings and meanings of shadows only exist in viewers' cognitive structures, Shadow Theory is established on the notion of separated semantic space and data space. W-tags anchor with the meanings as mental entities in specific perspectives and at specific levels of abstraction, then be attach to shadows that are collected from various sources in information ecosystems. In this way, data integrations are performed in semantic space due to their meanings, not constrained in data space by their logical representations.

We can call this concept as **meaning independence**, built on top of the concept of data independence for Relational Model. Data independences proposed by Codd [65] establishes the foundation for Relational Model such that data can be modeled and managed without knowledge of their physical structures. In a similar way, the concept of meaning independence is for data integration to be performed without the constraints of the underlying logical structures.

Readers may wonder that whether semantic space is the similar concept like views in the traditional levels of abstraction used for database [Silberschatz, 1997 #1574] physical level, logical level, and view level. They are very different due to the following reasons:

(1) **Constrained versus not constrained by logical representation**. Views depend on schema and are not independent from logical representation, but W-tags in semantic space can be totally independent from the logical representations of shadows in data space.

(2) **Single version of the truth versus multiple version of the truth.** Traditionally, the design of schema is an effort to consistently integrate the requirements of different views that it need to support (i.e. the hidden assumption for single version of the truth). W-tags in semantic space are designed to support multiple versions of the truth such that the semantic heterogeneity can be explicitly represented to users, to help them resolve difficulties in the real world (not to hide the real world difficulties under a faked unification in the model world).

(3) **Implicit versus explicit meanings to avoid semantic heterogeneity**. Meanings of views are represented by the same mechanism of schema, that is, by attribute name, view name, domains. It is designed mainly to reduce redundancy by storing "the same" data in a consistent way to avoid inconsistency; therefore it lacks explicit representation of meanings at data instance level and the criteria for "the same" versus semantic heterogeneity. W-tags in semantic space are designed to anchor with meanings explicitly at instance level, not at schema level.

---

[9] Note that we do not encourage the use of P-tags as the way to enforce data integrity; instead, templates of W-tags can serve as the integrity mechanism in semantic space. Those shadows failed to be attached with the combination of specific P-tags and W-tags are still shadows in theory, just like bad data is still data. It is up to the data modelers to determine if there is still any useful meaning and may attach such shadows with certain W-tags without P-tags.

(4) **Flexibility to adapt changes versus difficulty of schema evolution**. Evolution of views is bound with schema evolution, a known difficulty in Relational Model for decades due to inertia of data representations. However, semantic spaces are in human brains, so they continuously evolve as business semantics responds to their dynamic changing environments without waiting for the changes of logical representations in data space. The design of semantic space includes such flexibility that revision can be applied in semantic space without changes in data space. Users can create new W-tags for evolved meanings, and assign such news W-tags to shadows according to revised business requirements; further, users can use semantic relations between old and new W-tags as well as E(quivalence)-tags that we will explore in next section to quickly and properly adapt the changing business semantics.

The concept of meaning independence provides a foundation for using Shadow Theory to perform data integration. The basic ideas mimic how human make comparisons among different *things* reflected in their cognitive structures of the world. The logical representation in database is something external to human brains, hence it does not prevent human from making comparisons semantically directly through mental entities in their internal representations.

From this point of view, we believe that the key factor that why humans are still required in the process of data integration is due to the semantic space living in their brains. The research direction to fully automatically perform data integration is simply not possible, i.e. without semantic space, we cannot evaluate the meanings behind any data models for how similar of how differences the meanings are, or which is the same as which. Therefore, our objective is to help human to easier manage the current situations of data integration, to easier identify where semantic heterogeneity or structural heterogeneity exist (due to different ways of how human think, as well as due to the existing data models different people have designed).

In next section, we will move to the other characteristics of semantic heterogeneity that different representations for the same meaning. It is the keys for establishing mappings among different meanings that are now represented by W-tags.

# 4. Data Integration

In previous section, we have discussed how we could represent first characteristics of semantic heterogeneity by differences of W-tags. That is, the situation for different meanings for the same representation can be modeled through different W-tags attached to the same shadow, and there is only one meaning (as a mental entity in viewers' cognitive structure) anchored by each W-tag.

Now we can move to the second characteristics that different representations for the same meaning. That is, the mechanism to handle the situation that different data values with different logical structure can be treated as equivalent from some viewers' perspective since they represent the same meaning. Further, different meanings can be treated as the same due to their similarity, commonality, or other logic criteria. In other words, the challenge is that how we can model the following situations:

(i) Different shadows with different logical representations in data space attached with the same kind of W-tags (but different instances), can be treated as equivalent by algebra to support needed data operations.

(ii) Different shadows with different logical representations in data space attached with different kinds of W-tags can be treated as equivalent by algebra to support needed data operations.

The kernel of algebra we want to develop is that how different shadows can be compared semantically in semantic space without worrying about their logical representation in data space. In this section, we will first introduce E(quivalence)-tag, short as E-tag, as the bridge between mental entities with the details of Principle 4. Then we will introduce Principle 5 as the generic guideline for data integration process, and Principle 6 for what features of data model we need to support in order to help users to use the integrated data properly and efficiently.

## 4.1 E-tag: Equivalence of Shadows

Mapping is the key step in data integration to decide what data elements from different sources can be associated together and being treated as the same in order to fulfill the objectives of data integration. With the example of ECID integration, we raised the following question in section 2.3:

Question 3: **What is the nature of mapping?** In what sense can mapped data be treated as equivalence? Does such equivalence uni-directional or bi-directional?

Since we choose to model meanings as mental entities that only exist in viewers' cognitive structure, it is natural to answer the question as that mapping is performed among mental entities due to their meanings, not due to the logical representation of shadows in data space. In other words, no matter how different or similar two shadows may be, mapping can be done due to their meanings as mental entities, which are treated as the same through certain criteria chosen by data integrators.

Here, the term *bridging* can better describe our intension than the term *mapping* or *equivalence*, since *bridging* acknowledges the differences between mental entities, and the connection between them is for fulfill specific **functional objective** to access assets across the boundaries of different perspectives or different levels of abstraction.

This is similar to what researchers have pointed out that there is a need to recognize the difference between what we consider to be the same, and what the systems treat as the same [23]. First, in traditional approaches different shadows in data spaces are treated as different data elements physically and logically due to different database systems, and the way they can be treated the same is by exactly the same data values & logical structures (i.e. treated as the same by systems). Second, mapping is the mechanism to bypass the constraints of different logical representations such that user can treat the mapped data as the same semantically (i.e. considered the same by users).

However, in Relational Model, what users considered to be the same and therefore mapped together cannot be supported by the underlying Relational Algebra efficiently due to the different logical representations of these data. This is the exact weakness that we need to overcome: why not design a data model such that the database system can close the gap? In other words, let the system treats data elements as the same following how users consider the same semantically. Indeed, to help different users in information ecosystems with different criteria, we want the system to treat data as the same if users think the data have the same meaning in their view, no matter how different or similar their logical representation may be.

Further, different meanings as different mental entities may also be considered as the same by users from specific perspectives. The independent existence of different mental entities should also not prevent users to treat them as the same to fulfill certain functional objectives. Such mapping operations are common in human conversations, for example, when referring to specific objects like cars, people may have different mental entities like *vehicles* or *automobiles*. The communication can succeed as we have the capabilities to ignore the differences and treat them as the same under specific context of the conversation.

Therefore, we propose the following:

Principle 4. Equivalence between meanings (as mental entities from different viewers' cognitive structure) is a subjectively decision, and we can model such equivalence by E(quivalence)-tag, short as E-tag, with supporting evidences, just like a bridge to cross the boundaries of different perspectives.

We use the term E(quivalence)-tags, short as **E-tags**, to represent bridging between W-tags with these supporting evidences. Figure 17 illustrates what we described here with the example of legal entity-based ECID vs location-based ECID, represented as a different potions of elephants independently. The shadows as logical representation in data space may not share any common characteristics for their data values or logical structures, e.g. hierarchical versus graph structure. For what we can observe from reality based on specific perspectives, we may not even have any commonality or similarity. However, in terms of referring to the elephants as a whole, data integrators can map elephant heads to elephant tails and treat them as the same meaning in order to reach specific functional objectives. The mapping is not valid in any of the original data models as they cannot support different perspectives. Such semantic heterogeneity or structural heterogeneity that is due to missing information cannot be resolved without extra input from the reality, just like inconsistent 2D pictures of the same objects cannot resolve the issues without going to the 3D models with extra information.

We believe that bridging between meanings as mental entities is the best level to establish such equivalence without worrying their differences in logical representation or what can be observed from reality based on different perspective. That is, the W-tags that are attached to shadows in order to anchor meanings as mental entities uniquely are the best place we can build representation of equivalence.

Although the existence of such mental entities is subjectively due to the nature of the individual cognitive structures, bridging cannot be randomly performed; it requires certain evidences collected by the data integrators to support such bridges. Such evidences can also help other users to understand how and why these mental entities can be treated as equal. Evidences can be descriptions about who made the equivalence decisions at what time, or notes taken by conversations or interactions with the subject matter (e.g. phone calls with customers directly), or even specific data process that follows certain logic rules as system requirements.

In other words, we propose that proper bridging requires extra information, more than just the data provided from other systems or meanings provided from users of those systems. We use E-tags to represent the extra information in semantic space. Such critical information does not exist in data sources, and the meanings cannot be modeled in the original model due to the limitation of the original perspectives.

When evidences become no longer valid, we need to revoke the bridge of equivalence. To use existing bridges for a specific data integration project, there is a need to re-evaluate the evidences to determine if they can satisfy the specific criteria for the project. Even evidences may not provide all required information in order to support such evaluations, they do explicitly provide some clues about the base foundation how such decisions were made. This mechanism also reminds us that there are limitations for any equivalence decision, and inconsistencies or conflicts due to that semantic heterogeneity may not be able to be resolved in data model without their resolution in the real world.

The following rules are proposed for proper use of E-tags:

**E-tag Rule #1**. E-tags are attached to a pair of W-tags that uniquely anchor to meanings as mental entities in the different perspectives. The pair of W-tags is treated as the same for the perspective chosen in the integrated data model based on supporting evidences under specific criteria for reaching functional objectives.

**E-tags Rule #2**. For a specific data integration project, existing E-tags established by other data integration projects should be evaluated with consistent criteria to determine whether they can be applied within the chosen perspective of the integrated data model.

**E-tag rule #3**. When supporting evidences become invalid, E-tags need to be revoked.

Readers may sense the attitude we have towards mapping, i.e. to establish bridges between mental entities is not to perform scientific discovery for a universally accepted notion of equivalence, but to make a local **design decision** for what

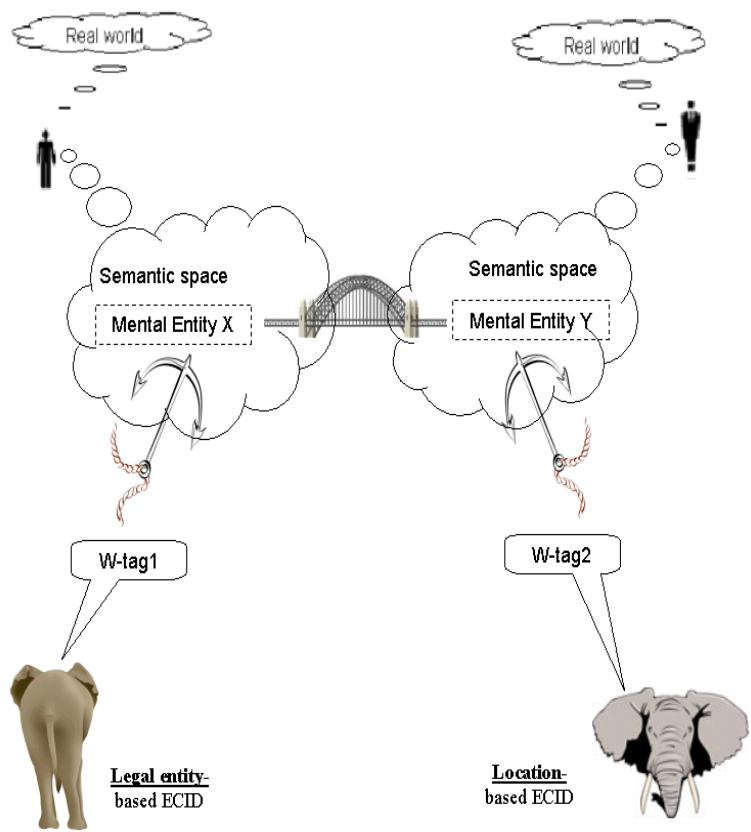

**Figure 17** Meanings as mental entities are the most appropriate level to establish bridges to model the notion of equivalence. Such bridges acknowledge the differences of different mental entities, and can cross the boundaries of semantic heterogeneities and structural heterogeneities. That is, no matter how different the shadows are in data space, or how different the subject matter may be observed from reality from specific perspectives, data integrators may make subjective decisions to view these different mental entities as the same, holding the same meanings in order to reach the functional objectives.

The diagram here illustrates the situation of ECID integration problem graphically. We can think legal entity-based ECID are like data collected based on observation of *enterprise elephants* from back, while location-based ECID are like data based on observation from front. There exists neither similarity among observation about the truth, nor commonality about their logical representation in data space. However, if the data integrators can have evidences to support bridges to associate heads to tails, they sure can treat elephants heads the same as elephant tails to support specific functional objectives, which are not supported in either of the original data models.

meanings as mental entities are treated as the same to achieve specific functional objectives (to associate data that is available in a specific perspectives or certain levels or abstraction). Under the context of supporting multiple versions of the truth, different data integration projects may have different functional objectives such that the criteria for E-tags may not be consistent universally. For example, ECID integration performed only for the merged domestic billing organizations may have very different criteria than ECID integration performed for supporting a specific product internationally. The former requires supporting invoice calculations in a consistent manner, while the later focus on integrating heterogeneous data to understand how customers use the specific products.

If someone wants to make the data integration process automatic without relying on human to make decisions, then this is the place to plug in mechanism to formally model how to make decisions. There are many different theories developed in related areas, for example, Decision Theory, Utility Theory, Probability Theory, and so on. Since our goal is to provide generic guideline for the overall data integration process, we will not further dive into this subject and let interested readers or data integrators to use their preferred approach for the needed design decisions. In this paper, we will simply assume that data integrators make the decision based on their knowledge about the meanings of the data and their desired functional objectives.

In next section, we will discuss the mathematical foundation to support E-tag operations. In section 4.1.2 we will introduce strong equivalence versus weak equivalence, and in section 4.1.3 we will review level shifting with the notion of E-tags.

Before we end this section, we will also provide a short comparison with Relational Mode, just like we have done in section 3.4.2 for W-tags. The purpose if to use Relational Model to explain to readers about the very different tradeoff we made for model commonality versus difference, as illustrated in Figure 1 in section 1.

In Relational Model, there are two mechanisms that can represent the same meaning explicitly when used properly:

(i) The same data values in the same domain. However, it is still not guaranteed that the same attribute values in the same table must represent the same meanings for all tuples within the same table, since users may overload the same table with semantic heterogeneity, as the example of column *status* we discussed in section 2.1.

(ii) The same set of attributes with the same corresponding data values and domains (for equivalence of two tuples). The notion of unique keys can use unique domain values to identify tuples uniquely, but the same meanings must have the same set of attributes in the tuples. In other words, the same meaning represented by different tuples cannot be treated as equivalent in Relational Algebra naturally.

That is, if different users want to use Relational Model to represent the same meanings, and expect their representations to be treated as the same by the Relational Algebra, the only native solution is to use the same data values, the same domain, and the same logical structure. This is enough for semantic homogeneous environment, since there is no need to worry the situation that different logical representations may have the same meaning, or different meanings may have the same logical representations.

This is the root issue for data integration when we need to deal with semantic heterogeneity. Semantically, the notion of similarity cannot be represented as similarity among data values, and Relational Algebra cannot support the subjective decision of equivalence due to the degrees of similarity. It left no choices but force data integrators to model similarity as relations between attribute values. Therefore, semantically equivalent but logically differently represented data must be processed through complex SQL. That is why in the ECID example, a simple question like the calculation for total number of semantically different enterprise customers (from all data sources aggregated together) is actually an extremely difficult task, with many different answers in reality.

Further, different logical structures for representing the same subject matter establish a natural barrier for data integration. With no exception, existing data integration approaches have to transform different logical structures to a common one in order to overcome such challenge. Even with modern software tools to help the transformation process, the fundamental tasks are the same and need human to make final decisions to evaluate whether the logical transformation make sense. No matter which generations of ETL[10], the logical transformation is all added up to the overall complexities for data integration. Readers can imagine the scale of complexity to manage different logical representations increases very fast, from example, an information ecosystem with 500 data sources, not only there are 500 different logical formats, but also there are multiple data exchange or interactions among these systems. The worst part is due to schema evolution, a known difficult issues for Relation Model for decades, that if one system need to revise its schema for certain reason, all of the related ETL that depend on the revised schema have to be adjusted, too.

Shadow Theory answers the issues by E-tags on top of W-tags.

(i) Within the same perspective, E-tags can be established between the same kind of W-tags but different W-tag instances to indicate that they represent the same meaning, no matter how different the logical structures of shadows can be in data space.

(ii) Between different W-tags from different perspectives, E-tags can perform the same functionality without limitations of the perspectives.

In other words, there is only one standard mechanism of E-tag to represent the same meanings. The notion of E-tag and W-tag is an effort to try to bypass transformation of logical representation in order to focus on semantics level. No matter what logical structures are provided by data sources, equivalence can happen and represented independently.

Furthermore, we want to develop algebra that only rely on W-tags and E-tags, such that W-tags with E-tags between them can be treated as equal naturally in the basic operations. In section 5, we will explore our proposed algebra to fulfill this wish, and we will discuss the details for how to represent similarity in next several sections.

Note that there are some costs we need to pay for this approach: we need to establish E-tags with some extra inputs, not like Relational Model to rely on the same data values. This is where the notion of supporting evidences kicks in; without evidences as the extra input, we have no foundation to build E-tags. It is not too bad actually; since this forces data integrators to collect information either as individual events or as some abstract logic rules based on business semantics. The good side is the evidences also expose the limitation for any subjective decision about equivalence,

In summary, the tradeoff we made between representing the same meaning and different meanings is that, we acknowledge the natural differences of everything in information ecosystems by

---

[10] The four generations include hand-coded scripts, automatically generated routines by ETL flow, engine-based ETL by meta-data as conversion rules, and Model Driven Architecture (MDA) to generate target schema or data mappings.

W-tags, from physical to logical to semantics. Then we rely on E-tags to recognize what we should treat as equal.

### 4.1.1 Mathematical Foundation to Model Equivalence in Semantic Space

With conceptual introduction of E-tag in previous section, now we can proceed to the needed mathematical foundation. When we introduced W-tag in section 3.4.2, we proposed to model meanings by the notion of area (instead of atomic data elements) as the ontological primitive units, and apply the properties developed in point-free geometry to support algebra we need. That is, each W-tag instance anchors a unique meaning as a mental entity existing in viewers' cognitive structure, and we model these mental entities as areas in semantic space. One of the advantages is that we can explain the concurrent different decomposition mechanism about the same *thing* is like different ways to decompose an area, and there is no limitation for when such decomposition must end.

Here we can explain the second advantage is that **equivalence of two mental entities can be modeled as overlapping of the two associated areas in semantic space.** Therefore, this advantage can help us to address the difficulties of supporting the rich varieties of similarity or commonality as different ways how areas can overlap with each other. In this way, the subjective decisions data integrators made to treat different mental entities as the same are like design decision to determine the degree of overlapping that can be classified as equivalence to serve as bridges for functional objectives.

Further, we can also explain the direction about equivalence that one meaning *A* can be treated the same as another meaning *B*, but not in the reverse direction. It is like the area of *A* is included in the area of *B* such that the associations about area *A* to other mental entities as areas are valid to area B, but not vise versa. We call such equivalence as **weak semantic equivalence**, modeled as inclusion relations among areas in semantic space. We use the term **strong semantic equivalence** for the case where two areas are exactly overlapped together in semantic space such that the association to other mental entities for one area can be transfer to the other in either direction.

The notion of equivalence by data values used in Relational Model is a special case of semantic equivalence. In such special case, meanings as areas in semantic space are represented as numbers in data space. As long as it is applied in semantic homogeneous environments, we can reduce the multiple dimensions[11] of semantic space into a single dimension of number as there is no need to model different perspectives, and the meaning of a number (in a domain) is always the same. It is like to simplify the notion of areas into square inches such that any areas with 4 square inches are the same with the specific domain.

With such region-based mathematical foundation, we can efficiently model the two major characteristics of semantic heterogeneity:

(1) Different meanings for the same representation:

---

[11] We do not limit semantic space must be two dimensions, and the notion of area can also be generalized from two dimension to more dimensions.

- For the same Shadows, different W-tags can be attached to anchor different meanings as mental entities uniquely.
- Further, W-tags capture not only the meanings but also the characteristics due to the viewers' cognitive structures that are the root causes of semantic heterogeneity.

(2) Different representations for the same meaning:

- For different shadows, E-tags established on top of their attached W-tags can indicate which different representations can be treated as with the same meaning.
- Further, E-tags can support the rich varieties of such subjective decisions about what are equivalent, and the direction for how equivalence can be applied.
- Combined with the W-tags, we can reach our goals that two shadows can be viewed as the same (i.e representing the same meaning) no matter how different their logical structures or data values are.

### 4.1.2 Strong vs weak semantic equivalence

In next section, we will explore the details for equivalence direction, the E-tags established as bridges between two W-tags. If the bridging is bi-directional, we call it as strong semantic equivalence; if the bridging is only one direction, we call it as weak semantic equivalence. .

(i) **Strong semantic equivalence**: the bridge of E-tag between W-tags are bi-directional. That is, from shadow *1*, we can follow its *W-tag1* to the meanings on the other side of the bridge *W-tag2* that is attached to shadow *2*. We can also perform in reverse direction.

Continuing the notion used in Figure 15, the top graph in Figure 18 illustrates strong semantic equivalence between *W-tag(EC1)* and *W-tag(EC2)*. We use a bi-directional double line between *W-tag(EC1)* and *W-tag(EC2)* to represent the strong semantic equivalence, denoted as *E-tag(W-tag(EC1) ⊇⊆ W-tag(EC2))*. The shadow *EC1* represents a legal entity-based ECID, hence the *W-tag(EC1)* represents its semantic relations in a hierarchy structure for its organizational structure based on legal entity perspective, illustrated as left side. The shadow *EC2* represents is a contract-based ECID, hence *W-tag(EC2)* represents its semantic relations in a hierarchy structure for the organizational structure based on contract perspective, illustrated as the right side.

Due to this strong semantic equivalence, we can infer the existence of the decomposition that *W-tag(EC1.2.1)* is part of *W-tag(EC2)* , illustrated as the uni-directional dashed line (1) in the upper graph (since *W-tag(EC1.2.1)* is part of *W-tag(EC1.2)*, which is part of *W-tag(EC1)*. Similarly, combined with semantic inheritance (discussed in section 3.4.3), we can infer the existence of that *W-tag(EC1)* is part of *W-tag(EC root2)*, illustrated as uni-directional dashed line (2), and *W-tag(EC1.2)* is also part of *W-tag(EC root2)*, illustrated as uni-directional dashed line (3).

(ii) **Weak semantic equivalence**: the bridge of E-tag between W-tags are uni-directional. That is, from shadow *1*, we can only follow its *W-tag1* to the meanings on the other side of the bridge *W-tag2* that is attached to shadow *2*, but we can not perform the same in reverse direction.

The bottom graph in Figure 18 illustrates weak semantic equivalence. We use a uni-directional double line from *W-*

*tag(EC2)* to *W-tag(EC1)* to represent the weak semantic equivalence, denoted as *E-tag(W-tag(EC1)⊆W-tag(EC2))* , i.e. the meaning as a mental entity anchored by *W-tag(EC1)* is bridged to the meaning as a mental entity anchored by *W-tag(EC2)*. We can think the weak semantic equivalence, *E-tag(W-tag(EC1)⊆W-tag(EC2))*, as the meaning of *EC1* can be viewed as subset as the meaning of *EC*, just like a normal decomposition. That is, although the weak semantic equivalence is classified as horizontal semantic equivalence we have mentioned in section 3.4.3 since it is across two perspectives (left hierarchy of legal entity-based ECID, and the right hierarchy of contract-based ECID), but the nature of weak semantic equivalence makes it also like vertical semantic equivalence in which two W-tags at different levels of abstraction can be treated as the same.

In the lower graph, the uni-directional dashed line (1) represents a decomposition that *W-tag(EC1.2.1)* is part of *W-tag(EC2)* due to the weak semantic equivalence *E-tag(W-tag(EC1)⊆W-tag(EC2))*. Similarly, combined with semantic inheritance, we can infer the existence of that *W-tag(EC1)* is part of *W-tag(EC root2)*, illustrated as uni-directional dashed line (2), and *W-tag(EC1.2)* is also part of *W-tag(EC root2)*, illustrated as uni-directional dashed line (3).

Readers may notices that the example results of the decomposition illustrated as uni-directional line (1), (2), and (3) are essentially the same in upper graph and in the lower graph. It is exactly our point that strong semantic equivalence and weak semantic equivalence can both helps us to read our function objectives in these examples in Figure 18: to map legal entity-based ECID into contract-based ECID hierarchy such that legal entity-based ECID is part of contract-based ECID semantically. Strong semantic equivalence can also infer some decomposition in the reverse direction, but weak semantic equivalence cannot.

### 4.1.3 Level Shifting due to Semantic equivalence

In section 3.4.3, we have introduced the notion of level shifting in semantic space. Here, we can review the definition from the notion of semantic equivalence. The purpose we need to model level shifting is that data integration in information ecosystems is performed through various kinds of data exchanges with different systems. It is like human communication process that people may refer *things* based on their preferred levels of abstraction. To successfully understand the meanings, human can easily shift the level of abstraction and treat meanings at different levels as the same based on certain subjective criteria.

Therefore, level shifting is the application of semantic equivalence that happens between meanings at different level of

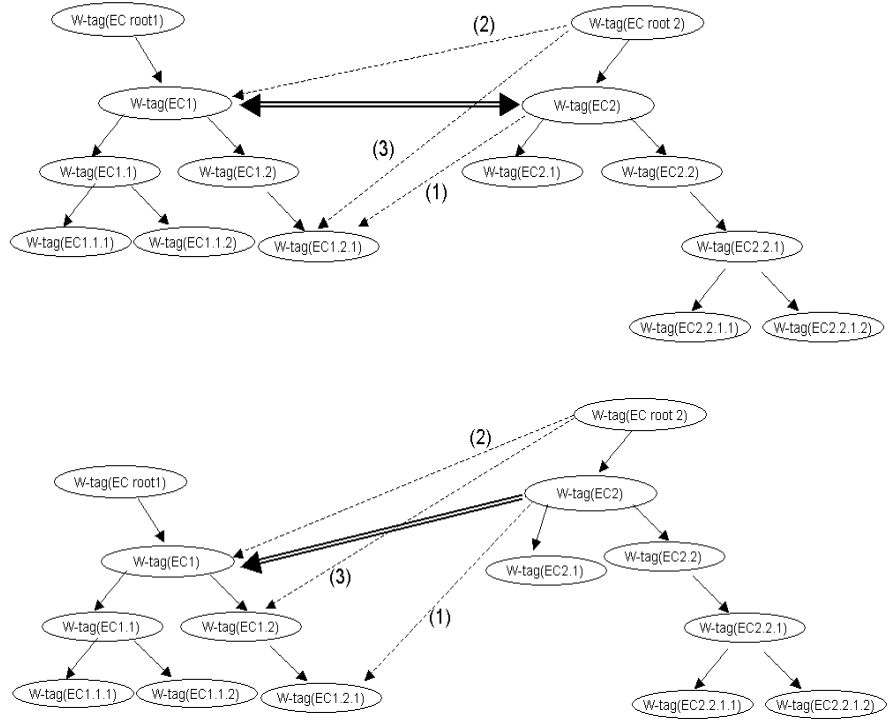

**Figure 18. Strong semantic equivalence** (top) versus **weak semantic equivalence** (bottom). *EC1* represents a legal entity-based ECID and its W-tag is in a hierarchy representing the overall company structure in legal perspective. *EC2* represents a contract-based ECID and its W-tag is in a hierarchy representing the overall company structure according to contract perspective. Strong (weak) semantic equivalence is represented as a bi-directional (uni-directional) double line between *W-tag(EC1)* and *W-tag(EC2)*.

Due to such bridging established by E-tag, we can infer that the uni-directional dashed line (1) representing a decomposition that *W-tag(EC1.2.1)* is part of *W-tag(EC2)*, and dashed line (2) representing a decomposition that *W-tag(EC1)* is part of *W-tag(EC root2)*, as well as that dashed line (3) representing a decomposition that *W-tag(EC1.2)* is part of *W-tag(EC root2)*.

abstraction. The distinction of strong versus weak semantic equivalence can help us to model this phenomenon to reach functional objectives. In the example discussed in [9], p.330 – p.333, that the integration of books from one perspective into publications from another perspective can be better classified as level shifting. Batani illustrated that during the integration process data integrators (or schema designers) can move the properties of books that are common to those of publications to the upper level of abstractions. However, the Relational Model and the Relational Algebra cannot tell when to treat books the same as publications since they are naturally represented as two different kinds of domains or tuples. When more complexities are added in the data integration, the challenge about original meanings will pop out. For example, if we need to integrate the table of e-media from third perspective with books and publications, we need to ask many questions that, should it be another kind of publications but at the same level of abstraction with books if books only contain physical books that are published? Or, if books already include e-media, but e-media may contain more than just books, should we treat e-media as lower level than books?

Our approach performs the integration in very different direction. First, W-tags of books, publications, and e-media are attached to shadows as data instances. Second, we do need to have a common schema to force a uniform logical representation. Instead, we build E-tags among these W-tags to recognize what should be treated the same (subjectively) by the algebra in a natural way. For example, a shadow tagged as *book* in perspective 1 may be identified as the same as the shadow tagged as *publication* in perspective 2 due to the same author, the same title, and the same publishers, we establish an E-tag with semantic equivalence between these two W-tag instances (as a book is a publication, but a publication may not be a book). If we also find a shadow tagged as *e-media* from the third perspective, we can have the flexibility to establish an E-tag with semantic equivalence between the W-tag *e-media* and the W-tag *book*, with the direction determined by which function objective of the data integration project.

Dependent on evidences can be identified, we can establish strong or weak semantic equivalence in this example. If weak semantic equivalence is established, then levels of abstraction are automatically aligned based on the evidences, for example, *E-tag(e-media $\subseteq$ book)* and *E-tag(book $\subseteq$ publication)*. If strong semantic equivalence is established, then we have levels aligned in both directions.

Since our goal is to support subjective decisions about what should be treated as the same in a natural way for the algebra, we do not want to establish rigid alignment between levels of abstraction at schema level. Instead, we provide the flexibility to data integrators to establish semantic equivalence, and level shifting is modeled to help the data operations with semantic inheritance.

On the other hand, data integrators may need to establish generic alignment between W-tags to establish & align levels of abstraction not at individual data instances, we propose the notion of template which is a decomposition structure in semantic space but the W-tags are not required to attached to shadows. Templates are more like an observed pattern for the semantic relations among W-tags, instead of a rigid schema concept in Relational Model. We will discuss the details of template in section 5.4 when we develop algebra for Shadow Theory.

In summary, based on semantic equivalence and semantic inheritance, level shifting along decomposition structure provides us the needed flexibility to

(1) Describe a shadow in semantic space as different meanings at different levels of abstraction.

(2) Help data integrators to find a proper place to establish bridges to cross the boundaries of semantic heterogeneity, such that E-tags happen at different levels to support the functional objectives of the data integration project.

### *4.1.4    Representing Similarity in Semantic Space*

In section 2.3, we have explained that similarity or commonality may help to identify potential candidates for mapping, but it cannot serve as the decisive criteria for data integration. That is, the mapping between data elements from different sources is a subjective design decision in the level of logical representation, so does the notion of similarity in the level of conceptual representation, which is based on what the different users can observe about the subject matters, plus their background knowledge and chosen perspectives. However, these two can be correlated, but not necessary be consistent with cause-effect relations.

Therefore, we propose to follow functional objectives instead of semantic similarity in order to decide mapping, i.e. what should be treated the same with proper E-tag and supporting evidences. The reason is obvious that if there exist multiple versions of truth and no one have the full knowledge about the reality, a tail of an elephant can represent the elephant, just like a head of an elephant, and there is no commonality or similarity between the tail and the head. If data integrators can identify supporting evidences that the tail or the head represent the same elephant in reality, they sure can be treated as representing the same meaning in terms of the functional objectives like counting the number of elephants..

One the other hand, we do not object to use similarity or commonality to help identifying potential mapping candidates. Here we will discuss how we can represent similarity properly such that we can have some tools to help data integrators to make their decisions.

For programming and data modeling, the traditional convention to model commonality and similarity is by hierarchical structures such that more common or similar properties are at higher levels. The philosophical foundation is based on the classical Aristotelian theory of concepts [66] in which a concept is defined via a set of necessary and sufficient properties. That is, the higher levels in the hierarchy hold the common properties that lower levels share; therefore, their relative distance in such hierarchies can represent similarity between two concepts. The implementation of such idea can be observed in many different fields like ontology, Object-Oriented Programming, taxonomy, or classification in daily lives. In section 3.4, we have proposed to use generic decomposition structures to model the semantic relations among W-tags, since meanings as mental entities can be decomposed in semantic space. This decomposition structure can also serve as a hierarchy structure in order to represent similarity. It is like Venn diagram to decompose areas into common and uncommon sub-areas that corresponds to different levels in the decomposition structure.

However, this approach suffers a fundamental weakness that we need to be able to specify explicitly the necessary and sufficient properties for concepts at each level in order to build hierarchy. In the context of data integration in information ecosystem, in which multiple versions of truth exist and no one has the full knowledge of reality, this is a luxury expectation and not practical approach. Therefore, we need to explore different ways for model similarity in order to help data integrators to identify potential mapping candidates.

In the book of [56], Gardenfors discussed a different way to model the notion of similarity. First, he described the difference between intrinsic and extrinsic representation (p.44) that is introduced initially by Palmer [67]. For example, the representation of human age by numbers is an extrinsic representation since the structure of the digit sequences does not have the same structure as the represented relation. However, if age is represented by height of rectangles, then the structure of the represented relation is intrinsic. That is, we can say that a representation is intrinsic if the relation has the same inherent constraints as its represented relation, and it is extrinsic otherwise (i.e. there is a need to have a set of rules to interpret the meanings or the sequence like number 5 is larger than number 3).

With this view, Relational Model is an extrinsic representation in nature since the basic principle is that "*associations between*

*relations are represented solely by values*" [22]. Therefore, we need a set of to interpret the meanings. We sure have difficulties to represent the notion of similarity in Relational Model, as the degree of similarity must be represented as numbers in extrinsic way, for example, number 5 means more similar then number 3 (as an attribute for a relation between two data values or tuples).

In an attempt to use intrinsic way to represent concept, Gardenfors proposed to use the geometry-based conceptual space as "*a set of quality dimensions with a geometrical structure*" (p.44 in [56]). One of the advantages in conceptual space is that similarity can be modeled as distance for direct comparison: the smaller the distances are between the representations of two subject matters, the more similar they are. This approach is also supported psychology literatures that similarity is an exponentially decaying function of distance [68-70].

This approach is consistent with the prototype theory of categorization developed by Rosch [71-73]. The main idea of prototype theory is that within a category of objects, certain members can represent this category better than others, for example, robins versus penguins for the concept of birds. The members that can represent this category the best are called prototypical members. This is contrary to the classical Aristotelian theory of concepts [66] in which a concept is defined via a set of necessary and sufficient properties.

Compared with Gardenfors' Conceptual Space, we have very different requirements for applying Shadow Theory to design a data model for the purpose of data integration. First, we need to manage semantic heterogeneity by explicit representation of the meanings and equivalence among these meanings. Second, we cannot assume universal accepted atomic data elements due to the support of multiple version of the truth; therefore, we need the use the notion of area to represent meanings that can be decomposed in different ways concurrently.

As a result, we propose to represent similarity by the notion of overlapping. That is, the degree of overlapping between two areas in semantic space represents how similar these meanings are. In one extreme, the two areas are exact the same; we use an E-tag with strong semantic equivalence to represent such case. To another extreme, one area is completely within another, it is represented as an E-tag with weak semantic equivalence. For other overlapping situations, it is up to data integrators to make subjective decisions about the thresholds about what degree of similarity can be considered as equivalence based on evidences they can identify. For example, if 95% of the area 1 is overlapped with area 2, one may decide that area 1 is considered equivalent to area 2 approximately (as weak semantic equivalence in one direction). If 80% of area 2 is overlapped with area 1, then one may decide strong semantic equivalence such that the equivalence is in bi-direction approximately.

Compared with Relational Model that relies on extrinsic representation for similarity, our approach is more like an intrinsic representation although we still need rules to interpret meanings. The advantage of using overlapped areas to represent similarity is that it provides data integrators a way to evaluate similarity based on meanings in the context of different business semantics, like the ECID used in billing, ordering, repair, or marketing application domains.

One weakness is that such measurement about overlapping is subjective, that is that is why we keep emphasizing that mappings is a subjective decision in the context of multiple versions of the truth, especially when no one have full knowledge of the reality. The same weakness also exists in Gardenfors' Conceptual Space to apply the notion of distance for representing similarity: measurement of distance is subjective, just like what is reported in [74].

The representation of overlapped areas is also consistent with Prototype Theory that robins can be viewed as have exactly overlapped areas in the concept of bird, while penguins have less overlapped areas. For example, in the ECID integration example, the concept of *enterprise customer* has very rich varieties from different perspectives. Legal entity-based ECID may be more like a prototypical member, while contract-based ECID can be another. However, their overlapped areas in semantic space are blurred with complexities of semantic heterogeneity. No matter how different they are, data integrators can use the notion of areas in semantic space to make their judgment about what should be treated the same in order to functional objectives of the data integration project.

Here is one scenario that different answers of ECID mapping may happen. Let's assume the data integration is to across the boundaries between legal entity-based ECID used for product A that is available only in US domestic, and contract-based ECID for a different product B that is available internationally. To fulfill functional objective 1, which is to calculate total revenue for all kind products in all areas, one data integrator may decide to simply establish a weak semantic equivalence between the legal entity-based ECID for US headquarter and the contract-based ECID for the international headquarter, such that total revenue is just a sum up of revenue for all sub-companies even they are semantic heterogeneous. For function objective 2 that is to promote sales of product A to those enterprise who already use product B internationally, such mapping is not meaningful since mix semantic heterogeneous ECID together in a hierarchy cannot help to identify which sub-companies in US that do not use product A.

In summary, we propose to use areas in semantic space to model meanings, and based on the concept of overlapped areas we can model how similar two meanings are, and data integrators can make design decisions about what meanings can be treated as the same subjectively. We realize the limitation is on the measurement for the degree of overlapping, since it is rooted at the fact there is no universal accepted ontological primitive units across all users' cognitive structures. This is the fundamental fact in the context of multiple version of the truth, we choose to compromise here and we already have quite some utilities to help data integrators for their jobs.

## 4.2 Meaningful Data Integration and Criteria for Evaluation

With the basic model utilities of W-tags and E-tags, we now can discuss meaningful data integration and criteria for evaluation. We will start with evaluation criteria first, then we will come to what shadow properties to be integrated and how.

### 4.2.1 *Evaluation criteria for data integration projects*

Here we will first have a quick review for the three common criteria proposed to evaluate data integration: completeness, understandability with clear semantics, and correction. We will explain the underlying assumptions of these proposal and suggest an alternative interpretation for these criteria.

(1) **Completeness.**

One way of defining completeness is that every property or characteristics carried by existing schema/model must also exist in the integrated schema/model [9] [48]. This definition is problematic in the context of information ecosystems when data from inconsistent or conflicting perspectives need to co-exist together, especially when there do not exist 1:1 consistent mapping. The root issue is that it violates the fundamental objective of traditional database design that focus on avoiding redundancy and inconsistency. As a result, the criteria of completeness for data integration itself lead to conflicts with the criteria of consistency for the underlying data model design where data integration is implemented.

Due to the fact that integration methodologies are constrained by the data models they support [9], the issue is due to the incompatible expectations at two different levels: (1) at database level, the traditional goal is to model data consistently without redundancy, (2) at data integration level implemented on top of database level, the struggles pop out if there are needs to allow inconsistencies or even conflicts co-exist together. The root can be traced back to the philosophical assumption about the single version of the truth. When it encounters with reality of multiple version of the truth, plus the fact that no one in information ecosystems can have the full knowledge about the reality.

Practically, the downstream systems are usually designed after the existence of upstream systems. Completeness of data integration assumes all of the required information is available in the data sources such that if data integration is performed, then the downstream systems should be able to fulfill all of its local needs. This assumption is valid if the integrated database is to replace the existing data source, but not valid when both the data sources and the integrated database co-exist for different satisfying different business requirements. That is, if a data integration project is evaluated as with completeness from the view of upstream systems, it may fai to satisfy the needs of downstream systems.

For Shadow Theory, we propose another way to define completeness to evaluate data integration project. The completeness should be limited to the chosen perspective(s) such that semantic heterogeneous data is integrated into meanings as mental entities that can be recognized to the maximum extents the chosen perspectives. For the ECID integration example, if legal entity-based ECID is the chosen perspective, then all other different kinds of ECID should be mapped to legal entity-based ECID since location and contract cannot be recognized as an enterprise customer identifier. That is, the criteria of completeness do not mean every property in different data sources can be integrated together; instead, the properties that can be integrated is limited by the expressive power of the chosen perspective for the data integration.

It is actually common sense that different perspectives help us to have wider / deeper understanding about the subject matter, and some properties from a perspective may not be able to be expressed in the second perspective. The key is not completeness, but rather the objective of the data integration about what kind of properties should be integrated. This is exactly what we will address after discussing the criteria of evaluating data integration projects.

(2) **Correctness**.

One way to define correctness of data integration is that source data and the integrated data are mapped precisely such that it can be evaluated based on query against the original sources and integrated data model with exactly the same answer. Such query answerability is proposed as one of the important criteria for data mapping in [43], and further used to evaluate model management composition operator [35]. However, for data integration in information ecosystems, we have multiple data sources provide similar or inconsistent data about the same subject matter. There is little chance to run the query in the integrated data model with results back exactly the same as run the same query in every one of the original data sources, since subjective filtering or data selection criteria is performed during the integration process. In addition, the original data is often mixed with extra information from different sources such that it is difficult to reverse the mix to come out with original ingredients.

A different approach is proposed in [32] such that the focus switches from queries to data instance space by definition in terms of the schema mappings alone, without the need of reference to a set of queries. However, it suffers the same issues as query answerability based approach for data integration in information ecosystems. In addition, in section 4.1 we have explained the nature of mapping between data instances from different data sources, that it is a subjective decision made by different data integrators for achieving different functional objectives. When there are multiple answers, how can one evaluate which one is correct?

Further, both of the approaches also have a fundament philosophy problem. That is, they simply accept the assumption that every property in the different data sources should be integrated. In terms of the three kinds of shadow properties we discussed in section 3, due to the subject matter, due to the wall-like of surface, and due to the projection process, they do not consider the possibility that data integration project may only want to integrate portion of the properties modeled in data sources.

Therefore, we believe that to make evaluation about correctness, the pre-requirement is clear specification for what kind of characteristics should be integrated. This needs a different methodology to investigate the meaning of the data, and the three kinds of shadow properties can serve this purpose: characteristics due to the subject matter, due to the wall-like surface, and due to the projection process.

Only if we have clear scope of the meanings to be integrated, then we can proceed to next step to evaluate correctness based on the specific business semantics that the data integration project is for. That is, we need to identify the mental entities in the users' cognitive structures as the foundation to make evaluation. This also includes how the users expect to use the meanings in the integrated database. Note that the users are the one who are the business owners of this data integration project, not the one who originally create the data in the source data models. This is critical difference between these two and sometimes they can be mixed together for different reasons. For example, data integration may be for replacing existing several systems, then the users include the all of the users of the different original systems. If the data integration is for designing a new system to co-exist with existing data sources, then the users must be the one who will use the new system, not the one who use the original data sources.

Next, the subjective mapping decisions about what can be treated as the same semantically. That is, the criteria for establishing E-

tags between those W-tags from the original sources. Similarity or commonality with various degrees may help to identify potential candidates, but they are not the deterministic criteria. The most important factor is that no matter how different the original meanings or data could be, they can be treated as equal if data integrators can collected proper supporting evidences. For example, the correctness of integrated enterprise customer data for billing applications may be different then that by repair applications, again different than what the enterprise customer themselves in terms of their own legal or financial organizational structures.

In summary, our proposal to evaluate correctness include the following criteria:

(i) Clear specification about the objective of data integration for what kind of properties should be integrated. The three kinds of shadow properties can help data integrators perform such analysis.

(ii) The foundation of business semantics to evaluate the integrated data semantically. That is, the desired W-tags that anchor the mental entities in the users' cognitive structures as the criteria to meanings represented in the integrated data.

(iii) Criteria for the subjective mapping decision, including similarity or commonality with various degree, and especially the supporting evidences.

In this way, the correctness is limited to the semantic space of the users, within the objective of the data integration projects. We will further discuss the details in section 4.2.2.

(3) **<u>Understandability with clear semantics</u>**.

Expecting the integrated data representation to be understandable by users with clear semantics to support application use seem a reasonable request [9, 43]. However, in large-scale information ecosystems where there exist semantic heterogeneity and multiple version of the truth, there are practical issues about this simple request. That is, no one have the full knowledge of the reality, as well as no one can understand the full details from every data model. Any user of the integrated data has limitation to understand the meanings due to his chosen perspectives, level of abstraction, or ontology; therefore, understandability with clear semantics is a very subjective judgment made by users just like students with different background to evaluate text books designed for college versus designed for elementary school.

In addition, the majority of current systems are based on Relational Model, which we know from experience that there is no easy way to understand complex schema design without asking human to explain and interpret the associated business semantics. When the data model itself does not contain explicit representations for clear semantics to help users to understand their meanings, it is also difficult for data integration that is built on top of data model.

As we have explained in section 2.2, the drawbacks for relying on documentations to help users understand include the gap between the data model and what is describe in the documents, users' subjective perspective may prevent their understanding, as well as lacking of comprehensive documents for every perspective in information ecosystems. Further, to model business logic rules is also important in order to help users navigate in the jungles of different semantics space. Specifically, it will be very helpful for users if the integrated databases can interact with users to explain the meanings of data, where the original data are from, how the integrated data are mapped, what events happen, which process perform the change, and why the processes do so (i.e. following which business logic rules), and even detect inconsistencies or conflicts from different perspectives.

Therefore, understandability with clear semantics is not an easy to do request, although it should be a minimum request, just not possible for Relational Model to help when we need to manage semantic heterogeneity and support multiple version of the truth. Our proposal is that it should be the data model's responsibility to explain itself, not the criteria to evaluate schema design for data integration. That is, we propose the data model should have the mechanism to explicitly represent meanings as mental entities from viewers' cognitive structures, not just the logical representations and leave semantics in the design documents. The meanings should be able to evolve when there is any change in the model design. This is exactly why we propose W-tags and E-tags: for users who already have the foundations for the mental entities in their cognitive structures, descriptions of W-tags should be able to uniquely anchor the specific meanings as mental entities, for those who do not have such mental entities, description of W-tags may be able to help them to understand the basics of the meanings.

As for E-tags, supporting evidences in the format of business rules or interactions with the reality should help to explain why the different W-tags could be treated as equivalent semantically. Even it may not be able to answer any kind of questions from all aspects, there should have minimum clues for how to further investigate how the E-tags are established.

Our goal is that the integrated data model should be able to explain what happened in database automatically, and provide logic reasoning like what knowledgebase can perform. For ECID example, users should be able to ask the databases questions like: what is W-tag *legal-entity-based ECID*? How one can associate from *legal entity-based ECID* to *service accounts* with associations to *contract-based ECID* only? or, why the *ownership* is no longer valid?

### 4.2.2 *Steps to integrate shadow*

With the alternative interpretations of evaluation criteria proposed in previous section, now we can go back to the question we raised in section 2.3:

Question 4: <u>What kinds of characteristics of the data should be integrated and re-use in the local model?</u> How data integrators should perform their design activities such the integrated data can be meaningful for users and re-useable again later for different needs? And how we should handle the inconsistencies or conflicts?

Our proposal can be summarized as the following principle.

Principle 5. Meaningful data integration should be performed only with required shadow properties, and the scope of the subjective equivalence decisions should be explicitly represented with meanings of the data.

The generic steps for data integration are discussed as the following. Note that these are generic steps, not the exact sequence to perform. The process can be iterative due to the incremental understanding about the meanings from data sources and about the business semantics/operations of the integrated data.

|  | C1. Due to subject matter | C2. Due to wall-like surface of system requirements | C3. Due to projection process |
| --- | --- | --- | --- |
| Only subject matter | V |  |  |
| Subject matter and system requirements | V | V |  |
| Everything | V | V | V |

**Figure 19. Three examples of chosen shadow properties to be integrated.**

*I. Clarify what shadows properties to be integrated*

First, let's explain about the chosen shadow properties based on the three kinds of shadows characteristics: due to the subject matter, due to the wall-like surface of system requirements, and due to the projection process. Figure 19 illustrates three possible examples for shadows properties to be integrated. The first one is for the situation that downstream systems will co-exist with upstream systems in the information ecosystem, but the functional objective is only to have better understanding about the subject matter. The characteristics due to wall-like surface of system requirements (of the upstream systems) are not important, as there is no dependency between the business operations in the upstream and downstream. So does the projection process that downstream systems may not need to include such characteristics in the scope of their data integration. For example, whether upstream systems model an ECID as an entity, an attribute of an entity, as a relationship, or as an attribute of a relationship can be ignored such that every shadow is converted to a generic representation for local use.

For the ECID data integration example, one scenario for this case is like a data warehouse as the downstream to collect customer data from various sources. The characteristics due to the business requirements in upstream systems are not importance and can be filtered out. For example, the different lifecycles as status of the ECID, there may exist 2 stages for legal entity-based ECID ("*A*" is for active legal entities and "*I*" for inactive legal status), 3 stages for contract-based ECID ("*A*" stands for active contract status, "*P*" stands for pending status, and "*I*" stands for inactive contract), 4 stages for location-based ECID ("*A*" represents active services at the specific location, "*D*" represents that the services are disconnected but may be reconnected later, "*I*" represents the services are connected but not in use, and "*F*" represents that the customer was at final status like a logical deletion). The data warehouse may not want to integrate these various kinds of status into its data model, and they can filter out the unwanted shadows before any data process, for example, only keep legal entity-based ECID with status = '*A*', contract-based ECID with status = '*A*', and location-based ECID with status in ('*A*', '*I*', '*D*').

One example for the impacts is that, the equivalence of ECID between legal entity-based ECID, location-based ECID, or contract-based ECID simply does not consider the differences of these two characteristics in their original sources. That is, in terms of Entity-Relationship Model, the classification of entity, relationship, and attribute is ignored since the objective of the data integration does not include the characteristics due to projection process. The different lifecycles due to business operations in the data sources as system requirements are also ignored due to the same reason.

The second case is also common in information ecosystems that downstream systems the secondary kind is common such that downstream systems need to integrate both characteristics due to the subject matter and due to the system requirements of the upstream systems. One scenario for the ECID example is for a portal to integrate enterprise customer data for the purpose to entitle users to access specific information like invoice, tickets, or orders, which are provided from different data sources directly. The different lifecycles are critical as the users can have different privileges in accessing the associated information.

One example for the impacts for the second case is that, equivalence between these different ECID has to consider their associated status, i.e. lifecycles in the original data sources. In other words, although the lifecycles are not related to the subject matter (i.e. enterprise customers in reality), but rather due to the business semantics/operations that are supported by the data sources, we have to include them in the design decisions for equivalence. Such mapping is not just to identify which legal entity-based ECID represents the same thing as which location-based or contract-based ECID, but also alignment of business operation process due to the different ECID. For example, a contract-based ECID with status '*A*' (representing active contract status), which cannot be mapped to a location-based ECID with status '*F*' (representing final status like logical deletion) since the business semantics of the integrated data model requires that equivalence only can be between enterprise customer records in active business stages.

For the third case that all kind of shadows properties must be integrated, one example scenario is that the purpose data integration is to design a new system to replace the multiple sources. The difficulties are not only in data integration, but also in the changes of the whole information ecosystem since data flow from upstream systems to downstream systems will be changed, too. This can be driven by changes for the associated business operations like system consolidation after M&A activities (e.g. consolidate several billing systems into a single one), or triggered by the needs of cost reduction in information ecosystems such that business operations are passively revised. Existing information from every perspective needs to remain within the same data model, with some possibility for new design to improve efficiencies.

Inconsistencies or conflicts due to different perspectives are the major challenges. For the ECID example, all of the three kinds of shadow properties may impact the decision about equivalence. The same shadows that are viewed entities in one perspective but as relationships in another have to co-exist in the same database.

Mapping between different kinds of ECID with considerations of their lifecycles in business operations need to remain, too. Therefore, the results of data integration are corrected for the specific context, may not be generic true for other situations like only considering the subject matters. This is one common reason that why we reported that there often exist multiple versions of the truth for the similar data integration projects performed by different departments in section 2.3.

In summary, the purpose of proposing three kinds of shadow properties is to serve as a generic guideline to help data integrators to clarify what kind of information should be integrated. In practice, there could exist difficulties for such classification since data as shadows are usually mixed together such that characteristics due to projection process can be aggregated with characteristics with system requirements, or the subject matter is hidden behind the system requirements that we cannot separate them. Hence, it is really like a checklist for clarifying the objectives of data integration to avoid the common failures due to lacking of clear and consistent data integration objectives. Data integrators can use this list to ask themselves the questions at any step: Is this shadow likely due to the subject matter that should be integrated? Or the shadow properties that causing inconsistencies or conflicts are due to projection process that we can ignore?

### II. Explicit representation for the meanings in their original perspectives

Next step is to establish explicit representation for meanings in the data sources that are within the scope of data integration. We need to proper model such meanings by W-tags and establish semantic relations among them. The design of W-tags is for such purpose to recognize different meanings as mental entities from different viewers' cognitive structures (i.e. different perspectives). In order to distinguish with meanings that are in the integrated data model, we will call these W-tags as **source W-tags**.

The objective for this step is to serve users from any involved business perspective of the data sources such that their business semantics can be represented explicitly. This is an incremental and dynamic process, one reason is that business semantics will continue evolve without waiting data models to catch up, the other reason is that data integrators' understanding about he specific business semantics may grow deeper and wider gradually.

This is the critical step for managing semantic heterogeneity, especially for the part that different meanings may be with the same representations such that we need to use different W-tags to anchor different meanings uniquely. We do not worry about inconsistencies or conflicts at this step, as the decision about equivalence among meanings of shadows will be in the fourth steps.

This is similar to the spirit of "pay-as-you-go" proposed in [75] since we only pay attention for the critical portions for data integration, and allow co-existence of inconsistencies. However, there is a big difference in later steps as we are not satisfied with statistics or probability, we need some explanations about the subjective decisions for mapping, and an evidence-based mechanism to support such data integrators' choices of equivalence.

Note that we do not mean such explicit representation must be formal like ontology or top-down object-oriented class hierarchy. Our focus is on the must useful information; therefore the semantic relations among W-tags do not need to be complete. Only those useful ones should be included, and we do not need to build upper (or top level) ontology unless it is essential in the business semantics. This is similar to middle-out approach [76, 77] to start with the terminology most common used. For example, we only need a W-tag associated to the mental entity of "*legal entity-based enterprise customer*" instead of a full taxonomy of W-tags associated with all kinds of legal entities. Later when we at the stage to make decisions about equivalence, we have the chance to generalize or specialize associated concepts in order to align with other mental entities in different perspectives.

### III. Identify the desired business semantics

The next step is to explicitly represent the business semantics of the integrated data model. That is, we will use W-tags to capture the meanings as mental entities in the viewers' (of the integrated data model) cognitive structure. To distinguish with source W-tags, we will call these W-tags as **target W-tags**.

This is the second point we have mentioned about the criteria of correctness: based on the business semantics that data integration project should try to satisfy, we can evaluate whether the mapping decisions made by data integrators are correct or not. For the ECID example, if the objective of data integration is for portal to display such information to external users such that they can recognize themselves (as enterprise customers), and it is these users who can evaluate the correctness of data integration. If M&A happened between two banks $X$ and $Y$, then it is up to the employees of the merged banks to provide the correct information about their new names, contacts, and organizational structures. Whether the ECID is legal entity-based, location-based, or contract-based is not important for these users since such differences are internal issues in service provider side.

If the objective of the data integration in a data warehouse is to provide an aggregated report for marketing or sales to prepare promotion campaigns, then the correctness should be evaluated as the business entities that campaigns are targeting at: could be a combination of legal entity-based ECID, location-based ECID, or contract-based ECID which satisfy the specific criteria. In this scenario, removing duplicates is the critical part that the associated legal entity-based ECID, contract-based ECID, or location-based ECID should be group together and count as a single enterprise.

### IV. Make subjective decisions on semantic equivalence

With source W-tags and target W-tags available to data integrators, now the challenges are how we can establish semantics relations among them in order to make decisions about equivalence represented by E-tags: which target W-tags are mapped to which source W-tags based on what criteria. This is the place that we can generalize or specialize concepts in order to align different levels of abstractions. Meaning equivalence, including strong and weak, vertical or horizontal, and level shifting, are the available utilities to describe the various possible mapping decisions here.

As we have explained in section 4.1, these are subject design decision made by data integrators, and we believe many existing data integration algorithms or techniques can be converted into the new data model foundation to help data integrators for this step. For this paper, our main objective is to explain the need of such new data model foundation. It is like a platform that we can bypass the limitations of logical representations and focus on managing semantic heterogeneity. Therefore, we will not discuss specific algorithms for how to make such decisions here; instead, we will use an example to illustrate the generic steps.

Figure 20 illustrates shadows collected from three different sources. The hierarchy on the top left (yellow area) represents the organizational structure for a specific customer based on legal entity-based ECID. These shadows are provided source 1, which is in the information ecosystem of service provider B. The hierarchy on the top right (green area) represents the organizational structure for "*the same*" enterprise customer based on location-based ECID. These shadows are provided source 2, which is in the information ecosystem of service provider G. By "*the same*", we mean data integrators have some common sense understanding based on the marketing brands of this enterprise customer, but not yet have the details or evidences about the exact mapping. The hierarchy on bottom right (blue area) represents the structure of service accounts collected from source 3, which is like source 2 also in the information ecosystem of service provider G.

Following the generic steps we recommend here, data integrators have clarify the overall business objective is to calculate total revenues for large enterprise customers for marketing analysis. Therefore, the criteria for what should be integrated among different kinds of ECID focus only on the subject matter, the characteristics due to different business requirements in the data sources or due to projection process can be ignored.

In the second step, different kinds of W-tags are captured for each perspective as the source W-tags, and semantic relations between these W-tags are also established based on information provided

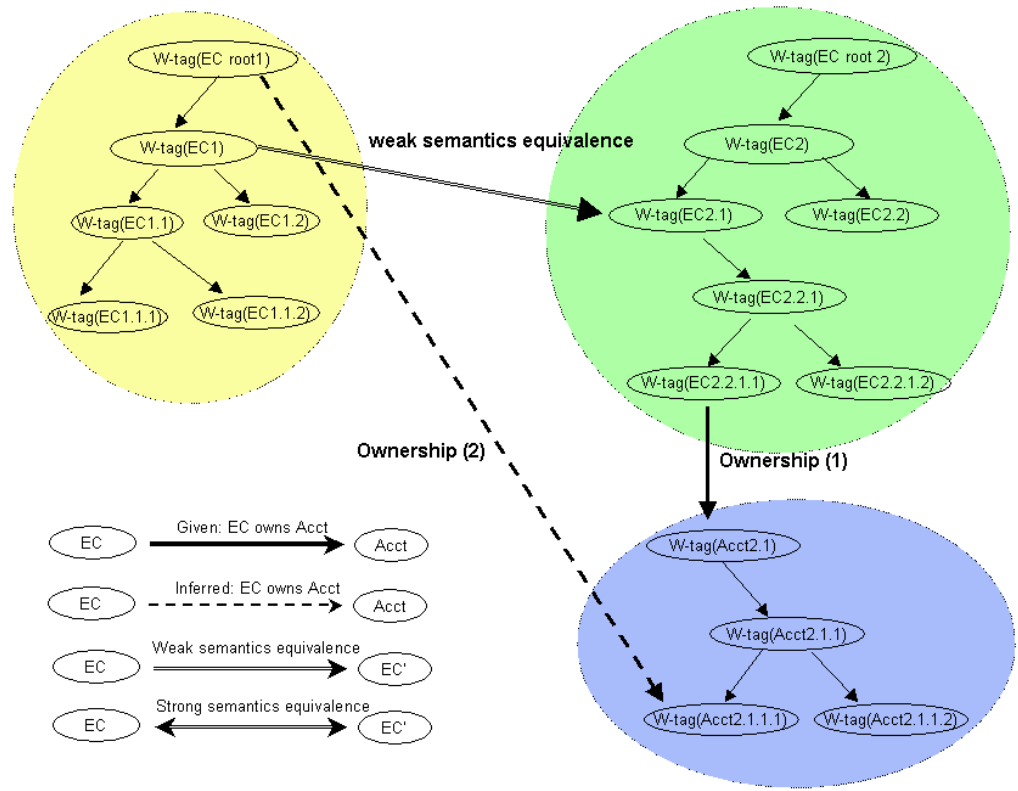

**Figure 20.** Example for generic data integration steps.

I. **Clarify what shadows properties to be integrated**: The overall business objective is to calculate total revenues for large enterprise customers for marketing analysis. Only the characteristics due to subject matter of enterprise customer need to be integrated; those due to system requirements or projection process in upstream/source systems are ignored.

II. **Explicit representation for the meanings in their original perspectives**: The identified source W-tags are *legal entity-based ECID* (yellow area, from source 1), *location-based ECID* (green area, from source 2), *service account* (blue area, from source 3), and *ownership(1)* from source 4.

III. **Identify the desired business semantics**: The target W-tags include *generic service account* which carries individual invoices to calculate total revenue, *generic enterprise customer* which is based on *legal entity-based ECID* and it can includes *location-based ECID* if properly aligned, *generic ownership* which is an *ownership* with *generic enterprise customer* and *generic service account*.

IV. **Make subjective decisions on semantic equivalence**: Assuming data integrators can find supporting evidences for the weak semantic equivalence between *W-tag(EC1)* and *W-tag(EC2.1)*, we can achieve the functional objective of this desired W-tag *ownership(2)* after a sequence of operations of semantic inheritance and semantic equivalence from *ownership(1)*. Hence, total revenue can be calculated from the invoice of *generic service account* that is part of the *generic ownership* with *W-tag(EC root1)*.

by individual sources. Since the focus is only on the subject matter, only W-tags representing the legal entity-based ECID and the location-based ECID are collected as the diagrams illustrated. The other meanings associated to the same shadows are ignored.

For the third step, data integrators need to identify what are the target W-tags. Since the overall goal is to calculate total revenue for enterprise customers, the functional objectives are to identify all service accounts that carry individual invoices for specific service products. Therefore, the first kind of target W-tags is the *generic service account*, which is a generalization for all kinds of service accounts with invoices.

Next, since different kinds of ECID are already associated with service accounts in the information ecosystems, data integrators may decide to aggregate service accounts through the help of ECID. Hence, the second kind of target W-tags is *generic enterprise customer* that satisfy the need.

With these two target W-tags *generic service account* and *generic enterprise customer* identified, we need a third target W-tags *generic ownership* which represents the meaning that a *generic enterprise customer* owns the *generic service accounts* like what is displayed in Figure 14 (2a) and (2b).

In the fourth step, data integrators need to make design decisions about how to interpret target W-tags with the source W-tags to achieve the overall goal of the integration. The functional objective is to organize all *generic service accounts* into a structure in order to calculate the total revenue; that is, to establish *generic ownership* whose sub-component of *generic enterprise customers* can shift levels to top in order to add all invoices for each *generic service account* as the total revenue for this enterprise.

Assuming another data source 4 in the information ecosystem of service provider G can provide the needed source W-tags of *ownership* between location-based ECID and service accounts. For example, in Figure 20, the solid line arrow with W-tag *Ownership(1)* represents the meaning as a mental entity that the *W-tag(EC2.2.1.1)* owns *W-tag(Acct2.1)*. Again, we need to remind readers that we use a single generic representation for W-tags, and different graphical representation (an arrow versus a circle) is simply to reduce the complicities in the diagrams for easier understanding. We do not imply that this W-tag *Ownership(1)* is a relationship type like in Entity-Relationship Model.

Let's assume the data integrators decide that the best fit the *generic enterprise customer* is the legal entity-based ECID, and the location-based ECID can become a valid *generic enterprise customer* only after been properly integrated to a legal entity-based ECID. Therefore, the challenge to calculate total revenue is how data integrators can establish bridges between legal entity-based ECID and location-based ECID such that the *generic ownership* can be established from service accounts to the top level of the legal entity-based ECID.

In this example, we assume data integrators can find supporting evidences for the weak semantic equivalence between *W-tag(EC1)* and *W-tag(EC2.1)*, illustrated as the double line arrow in Figure 20. Intuitively, the graph illustrates the final desired target W-tag *ownership(2)* is established for *W-tag(EC root1)* and *W-tag(Acct2.1.1.1)*, represented as the dashed arrow in the diagram.

There is a sequence of operations due to semantic inheritance and semantic equivalence in order to achieve the functional objective of this desired W-tag *ownership(2)*. We will not go through the step-by-step inference details as it should be straightforward in graph from *ownership(1)* to *ownership(2)*.

The main point for this example is to illustrate the generic steps based on Shadow Theory for data integration in information ecosystems. We do not discuss specific algorithms for involved design decisions, since this is a like a new platform that users can developing their own algorithms or bring their existing techniques into this new modeling approach.

As the answer for question 4, we have not yet addressed the issues about how to manage inconsistencies or conflicts in these generic steps. In next section, we will include this topic in our answers for the next question about for how to help users understand the data and what happened in databases. Remember that our goal is not to resolve inconsistencies or conflicts in the real world during the data integration process; instead, our goal is to help users recognize inconsistencies or conflicts in the real world, include them in the data models for user to understand, such that users can manage these issues or even fully resolve them in real world first. The data models should always reflect the situations whether issues are resolved or not.

## 4.3 Data model features for helping users to understand their integrated data and use it properly

Next, we need to think about the issue that the advantage of data model simply does not include helping users to easily understand what happen to their data, what logic rules the application programs follow to perform the changes, why errors happen, and where the inconsistencies or conflicts are.

The first reason is that this is traditionally viewed as the responsibility of documentation, something out of the scope of data model. In section 2.4.3, we explained that the current data models rely on documentation to help users to understand the meaning of data, and there always exist gaps between documentation and latest data in the data models. When there are multiple perspectives co-existing in information ecosystems, lack of comprehensive documentations makes users especially difficult to understand meanings for the rich variety of data flying around.

The second reason is that only a small portion of business logic rules can be implemented in data model level. For information ecosystems, the situation is even worse that when schema is overloaded with semantic heterogeneous data, the capability for data model to express business logic in schema is lost. The results are that the majority of cause-effects happened in information ecosystems are implemented in application programs somewhere in a procedure-oriented style. Even with workflow or business rule engines available at application level, it is a very manual intensive process to rely on human who understand these logics to explain everything.

Hence, we raised the question 5:

Question 5: <u>How should data models help users to understand the meanings of data?</u> For data integration, can data models help users to recognize the problems due to semantic heterogeneity such that users can manage inconsistencies or even resolve conflicts in their business operations first?

For helping users to navigate among different semantics space in the jungles of information ecosystems, the most important feature needed for a data model is something like GPS. Our proposal is that we should extend the responsibilities of a data model to include providing interactive explanations for users about the meanings of data, where the original data are from, how the integrated data are mapped, what events happen, which process perform the change, why the processes do so (e.g. following which business logic rules), and even to detect inconsistencies or conflicts from different perspectives for the same shadows. This can be summarized as the following principle.

Principle 6. To helps users to understand and use integrated data properly, data models need some features to explain the meanings of data, including modeling perspectives, business logic rules, and the criteria for decision decisions made for semantic equivalence.

Here we may need some clarification for the terminologies. We follow the common usage of the terms about a database, which is a collection of data, and a data model, which is a collection of conceptual tools for describing data, data relationships, data semantics, and data constraints [39]. Specifically, we have two levels of meanings for data models here:

(1) The design of a generic data model like the Relational Model with algebra/calculus as the foundation for its operations.

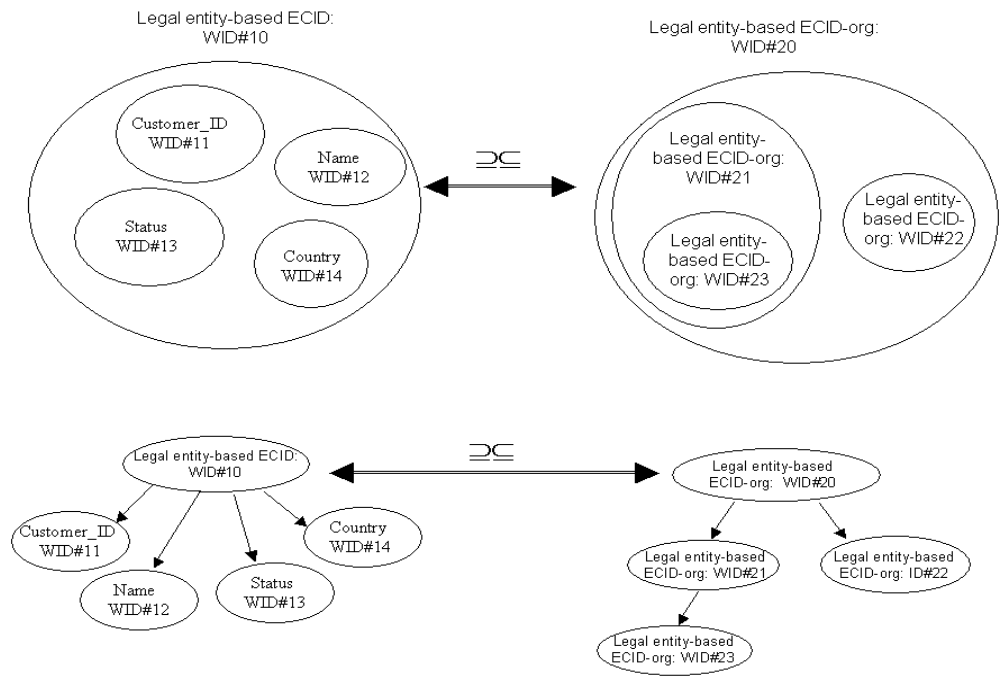

Figure 21. An example with Shadow Theory-based approach to model the meanings of a traditional design through primary keys/ foreign keys in Relational Model. The top left graph illustrates the decomposition of *legal entity-based ECID* with the meanings of its attributes as the sub-area in semantic space. The top right diagram illustrate the modeling of organizational structures such that any sub-area of *legal entity-based ECID* is again another *legal entity-based ECID.* The bottom graphs use hierarchies to represent the same.

(2) The generic principles to design the data models for specific applications, for example, enterprise customer data for used in a data warehouse or an enterprise portal.

Readers may wonder why the features we proposed here is not for the level of DBMS, Database-Management Systems, which is defined as a collection of interrelated data and a set of programs to access those data. That is, such automatic explanation mechanism can be viewed as some application programs to provide descriptions for users, and in theory, such layer can be added to any kind of data models.

The reason is that these features are at conceptual level based on the notion of Shadow Theory: the explanations about the meanings of shadows are for the mental entities in viewers' cognitive structures. The meanings already exist with or without DBMS, and application programs cannot access these meanings directly. Further, data as shadows are not required to be stored only in DBMS; they can in files or any kind of systems.

The key difference is that data in storage is not and should not be the center, especially for information ecosystems in which there are different storages holding the same or similar data. What we propose here is to adjust our position such that the meanings as mental entities in viewers' cognitive structure are the center for the modeling. That is, the business semantics for how the specific users follow and perform their business operations is the specific cognitive structure holding meanings, not the database as the tool to help them. This is especially important for information ecosystems in which there are different groups of viewers for the same shadows, with different ways to interpretations. Even data may seem similar, the semantic heterogeneity hidden beneath is what we need to identify, and the W-tags and E-tags proposed in previous sections are for such purpose.

Now, we will explain the details for this principle to help users understand their data.

### 4.3.1 Dynamic meanings and templates

W-tags and the semantic relations among them are the foundation for providing explanations about the meanings of shadows. In section 3.4, we proposed a generic decomposition mechanism for meanings in semantic space for representing the semantic relations among W-tags, and it can also function like templates such that we can even model the situations when we do not have all of the related data. In other words, we can express the meanings of existing shadows, and we can also express the associated meanings (as mental entities) that should also exist even we do not have the corresponding shadows.

When the same shadow is attached with different W-tags from the same or different perspectives and these W-tags have different semantic relations with other meanings, we can use E-tags to represent the situations and provide explanations to users. For example, Figure 21 illustrates a case to use Shadow Theory-based approach to model the meanings in a traditional design through primary keys/ foreign keys in Relational Model. The W-tag *legal entity-based ECID* is the specific ECID is modeled from legal entity perspective.

There are two ways to decompose *legal entity-based ECID* in the original data source:

(1) Originally modeled as primary keys: Each shadow attached with *legal entity-based ECID* is a tuple with attributes like Customer_ID (the primary key), Name, Status, and Country. Therefore, we treat the meaning of the shadow as a heterogeneous decomposition such that the *legal entity-based ECID* is decomposed into sub-meanings associated to its attributes, as illustrated by the top left diagram in Figure 21. I.e. the area that represents *legal entity-based ECID* in semantic space, WID#10, has sub-areas for the meanings associated to Customer_ID, Name, Status, and Country, and these sub-areas hold the characteristics describing the overall *legal entity-based ECID*.

(2) Original modeled as foreign keys: For modeling the organizational structure, the original data source established relations between *legal entity-based ECID* through foreign keys. Such relations can be viewed as the area of *legal entity-based ECID* in semantic space, WID#20, has sub-areas WID#21, WID#22, WID#23, and each sub-area is again *legal entity-based ECID*, as illustrated in top right diagram in Figure 21.

Obviously, the two different decompositions are for the same meaning as mental entities within the same perspective. Since our W-tag Rule # 8 requests that we need to use different W-tags, we will use *legal entity-based ECID-org* as the W-tag for the second decomposition (top right diagram in Figure 21).

Next, we use an E-tag to represent the strong semantic equivalence between *legal entity-based ECID,* WID#10, and *legal entity-based ECID-org,* WID#20. Since it is due to two different W-tags anchoring with the same meaning as a mental entity in the viewers' cognitive structure, the required E-tag are like establishing a **synchronization point** to bridge between different decomposition mechanisms. This corresponds to the primary key / foreign key relations in Relational Model when the foreign keys are used to describe how the same things can be decomposed in different way.

The purpose of this example is to show that we can use W-tags and E-tags to explain dynamic meanings for shadows, and the decomposition structures of W-tags are like templates in semantic space. If we can establish E-tags between W-tags with different decomposition structures, we can perform many useful inferences based on their semantic relations even we do not have all of the detail data, i.e. we do not need all of the sub-W-tags be attached to the associated sub-shadows.

Compared with documentation, one of the major advantages to provide automatic explanations based on W-tags and E-tags is due to that W-tags and E-tags are integral part of the data model itself, such that they can always evolve together with changes of the data model. In this way, we can eliminate the gap between documentation and the actual data model.

System analysts can interact with the "live" data models to know current meanings already in design, and analyze business semantics by comparison with current ones for needs of new projects. Such concept definitions that are usually delivered in requirements documents can now be alive and serve as the definition of W-tags, E-tags, and P-tags.

Compared with XML or other approaches for self-described data, the difference is that we do not try to provide the meanings with data instances together; instead, our focus shifts to the chosen perspectives in semantic space since W-tags and E-tags built on top are for meanings as mental entities living only in viewers' cognitive structures. Our goal is to support dynamic meanings with multiple versions of truth in such context, not to define static meanings for data living in storages.

One important application is for supporting incremental design changes that are demanded by agile software development. It is due to that we can by pass the know schema evolution difficulties: we do not have rigid logical structure that shadows must fit into. W-tags and their decomposition structures can be added or revised in semantic space independently from shadows in data space. And the semantic relations among W-tags from different perspectives can also evolve independently from each other. When business semantics evolves, or when there is progress for further understanding of the data, W-tags and E-tags can be revised while data may be kept the same. In this way, we can also eliminate the gap between the evolution of business semantics and the semantics represented in the data model.

### 4.3.2 *Dynamic Behavior and Business Logic*

In addition to dynamic meanings for shadows, we also need to help users to understand the meanings for observed dynamic behaviors of shadows. These include the tracking of data changes, explanations for why data change happened, and the business logic rules implemented in the application programs to perform changes from different perspectives.

Traditionally, business logic is partially implemented in data models (i.e. referential integrity in schema level) and partially implemented in application programs. For information ecosystems, the situation is even worse as schema may be overloaded with semantic heterogeneity, which further reduces the capability for the data model to express business logic in a declarative way. The result is that most of the business logic rules can only be implemented in application programs in a procedure-oriented way, hidden somewhere in the jungle of information systems (see section 2.4.3). The consequence is that, for users to understand what happened to data, why and how, they have to rely on human developers to hunt the explanations in the applications programs in the jungle of information ecosystems.

In Shadow Theory, applying business logic to perform operations for shadows in data space should correspond with behavior patterns of mental entities that interact with each other in semantic space. We should track not only changes in data space but also the meanings of changes in semantic space. Collecting meanings of changes can help users to understand and manage their data, especially for integrated data that changes can happen in the data sources or in local operations.

Under such context, our proposal is to request application programs that perform any data operations to change data (i.e. insert, update, delete) in the local integrated database directly must provide explanations for their actions. Since the business logic is for data operations under specific chosen perspectives, the explanation should also be limited to the corresponding W-tags and E-tags within the same perspectives for consistency.

To further organize these explanations, we can even ask application programs to register their business logic rules such that they can refer to these registered business rules in their provided explanations. One way to implement this idea is to enforce such registration process with authentication for access to database. Instead of current ways to grant database insert, update, delete permissions without checking what business logic will be performed, we can grant database access with permissions to only perform specific business logic which is pre-defined.

If the changes are made in upstream systems and then the data flow to local database, we can refer users to ask upstream systems for explanations. For example, for the application program to load *legal entity-based ECID* from upstream systems into the integrated database, it can use a specific process identifier to log in to the database such that each allowed data operation is registered with proper explanation for what the business logic is.

Combined with automatic tracking about what the application programs performed, dynamic meanings and templates we mentioned in previous section, the database system can provide users an overview for what meaning(s) the data represent, what happen to data, when the changes are applied, which process or users performed the operations, and the business logic rules these changes try to follow. Obviously, the quality of the explanations and how users can understand them depend on the individual data modelers and integrators.

Our objective about modeling business logic is limited to only helping users to understand the meanings of dynamic behaviors of shadows, not to build a global model of all business logics performed in the information ecosystem. This is due to some fundamental constraints, including the difficulty for establishing any global consistent modeling; if we cannot even get homogeneous meanings defined for data to be integrated, a consistent global model for business logic that operates on these heterogeneous data is even harder to achieve.

One level deeper, our goal is to reduce the complexity for data integration. In general, tracking data changes and their associated meanings can be performed at application level or at data model level. What we proposed here is to move this kind of functionality into the data model layer in order for users to focus on managing semantic heterogeneity. If the database system can automatically trace data changes with explanations, then data integrators can jump into the root difficulties due to semantic heterogeneity (opposite to current situation that data integrators have to dig meanings of data and business meanings in the jungle of information ecosystems in order to see where semantic heterogeneity is).

### 4.3.3  Data Integration: Understandability with Clear Semantics

In section 2.4.3, when we discuss the criteria of understandability with clear semantics to evaluation data integration project, we suggest this should be evaluated against the fundamental data model first. Further, we raise the question 5 and propose that the data model itself should include utility to help users to understand the meanings of data in semantic space, not just to manage data space. Such automatic explanation mechanism for shadows includes dynamic meanings and templates (section 4.3.1), dynamic behavior and business logic (section 4.3.2), and the explanations for data integration designs corresponding to the criteria of understandability with clear semantics.

The most important part is to help users understand the design decisions about what shadows are treated as the same due to that their meanings as mental entities in viewers' cognitive structures are determined to equivalently represent the same *thing* based on specific criteria and supporting evidences. That is, to explain to users how and why different shadows can be mapped together with strong or weak shadow equivalence, and how the semantic relations for the individual shadows from different perspectives can (or cannot) work together.

For example, Figure 22 shows an example for mapping between *legal entity-based ECID* (that is illustrated in Figure 21) and *contract-based ECID*. For each perspective, we have strong semantic equivalence to represent their different decomposition structures, traditionally modeled as the relations between primary keys and foreign keys, illustrated as WID#10 $\supseteq\subseteq$ WID#20 (in the top graph), and WID#30 $\supseteq\subseteq$ WID#40 (in the bottom graph) respectively.

Let's assuming that data integrators have identified evidences to support weak semantic equivalence WID#20 $\subseteq$ WID#40 such that the *contract-based ECID-org* WID#40 can be viewed as the same as *legal entity-based ECID-org* WID#20, but not vice versa. In other words, in terms of the organizational structures, these two W-tags associate to the same enterprise customer.

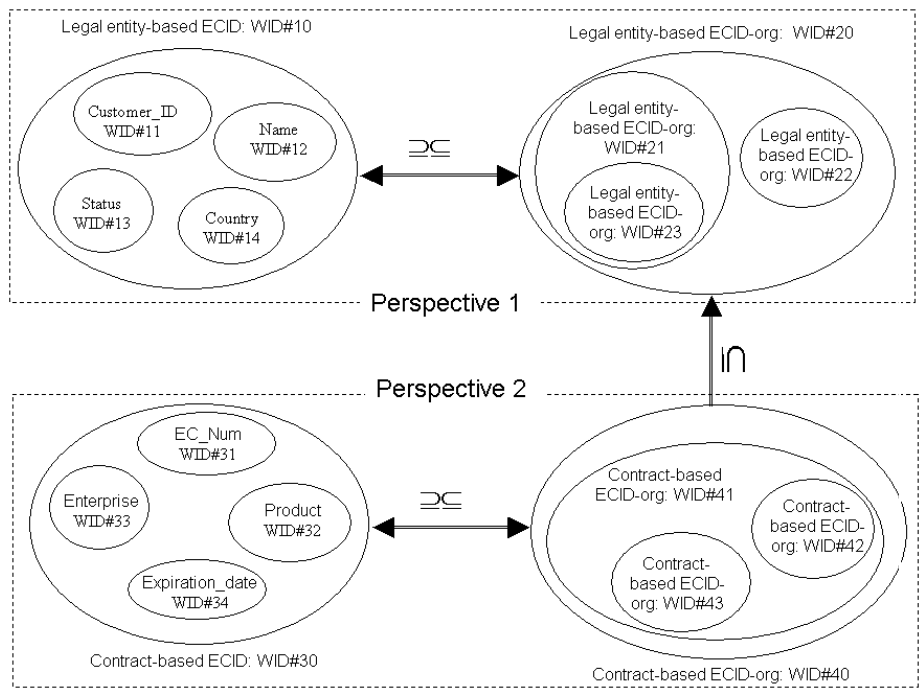

**Figure 22.** Continuing the example in Figure 21, this diagram illustrates that data integrators establish weak semantic equivalence WID#20 $\subseteq$ WID#40 across the boundaries of perspective 1 legal entity-based ECID and perspective 2 contract-based ECID, such that the *contract-based ECID-org* WID#40 can be viewed as the same as *legal entity-based ECID-org* WID#20, but not vice versa.

Readers may observe that by combining the strong semantic equivalence we have for individual perspectives, we can infer that *legal entity-based ECID* WID#10 $\subseteq$ *contract-based ECID* WID#30. Traditionally, data integrators may try to find the mappings between their different decomposition structures, e.g. Customer_ID WID#11 is mapped to EC_NUM WID#31 (since they are the unique keys), NAME WID#12 is mapped to Enterprise WID#33. However, there are some attributes we cannot find correspondences, for example, the PRODUCT WID#32 (which is a required foreign key) of *contract-based ECID* is not a notion existing in the perspective of legal entity-based ECID. In one way, people may explain that *legal entity-based ECID* can have any kind of products so they can be mapped like another weak semantic equivalence; however, since *legal entity-based ECID* only exists in the information ecosystems of service provider B, which does not include any of the PRODUCT for *contract-based ECID* existing in the information ecosystem in service provider W.

This example illustrates how semantic heterogeneity can be hidden under structural heterogeneity. Just like certain concepts cannot be translated precisely between different languages or different cultures, meanings as mental entities may not be able to be mapped precisely between different perspectives due to the limitations of the cognitive structures. However, when data integrators make subjectively design decisions to achieve the required functional objectives, we need to carefully include the differences of perspectives in our data model.

In this example, each semantic equivalence, WID#10 $\supseteq\subseteq$ WID#20, WID#30 $\supseteq\subseteq$ WID#40, WID#20 $\subseteq$ WID#40 should be explained to users with supporting evidences and the chosen criteria. For example, strong semantic equivalence WID#10 $\supseteq\subseteq$ WID#20 and WID#30 $\supseteq\subseteq$ WID#40 are due to their synchronization points that the different W-tags are attaching to the same shadows from their data sources. The evidence for the weak semantic equivalence WID#20 $\subseteq$ WID#40 could be the same address, the same contact person, and interaction with the enterprise customer directly to confirm these two ECID are for the same *thing* in reality. The chosen perspective for the data integration is based on legal entity-based ECID such that other kinds of ECID must be properly aligned in order to be used to serve any application purposes. In this way, we can explain to users about the design decision for how and why different shadows can be mapped together with strong or weak shadow equivalence.

Next, we need to explain to users that whether the different semantic relations associated to different W-tags can or cannot work together consistently. In the example, we can see that semantic equivalence is not transitive when we establish E-tags across the boundaries of different perspectives. That is, if *A* is mapped to *B*, *B* is mapped to *C*, we cannot simply infer that *A* can be mapped to *C* without considering the differences among their perspectives. In 5.2, Property 6, we will revise the transitive property based on point-free geometry with extra criteria to help determining when we can apply the common sense that if *A=B*, *B=C*, then *A=C.*

Further, when we discuss the logic foundation for supporting multiple versions of the truth in section 6.1.2III, we will discuss this issue in further details, which is a common trap we observed in data integration. It is a common trap because that there is no mechanism in Relational Model to give users a warning that they perform mappings across the boundaries of logical consistency (i.e. functional dependency). Since the available operators of Relational Algebra cannot support equivalence of heterogeneous data of *A, B, C* to be treated as the same (by their meanings), they can only be modeled as normal attributes in tables, i.e. *A* and *B* as attributes in a table mapping table *X*, while *B* and *C* in another mapping table *Y*. *X* and *Y* are just like normal relations, users can just join them together to get relation from A to C, which may break function dependency and thus not the real answers they need.

Back to the example in Figure 22, There is another sign indicating semantic heterogeneity: we can infer that *legal entity-based ECID* WID#10 $\subseteq$ *contract-based ECID* WID#30, but we cannot have a way to map the required component PRODUCT WID#32 of *contract-based ECID* since such notion does not exist in the perspective of legal entity-based ECID. This is the critical characteristic due to the chosen perspective, and such special factors make the mapping *legal entity-based ECID* WID#10 $\subseteq$ *contract-based ECID* WID#30 as a one-way bridge crossing the boundaries of different perspectives.

In one hand, this example shows the advantage of Shadow Theory approach that by mimicking human's behaviors, semantic equivalence can be established without being constrained by structural heterogeneity and semantic heterogeneity. On the other hand, this process ignores the differences that may be critical for certain functional objectives. Therefore, we need to help users to understand the limitation. It is like the typical example for Prototype Theory [71-73] that robins and penguins are both in the category of birds, one can fly and the other cannot. In the circumstances that we focus on their similarities and ignore the factor of flying capability, we may treat them equal semantically (i.e. one may model enterprise customers like robins, and another may model the same subject matter like penguins, then flying capability is like a special property of enterprise customers). But in other circumstances where flying capability is critical, we cannot treat them as the same if one can fly and the other cannot.

That is why we treat E-tags as bridges to cross the boundaries of semantic heterogeneity and structural heterogeneity, but whether such bridges are available to reach the goal of specific data integration projects depends on the supporting evidences and criteria chosen by the data integration. Data integrators have to specify the required characteristics explicitly in order for the data models to explain to users.

Since we implement business logic rules explicitly on W-tags, we can implement most of business logics in the data model level. Data integrators can specify the rules with only the specific W-tags, and these rules should not be applied outside of the original perspective. For example, in Scenario 3 we have discussed in section 2.4.3, there are two kinds of business logic rules: (1) Service provider B expects that every service account must have exactly one legal entity-based ECID to be the owner for financial responsibilities. (2) Service provider G expects one, or multiple location-based ECID to be the owner of a service account. Let's assume that data integrators can establish weak semantic equivalence between legal entity-based ECID and every location-based ECID, i.e. every location-based ECID is mapped to a legal entity-based ECID. If we allow the business logic rules to be applied to different perspectives, what may happen is that a service account in provider G that are shared by multiple location-based ECID can be owned by multiple legal entity-based ECID due to this mapping.

# 5. Algebra for Shadow Theory-based Data Model

In this section, we will propose algebra and calculus for Shadow Theory to support data operations in information ecosystems. We will start with the definition of basic structure in section 5.1, mixed with terminology and comparison with Relational Model in order to help explain related concepts. We will discuss the mathematical properties due to our needs to model meanings of shadows as regions (instead of points) in semantic space in section 5.2. That is, meanings as mental entities in viewers' cognitive structures are treated as regions such that we can decompose the same regions in different ways concurrently.

Since mental entities exist only in viewers' cognitive structures, we use W-tags to uniquely anchor with them. Each instance of W-tag has a unique W-tag instance ID (short as WID). Different instances of the same W-tags in semantic space carry the same ontological information and associations (from its templates), but correspond with different shadows in data space. The purpose of WID is to uniquely identify the specific instance of a meaning (as a mental entity), even when we may no longer have the shadow values.

In section 5.3, we will first illustrate an example with different kinds of ECID, and use them to illustrate the basic operations in semantic space as the algebra we proposed to support Shadow Theory. Although these operations are for semantic heterogeneous data, they can also be applied for semantic homogeneous data in order to be backward compatible with Relational Model. In section 5.4, we will discuss how to use templates and simulated schema to simulate relational schema.

## 5.1 Basic structure

We use the term **data space** to represent the set of all possible data as shadows. We use the term a **semantic space** to represent the set of all possible meanings as mental entities in viewers' cognitive structures.

A shadow $S_a$ is a data record in data space, which can represent multiple meanings in semantic space. For each different meaning in semantic space, we use a unique instance of **W(hat)-tag**, short as **W-tag**, from perspective $P$, to anchor the mental entity that exists in viewers' cognitive structures. If there are multiple meanings represented by the same shadow, then we need multiple W-tags attached to the same shadow.

In general, each W-tag carries ontological information indicating what *thing* the shadow is based on the specific perspective $P$. We assume all W-tags within the same perspective $P$ are consistent, or at least not in conflicts. That is, two perspectives $P$ and $P'$ can be consistent, inconsistent, or in conflicts due to the W-tags they have or their different semantic relations among W-tags are consistent, inconsistent, or in conflicts.

Further, to avoid confusion, each instance of W-tag carries a unique **W-tag Instance Identifier** (WID) that can exists even when the shadow is physically removed from storage; i.e. the meaning as mental entity can exist even without the associated logical representation. We will use WID in semantic space to support data operations we need, as an alternative for unique keys in data space that suffer many issues in information ecosystems (as we have described in section 2.1).

Here we use $W\text{-}tag^P(S_a)$ to denote the W-tag from perspective $P$ that is attached to shadow $S_a$. We call the W-tag is attached to the shadow since the associated meaning is not constrained by the logical representations of the shadow. That is, we use W-tags to represent any kind of meanings, and use shadows to represent data in any kind of logical representation. Our goal is to properly represent the **basic properties of semantic heterogeneity** such that there exist

(i) Different meanings for the same representation, and

(ii) Different representations for the same meaning.

Since we require every shadow has at least one W-tag, and any instance of W-tag is assigned with a unique WID, "Different representations for the same meaning" does not mean different shadows have the same WID. Instead, we define **semantic equivalence** for different meanings as mental entities that are treated as the same based on certain criteria and supporting evidences. We use **E(quivalence)-Tags**, short as **E-tags**, to represent semantic equivalence.

Therefore are two kinds of semantic equivalence, strong versus weak. For **strong semantic equivalence**, denoted as $W\text{-}tag^P(S_a) \supseteq\subseteq W\text{-}tag^Q(S_b)$ for shadow $a \mathrel{!=} b$, it indicates that the anchored mental entities are either (1) exactly the same one if within the same perspective, i.e. $P = Q$, or (2) treated as the same from different perspective, i.e. $P \mathrel{!=} Q$, such that we can bring certain semantic relations of $W\text{-}tag^P(S_a)$ in perspective $P$ to be used with semantic relations of $W\text{-}tag^Q(S_b)$ in perspective $Q$, and vice versa.

For **weak semantic equivalence**, denoted as $W\text{-}tag^P(S_a) \subseteq W\text{-}tag^Q(S_b)$ for shadow $a \mathrel{!=} b$, it indicates that the anchored mental entities are treated as the same only from one direction: $W\text{-}tag^Q(S_b)$ can be treated the same as $W\text{-}tag^P(S_a)$, but $W\text{-}tag^P(S_a)$ cannot be treated as the same as $W\text{-}tag^Q(S_b)$. Hence, we can only bring certain semantic relations of $W\text{-}tag^Q(S_b)$ in perspective $Q$ to be used with semantic relations of $W\text{-}tag^P(S_a)$ in perspective $P$, but not in the reverse direction.

Compared with Relational Model, the fundamental difference is that semantic equivalence can happen

(i) Between two attributes even if they hold different data values (from different domains) or different data types, or

(ii) Between two tuples even if they have different attributes, different data values (from different domains), or different data types, or

(iii) Between two different sets of tuples which are connected through foreign key and primary key relations respectively to form different logical structures like graph or hierarchy.

We use the term **semantic relations** to describe any relations between W-tags in semantic space, and the term **decomposition** to describe the generic relations between shadows in data space. Any shadow can be decomposed into components by different ways concurrently, and each component is again a shadow. We use a **decomposition structure**, $D^P(S_a)$, to represent a specific decomposition for shadow $S_a$ decomposed into a set of sub-shadows $(s_1, s_2, …, s_n)$ in perspective $P$. Note that the decomposition is not limited to one level, like a tuple is decomposed into a list of atomic attribute values in Relational Model. A decomposition structure can be a hierarchical structure such that sub-shadows are grouped in certain ways, and the sub-shadows cannot participate multiple groups or at different levels in such hierarchy structure. We can use different decomposition structures to model the situation if there is a need for any sub-shadow to be in multiple groups or at multiple levels concurrently.

Since $s_1, s_2, ..., s_n$ are observable data values for describing shadow $S_a$, we call them as **observable** of shadow $S_a$. Note that observable is like a role assignment for sub-shadows, and it differs from attribute of a tuple in Relational Models in

(i) Attributes are assumed to be atomic to avoid nested relations, sub-shadows as observable can be further decomposed without constraints.

(ii) Observable simply describes the role the sub-shadows take in describing the parent shadow; its existence can be independent from the existence of the parent shadows. Focusing on the child W-tags, this corresponds to the primary key / foreign key relations in Relational Model when the foreign keys are used to describe components of another shadow.

(iii) Since observable is again a shadow that multiple W-tags can be attached with, the role assignment also establish semantic relations between its W-tags and the W-tags of its parent shadow.

Corresponding to the decomposition in data space, we have semantic relations between W-tags in semantic space. For the decomposition structure within the same perspective $P$, $D^P(S_a) = (s_1, s_2, ..., s_n)$, we have the semantic relations $W\text{-}tag^P(s_j) \subseteq W\text{-}tag^P(S_a)$, where $i = 1, 2, ... n$. We can think this is a hierarchical structure and $W\text{-}tag^P(s_j)$ is at **lower level of abstraction**, compared with $W\text{-}tag^P(S_a)$.

If $W\text{-}tag^P(S_a)$ and $W\text{-}tag^P(s_j)$, $i = 1, 2, ... n$ are the same kind of W-tag (i.e. be classified as the same kind of *things* in ontology), then it is a homogeneous decomposition. Otherwise, it is heterogeneous decomposition. Although such distinction is defined for decomposition structures within the same perspective, it can be generalized when different perspectives are involved, for example, one of the sub-W-tags is in a different perspective.

The correspondence from semantic relations between W-tags in semantic space to decomposition in data space is not always required. That is, the database may not have all of the shadows corresponding to every mental entity existing in viewers' cognitive structures. This is due to the practical constraints of a single database that not directly related data is not required to be included. In other words, we can have the semantic relations $W\text{-}tag^P(s_j) \subseteq W\text{-}tag^P(S_a)$, where $i = 1, 2, ... n$, for the decomposition structure $D^P(S_a) = (s_1, s_2, ..., s_n)$, but we may not have all of the shadows $s_1, s_2, ..., s_n$ stored in database.

Readers may notice that we overload the symbol $\subseteq$ to represent either (1) weak semantic equivalence, or (2) the semantic relations corresponding to decomposition in data space. We need to highlight the difference that we use the symbol $\subseteq$ for weak semantic equivalence is when the different W-tags are in different perspectives, while for semantic relations corresponding decomposition, the W-tags are in same perspective.

The purpose of overloading is that we can have a generic representation for both semantic heterogeneous and homogeneous environments by using regions as the only ontological primitive units in semantic space. That is, the symbol $\subseteq$ represents the inclusion relations between regions in semantic space, no matter the regions are in the same or different perspectives. In other words, the basic concept for weak semantic equivalence is that a region in one perspective is included in another region from a different perspective, while the semantic relations corresponding to decomposition in data space indicate both regions are in the same perspective.

In a similar way, we also overload the symbol $\supseteq\subseteq$ to represent either (1) strong semantic equivalence when the W-tags are from different perspectives, or that (2) the different W-tags anchor with the same meanings as mental entities in the same perspective. Again, the purpose of overloading is that we can have a generic representation for both semantic heterogeneous and homogeneous environments by using regions as the only ontological primitive units in semantic space.

That is, when the symbol $\supseteq\subseteq$ is applied to W-tags within the same perspective, it indicates that the anchored meanings as mental entities are exactly the same one, however the users' cognitive structures allow them to be decomposed in different ways for reasons. Since it is due to two different W-tags anchoring with the same mental entity, the required E-tag are like establishing a synchronization point to bridge between different decomposition mechanisms. Focusing on the parent W-tag to be decomposed in different ways, this corresponds to the primary key / foreign key relations in Relational Model when the foreign keys are used to describe how the same meaning can be decomposed in different way.

When the symbol $\supseteq\subseteq$ is applied to W-tags from different perspectives, it indicates that the meanings as mental entities in semantic space exactly overlap or overlap to a degree such that the data integrators can treat them as the same subjectively. Due to the different perspectives, W-tags can be decomposed in different ways concurrently, and the associated E-tags with supporting evidences help users to understand about the decisions why they are treated the same even under structural heterogeneity.

With the above descriptions for the basic structure, our goal is a generic representation that meanings as mental entities are represented by only one kind of primitive unit, regions. Readers may wonder why it is important is to explicitly identify the boundaries of perspectives when we use these symbols. The reason is that it indicates the boundaries of consistency, semantic homogeneity versus semantic heterogeneity. It is critical for us to establish a data model to support multiple versions of the truth. Once out of the boundary of a perspective, data integrators need to be careful when they apply logic to make further inferences on their mapping. That is, the subjective decision made by one data integrator about the semantic equivalence between two W-tags may not be acceptable by another data integrator. In next section, we will discuss the required properties revised based on the original work from point-free geometry.

## 5.2 Regions as the Only Primitive Units

In section 3.4.1, we proposed to use regions as the primitive units to model meanings as mental entities in semantic space. The main reason is that we cannot have assumption of atomic elements, since any meaning that is assumed as atomic may be further decomposed (homogeneously or heterogeneously) in different perspectives. Without this assumption, we simply cannot build our data model based on First Order Logic like Relational Model did. Therefore, we need a different mathematical foundation, and point-free geometry initiated by Whitehead in [57, 58] can serve for our purpose. Here we will review the basic properties that we need to support data operations for Shadow Theory.

The basic idea is that each meaning is a unique mental entity from a specific perspective, and we can model a meaning as a region in a specific section of semantic space that corresponds to the chosen perspective. Any region can be decomposed in different ways concurrently, and the decomposition into sub-regions can continue without limitation. Further, regions can overlap with each other in many different ways, for example, one region includes another, two regions partially or completely overlap with each other, or the ratio of the overlapped areas to the regions themselves are high enough such that we can treat the two regions are the same.

Now we will use the terminologies and symbols from previous section to describe required properties in semantic space as the following. The identifiers denoted at the end of each property refer to the original axiom summarized as Definition 2.1 in [78]. For W-tags, $W\text{-}tag^P(s_j)$, $W\text{-}tag^Q(s_j)$, $W\text{-}tag^R(s_k)$, which anchor with unique meanings as mental entities in specific viewers' cognitive structures, denoted as perspective $P, Q, R$, the fundamental primitive binary relation is inclusion, denoted by "$\subseteq$", and we have the following properties in semantic space:

**Property 1.** Reflexive, Def2.1–(i).

$\forall\ W\text{-}tag^P(s_i),\ W\text{-}tag^P(s_i) \subseteq W\text{-}tag^P(s_i)$.

**Property 2. Weak semantic equivalence** across the boundaries of individual perspectives. E-tags are required if the inclusion is across the boundaries of individual perspectives, and we need some supporting evidence(s) to establish E-tags.

$\forall W\text{-}tag^P(s_i)\ \subseteq\ W\text{-}tag^Q(s_j),\ \exists E\text{-}tag(W\text{-}tag^P(s_i) \subseteq W\text{-}tag^Q(s_j))$, where $i\ !=\ j, P\ !=\ Q$.

**Property 3.** No upper bounds, i.e. any meaning as mental entity may be used as observable for another meaning within the same or different perspectives. Def2.1-(iv).

$\forall\ W\text{-}tag^P(s_i), \exists\ W\text{-}tag^Q(s_j),\ W\text{-}tag^P(s_i) \subseteq W\text{-}tag^Q(s_j),\ i\ != j$.

**Property 4.** No lower bound, i.e. any meaning as mental entity can be further decomposed in the same or different perspective. Def2.1-(iv).

$\forall\ W\text{-}tag^Q(s_j), \exists W\text{-}tag^P(s_i),\ W\text{-}tag^P(s_i) \subseteq W\text{-}tag^Q(s_j),\ i\ !=j$.

**Property 5.** Strong semantic equivalence. Def2.1-(iii).

If $W\text{-}tag^P(s_i) \subseteq W\text{-}tag^Q(s_j)$ and $W\text{-}tag^Q(s_j) \subseteq W\text{-}tag^P(s_i),\ i\ !=j$,

then $W\text{-}tag^P(s_i) \supseteq\subseteq W\text{-}tag^Q(s_j)$ if $P = Q$, i.e. within the same perspective.

If $P\ !=\ Q$, then there are some subjective decisions needed to determine whether the meanings as mental entities can be treated as the same or not. Such decisions should be based on the supporting evidences of E-tags and the functional objectives of the data integration project. The $E\text{-}tag(W\text{-}tag^P(s_i) \supseteq\subseteq W\text{-}tag^Q(s_j))$ depends on the combined supporting power of the equivalence decision for $E\text{-}tag(W\text{-}tag^P(s_i) \subseteq W\text{-}tag^Q(s_j))$ and $E\text{-}tag(W\text{-}tag^Q(s_j) \subseteq W\text{-}tag^P(s_i))$.

**Property 6.** Transitive. There are some conditions for transitive property to be valid. Def2.1-(ii).

If $W\text{-}tag^P(s_i) \subseteq W\text{-}tag^Q(s_j), W\text{-}tag^Q(s_j) \subseteq W\text{-}tag^R(s_k),\ i\ != j, j\ != k$, $k\ !=i$, then $W\text{-}tag^P(s_i) \subseteq W\text{-}tag^R(s_k)$ if $P = Q = R$, i.e. within the same perspective.

Otherwise, then there are some subjective decisions needed to determine whether the meanings as mental entities can be treated as the same or not. Such decisions should be based on the supporting evidences of E-tags and the functional objectives of the data integration project.

(i) If $P = Q, Q\ != R$, then the $E\text{-}tag(W\text{-}tag^P(s_i) \subseteq W\text{-}tag^R(s_k))$ depends on the supporting power of the equivalence decision for $E\text{-}tag(W\text{-}tag^Q(s_j) \subseteq W\text{-}tag^R(s_k))$.

(ii) If $P\ != Q, Q = R$, then the $E\text{-}tag(W\text{-}tag^P(s_i) \subseteq W\text{-}tag^R(s_k))$ depends on the supporting power of the equivalence decision for $E\text{-}tag(W\text{-}tag^P(s_i) \subseteq W\text{-}tag^Q(s_j))$.

(iii) If $P\ != Q, Q\ != R$, then the $E\text{-}tag(W\text{-}tag^P(s_i) \subseteq W\text{-}tag^R(s_k))$ depends on the combined supporting power of the equivalence decisions for $E\text{-}tag(W\text{-}tag^Q(s_j) \subseteq W\text{-}tag^R(s_k))$ and $E\text{-}tag(W\text{-}tag^P(s_i) \subseteq W\text{-}tag^Q(s_j))$.

**Property 7.** Given an inclusion relation between two regions, there exist another region to be between these two regions. Def2.1-(v).

$\forall\ W\text{-}tag^P(s_i), \forall\ W\text{-}tag^R(s_k),\ W\text{-}tag^P(s_i) \subseteq W\text{-}tag^R(s_k)$,

$\exists\ W\text{-}tag^Q(s_j)$, such that $W\text{-}tag^P(s_i) \subseteq W\text{-}tag^Q(s_j) \subseteq W\text{-}tag^R(s_k)$.

This is a different application of weak semantic equivalence, and it provides the foundation to support multiple answers for data integration (i.e. different users can make different decisions about what meanings as mental entities can be viewed as the same to achieve their functional objectives).

**Property 8.** Given any two regions, there exists a region that includes them both. Def2.1-(vi).

$\forall\ W\text{-}tag^P(s_i),\ \forall\ W\text{-}tag^Q(s_j),\ \exists\ W\text{-}tag^R(s_k),\ W\text{-}tag^P(s_i) \subseteq W\text{-}tag^R(s_k)$ and $W\text{-}tag^Q(s_j) \subseteq W\text{-}tag^R(s_k)$.

The application of this property is that we can use inclusion as the only primitive relations to describe any relations by W-tags, i.e. the relation between any two W-tags can be established by a third region that include these two.

**Property 9.** Given any two regions, there exists a region that includes them both. Def2.1-(vii).

If $\forall\ W\text{-}tag^P(s_i),\ W\text{-}tag^P(s_i) \subseteq W\text{-}tag^Q(s_j)$, and we have $W\text{-}tag^P(s_i) \subseteq W\text{-}tag^R(s_k)$, then we can infer $W\text{-}tag^Q(s_j) \subseteq W\text{-}tag^R(s_k)$.

## 5.3 Algebra for Shadow Theory-based Data Model

Based on the above properties of point-free geometry, we will propose algebra to support data operation. Before we introduce each operator, we will first discuss how to insert, update, or delete shadows in data space, and to assign W-tags, or E-tags in semantic space in this section.

In general, these Insert-Update-Delete operations can be classified according the questions we try to answer from the data model, including:

(A) **Who** performed the changes in either data space or semantic space? and **why** they make such changes (i.e. following which business operation rules) ?

- Process registration: Each application program needs to register with a PROCESS_ID in order for the data model to track who they are. Although this can be easily implemented as users authorization in database systems, we require this information at conceptual level in order to model that who does what in the jungle of information ecosystems.

- Description for business logic for any operation: For any tag assignment or shadow insert-update-delete operation, we need collect descriptions for the business logic in order to provide systematic explanations for users to understand why changes happen in the specific way. This is to support the extra responsibilities we propose in section 4.3.2 for understandability with clear semantics.

(B) **What** happen to shadows in data space? and **how** does it happened? What does it means in terms of meanings as mental entities in the specific viewers' cognitive structures?

- Insert new shadows with W-tag assignment
- Delete shadows
- Update existing shadows
- Assign Extra W-tags to Existing Shadows, revise W-tags, or delete W-tags
- Establish semantic relations among W-tags
- Assign E-tags to pairs of W-tags for semantic equivalence
- Establish Template For Decomposition Structure

(C) **When** and **where** (i.e. which upstream systems) did changes happen?

Now we will discuss the algebra, specifically we focus on the operators in semantic space. We will also include short comparisons when we introduce the properties of these operators and the differences compared with the corresponding operators in Relational Model.

The basic design principle for the algebra is **operation by meanings in semantic space**, i.e. operations by W-tags that anchor with unique meanings as mental entities in specific viewers' cognitive structures. Our goal is to support the characteristics of semantic heterogeneity unavoidable in information ecosystems, i.e. the same meaning can have different representations, and the same representation can have different meanings. In contrast, we can call the principle used in Relational Algebra as **operation by data values and their logical structure in data space**, which is really designed for semantic homogeneous environments where the same meaning should have the same representation, and the same representation should have the same meaning.

The advantage of using W-tags in algebra is that we enforce the operations to always performed with semantic heterogeneity, even we may have different data values and logical structure under the W-tags. By E-tags, we have only one standard but flexible mechanism to represent the same meaning, not like Relational Model relying on the same data values in the same domains and the same logical structure, still with no guarantee when overloaded with semantic heterogeneity. By different W-tags, we have only one standard mechanism to represent different meanings, existing as mental entities in different viewers' cognitive structures.

As we proposed in section 3.4.2, we can use W-tag Instance Identifier (short as WID) in semantic space as the alternative to unique keys in data space. The advantage of WID is that the origin of uniqueness is due to the unique existence of mental entities in specific viewers' cognitive structures. For a group of people with common mental entities, the uniqueness is due to the existence of such mental entities in the shared cognitive structure.

On the contrary, the uniqueness of unique keys in data space is due to their logical representation (i.e. schema) in a specific database.

This is the subtle by critical point we need in the algebra for managing the difficulties of semantic heterogeneity. The reason is that the reality of uniqueness in any model is not necessarily related with the uniqueness of the *things* as the subject matter in reality. The fact is, we cannot know what the *things* are without choosing a perspective implicitly or explicitly. The existence of uniqueness relies on the chosen perspective; hence, the uniqueness is constrained within the perspective and it cannot survive outside of the perspective.

Based on the above notion of W-tags, E-tags, and WID, we have the following operators in semantic space:

OP 1. Selection of existing shadows by their meanings

OP 2. Project shadows by their meanings into a different decomposition structure.

OP 3. Selection of shadows based on union of their meanings.

OP 4. Selection of shadows based on difference of their meanings.

OP 5. Selection of shadows based on intersection of their meanings

OP 6. Create meaningful shadows by combining existing shadows into newly defined decomposition structure.

OP 7. Bi-directional join operation in semantic space.

OP 8. Uni-directional join operation in semantic space

### 5.3.1 ECID Example

To better explain these operations with examples, we will need more details of the ECID data as the following.

**Scenario** 4. **Figure 23** illustrates the enterprise customer data received from four different sources. We use *P1* to represent the perspective of legal entity-based ECID from service provider B, *P2* to represent the perspective of location-based ECID from service provider G, and *P3* to represent the perspective of contract-based ECID from service provider W. *P4* indicates the perspective from integrated billing systems, in which the main focus is account and ECID is modeled as attributes of the account.

For simplicity, we assume that semantic heterogeneity only exists across perspectives, i.e., all data are modeled in a semantic homogeneous way within a single perspective. Therefore, the W-tags displayed in the diagram are the associated table names or column names in their original model (since only semantic homogeneous data are in these tables). The meanings of these data are explained as the following.

*P1: Legal entity-based ECID*

The first perspective *P1* is the legal entity-based ECID from service provider B. Assuming the source schema is not available, the customer related shadows provided for data integration is illustrated in Figure 23 *P1*. For each legal entity-based ECID, it has the following W-tags for its sub-shadows as observable:

(i) *Customer_ID*: The unique identifier to represent individual enterprise as a legal entity.

(ii) *Name*: The name of each legal entity registered with governments.

(iii) *Status*: The status for customer; A is for active customer, P is for pending customer, D is for disconnected customer, and I is for logical deletion such that the enterprise is no longer a valid legal entity.

(iv) *Country*: The country where this enterprise is registered in.

(v) *Parent*: The parent enterprise that this enterprise belongs to.

*P2: Location-based ECID*

The second perspective *P2* is the location-based ECID from service provider G. The customer related shadows provided for data integration is illustrated in Figure 23 *P2*. For each location-based ECID, it has the following W-tags for its sub-shadows as observable:

(i) *Customer*: The unique identifier to represent the enterprise customer at the specific location.

(ii) *Line1*: The name that the customer provided.

(iii) *Line2*: Street address.

(iv) *Line3*: Town or city.

(v) *State*: State or providence.

In addition, location-based ECID shadow can be grouped as ECID_Group for billing purpose. A mapping between ECID_Group and location-based ECID is also provided with Map_OID as unique key for each mapping record. For each ECID_Group, it has the following W-tags for its sub-shadows as observable:

(i) *Group_ID*: The unique identifier to identify each group.

(ii) *Group_Name*: The name of the group provided by customer.

(iii) *Bill_Address*: The address that customer want their invoices be forward to.

*P3: Contract-based ECID*

The third perspective *P3* is the contract-based ECID from service provider W. The customer related shadows provided for data integration is illustrated in Figure 23 *P3*. For each contract-based ECID, it has the following W-tags for its sub-shadows as observable:

(i) *EC_Num*: The unique identifier to the enterprise at the specific location.

(ii) *Product*: The service that enterprise customer subscribed.

(iii) *Enterprise*: The name of the enterprise used in the contract.

(iv) *Expiration_Date*: The date when the contract expires.

There is also a mapping shadow between parent and child contract-based ECID. The W-tags for the sub-shadows are the following:

(i) *Parent_Num*: The parent contract-based ECID.

(ii) *Child_Num*: The child contract-based ECID.

(iii) *Expiration_date*: The date when the mapping is no longer valid.

*P4: Integrated Billing_Account*

The fourth perspective *P4* is the integrated billing accounts across service provider B, G, and W for billing organization only. The integrated billing account shadows provided for ECID data

*P1*:
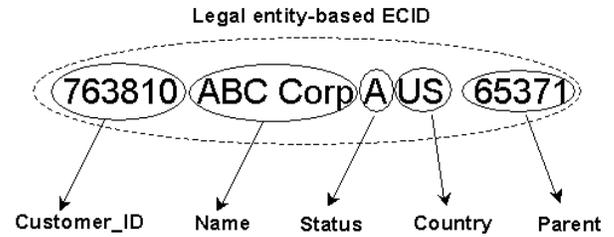

*P2*:
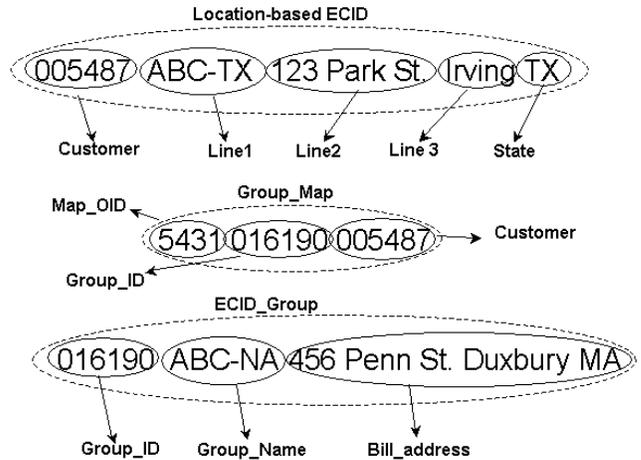

*P3*:
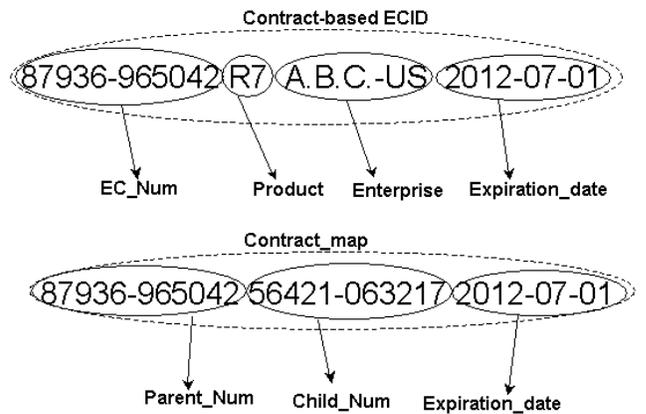

*P4*:
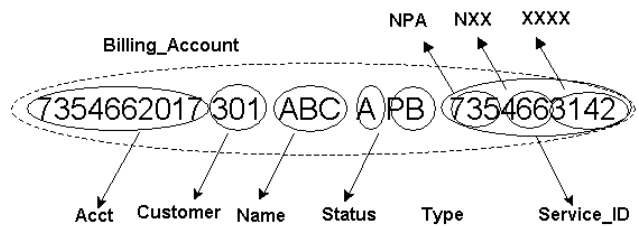

**Figure 23. Example data of ECID from different perspectives.**

integration is illustrated in Figure 23 *P4*. For each Billing_Account, it has the following W-tags for its sub-shadows as observable:

(i) *Acct*: The unique identifier for the billing account.

(ii) *Customer*: An identifier to distinguish different owners who own the same Acct at different time.

(iii) *Name*: The enterprise name provided by the customer for billing purpose.

(iv) *Status*: "A" indicates active status. At any times, there is only one record with active status for the same Acct.

(v) *Type*: The type indicator for Servgice_ID.

(vi) *Service_ID*: The alternative account key used in the original sources. If Type = PB, then it is a telephone number (W_tag = *P4*:Telephone_Number) following the format of North American Numbering Plan, NPA is area code, NXX is Exchange code, and XXXX is station code. If Type = WR, then it is a circuit identifier.

For illustration purpose, we can present these examples in XML such that XML data areas represent shadows in data space, and XML tag areas represent meanings as mental entities anchored by W-tag or E-tag in semantic space, and XML name space areas represent perspectives. For *P1: Legal entity-based ECID*, it is like the following XML:

<*P1:Legal entity-based ECID* WID=001>

    <*P1:Customer_ID* WID=002>763810</*P1: Customer_ID*>

    <*P1:Name* WID=003>ABC Corp</*P1:Name* >

    <*P1:Status* WID=004>A</*P1:Status*>

    <*P1:Country* WID=005>US</*P1:Country*>

</*P1:Legal entity-based ECID*>

WID represents the unique W-tag Instance Identifiers that system generated when the W-tags were assigned to the shadow. The decomposition structure is represented by a set of semantic relations between involved W-tags. For example, the decomposition structure for *P1: Legal entity-based ECID* includes the semantic relations that *P1: Legal entity-based ECID* (WID=001) as an area in semantic space includes the areas of *P1:Customer_ID* (WID=002), *P1:Name* (WID=003), *P1:Status* (WID=004), *P1:Country* (WID=005). Since the shadow value of *P1:Customer_ID* is modeled as the primary key to represent *P1:Legal entity-based ECID*, we can establish a strong semantic equivalence *P1: Legal entity-based ECID* (WID=001) $\supseteq\subseteq$ *P1:Customer_ID* (WID=002) to represent such relation in the decomposition structure.

With these examples, we can start the introduction for each operator with an explanation for its operation procedures, as well as a declarative definition.

### 5.3.2 Select Shadow by their Meanings

The select operation is to select areas in semantic space with their associated shadows in data space based on the specified decomposition structure(s) and perspective(s) through a set of predicates, and the predicates include descriptions of data values, meanings, and semantic relations between meanings. Institutively, we can use the diagram in Figure 24 to explain this operation.

For a given shadow $S_a$ in data space about some *thing* in reality, we may have different W-tags, e.g. $\textbf{\textit{W-tag}}^P(S_a)$ and $\textbf{\textit{W-tag}}^Q(S_a)$, attached to the same shadow due to different meanings from different perspectives. Therefore, there are different decomposition structures, $\textbf{\textit{D}}^P(S_a)$ and $\textbf{\textit{D}}^Q(S_a)$, with E-tag($\textbf{\textit{W-tag}}^P(S_a) \supseteq\subseteq \textbf{\textit{W-tag}}^Q(S_a)$) since they attached to the same shadow as a synchronization point to bridge the different decomposition structures for the same mental entity (to simplify our model, we request to allow only a single decomposition structure within a perspective, and when viewers allow different ways to decompose a meaning as a mental entity in their chosen perspective, we model them as two different perspective with a synchronization point for strong semantic equivalence).

This generic select operation can be used to select the shadow $S_a$ by either one of its W-tags $\textbf{\textit{W-tag}}^P(S_a)$ or $\textbf{\textit{W-tag}}^Q(S_a)$. Similar to the select operation in Relational Algebra, we use the Greek letter sigma($\sigma$) to denote selection, the predicate appears as a subscript, and the argument relation is in parentheses after $\sigma$.

For example, the following

$$\sigma_{W-tag^P(S_a)}(D^P)$$

indicates to select all sub-shadows within the given region of $\textbf{\textit{W-tag}}^P(S_a)$ in semantic space that is specified by perspective *P* following the decomposition structure $\textbf{\textit{D}}^P(S_a)$.

For the example in Figure 23, we can have

$$\sigma_{P1:Legalentity-basedECID}(D^{P1}) =$$

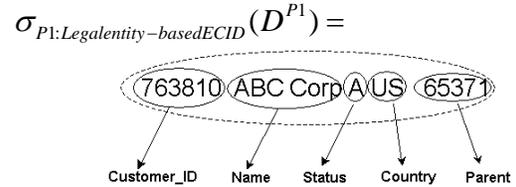

That is, we want to select all of shadows tagged with *Legal entity-based ECID* in perspective *P1*. This select operation retrieves all shadows that are attached with this kind of W-tag, with their sub-shadows and sub-W-tags according to decomposition structure $\textbf{\textit{D}}^P$.

Compared with the select operation in Relational Algebra, reader can see that the role of decomposition structure here is like the role of a table (a relation), which provides the basic structure for data operations. However, the difference is that decomposition structure is not a rigid logical representation constrained in data space here, but rather a representation for a set of flexible semantic relations between W-tags in semantic space. This is where the advantages are: the sub-shadows are regions that can be decomposed again in different ways (no assumption about atomic data elements), and we can perform the selection in a hierarchical way with unlimited levels of depth.

Another use of this select operation is to access those (sub-)W-tags exist only in one of the different decomposition structures. For example, let's assume that in perspective *P*, there is a sub-shadow $S_b$ such that $\textbf{\textit{W-tag}}^P(S_b) \subseteq \textbf{\textit{W-tag}}^P(S_a)$ due to its decomposition structure, and we cannot find any corresponding meaning in perspective *P* (as illustrated in Figure 24) We can use this select operation to select shadow $S_b$ by $\textbf{\textit{W-tag}}^P(S_b)$ with the given regions (i.e. $\textbf{\textit{W-tag}}^P(S_a)$) in semantic space that is specified by chosen perspective (i.e. *P*) following some patterns in available decomposition structures (i.e. $\textbf{\textit{D}}^P(S_a)$).

Here we need to explain the different role of predicates used in the selection operation. The predicates in Relational Algebra must be constrained by First Order Logic, (FO, i.e. variables can only in terms, not on predicates), while we allow predicates to be like in Second Order Logic (SO, i.e. variables can be in both terms and predicates). Since we use only W-tags for representing meanings, it is more like Monadic Second Order logic (MSO), an extension of FO by allowing quantified variables denoting set of elements.

This allows us to use the same W-tags for heterogeneous logical structures and non-atomic data elements that can be further decomposed in different ways even we cannot explicitly list all of its possible contents. When applied in XML queries and transformation, researchers reported the advantages of MSO include (i) No explicit recursions needed for deep matching, (ii) Don't-care semantics to avoid mentioning irrelevant nodes, (iii) N-ary queries are naturally expressible, (iv) All regular queries are definable [79].

However, more precisely, we can only say the predicates are MSO-like since we miss the atomic data elements and the contents of any set cannot be explicitly represented (i.e. a region can be viewed like a set of points, however, we cannot explicitly represent all of the points included or all of the possible ways of decomposition for this region). We can adapt Thomas's logical representation of MSO proposed in [80] with some revision, which has the same expressive power as traditional MSO, that all second order quantifiers are shifted in front of first order quantifiers, and first order variables are cancelled by simulating elements with singletons. That is, if lowercase variable $x$, $y$ represent atomic data elements, and uppercase variable $X$, $Y$ to represent sets (monadic second order objects), Thomas uses the atomic formula $\{x\} \subseteq X$ to replace $X(x)$ to indicate that $x$ is an atomic element in $X$, uses $Sing(X)$ and $Sing(X)$ to indicate that $X$ and $Y$ are singleton, and uses $X \subseteq Y$ to indicate an ordering relation $R(X,Y)$ that $X$, $Y$ are singletons $\{x\}$, $\{y\}$ such that $R(x, y)$ exists. The revision we need is to drop the atomic formula $Sing(X)$ and any first order variables since any W-tag for its associated shadow is modeled as a region in semantic space, such that we will never have any W-tag that is singular.

For the example in Figure 23,

$$\sigma_{P4:NPA(s) \wedge s=735}(D^{P4}) =$$

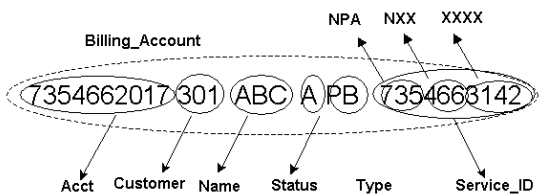

It indicates that we want to search any integrated billing account across service provider B, G, and W for billing organization that has area code 735. That is, we want to select all shadows attached with W-tag *Billing_Account* in perspective *P4* with the criteria that in the decomposition structure there exists a sub-shadow 735 attached with W-tag *P4:NPA*, no matter how deep it may be (it is at level 2 in this example).

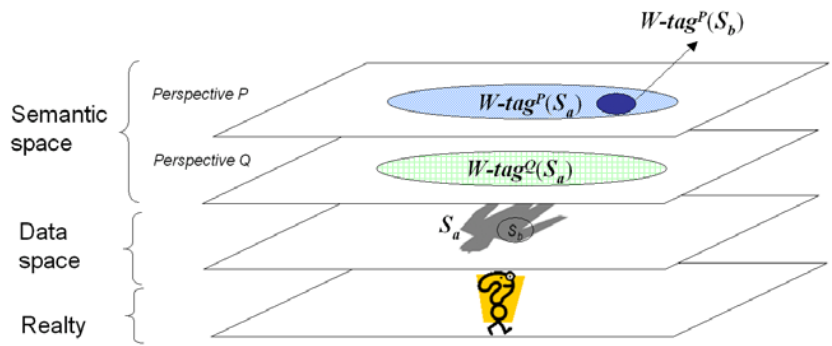

**Figure 24.** An illustration for select operation. Given a given shadow $S_a$ and its attached W-tags from different perspective, $W\text{-}tag^P(S_a)$ and $W\text{-}tag^Q(S_b)$, we can use W-tags to select the shadow or its sub-shadows $S_b$ existing only in the decomposition structure of perspective *P*.

If we review the select operation in Tuple Relational Calculus, the operation of $\sigma_p(r)$ is defined as the following:

$$\sigma_p(r) = \{t \mid t \in r \text{ and } p(t)\}$$

where $r$ is a relation with a set of attributes in data space, $p$ is a formula in propositional calculus, called the selection predicate, consisting of terms connected by : $\wedge$ (and), $\vee$ (or), $\neg$ (not). Each term is one of: <attribute> $op$ <attribute> or <constant> where $op$ is one of: $=, \neq, >, \geq, <, \leq$. The operations among attributes or constants are operations in data space, and the associated meanings in semantics space can be uniquely determined for semantic homogeneous environments.

In a similar way, we can define our select operation with W-tags instead of tuples, and MSO-like predicates instead of propositional calculus as the following:

OP 1. **Selection of existing shadows by their meanings anchored with W-tags.**

The select operation of $\sigma_f(D^P(S))$ is defined as the following:

$$\sigma_f(D^P(S)) = \{ D^P(S) \mid W\text{-}tags^P(S) \text{ and } f(s_a, W\text{-}tag^P(s_a)),$$
$$\text{where } s_a \text{ in } D^P(S) \}$$

$D^P(S)$ is a decomposition structure of existing shadow $S$, represented by a set of semantic relations through the $W\text{-}tags^P(S)$ and $W\text{-}tag^P(s_a)$ of its sub-shadows $s_a$ in perspective *P*. The formula $f$ is a MSO-like predicate, called the selection predicate, consisting of terms connected by $\wedge$ (and), $\vee$ (or), $\neg$ (not). The terms can be a mix of predicates in terms of data space or semantic space:

- $W\text{-}tag^P(s_a) \wedge (s_a \; op \; \text{<constant } S_x\text{>})$, the variable sub-shadow $s_a$ of shadow $S$ is attached by $W\text{-}tag^P(s_a)$, and it satisfies the predicate based on data values compared with constant shadow $S_x$. $op$ is one of operators in data space, including $=, \neq, >, \geq, <, \leq$.

- $W\text{-}tag^P(s_a) \; op \; W\text{-}tag^Q(S_y), >)$, the variable sub-shadow $s_a$ is attached by $W\text{-}tag^P(s_a)$, and the criteria is specified by semantic relation between $W\text{-}tag^P(s_a)$ and $W\text{-}tag^Q(S_y)$ in semantic space. The operator $op$ can be $\subseteq$ or $\supseteq$ for weak semantic equivalence, $\supseteq \subseteq$ for strong semantic equivalence. The $W\text{-}tag^Q(S_y)$ represents a W-tag with variable shadow $S_y$, or a W-tag with constant shadow $S_y$, which can also be specified by its W-tag Instance Identifier WID in semantic space.

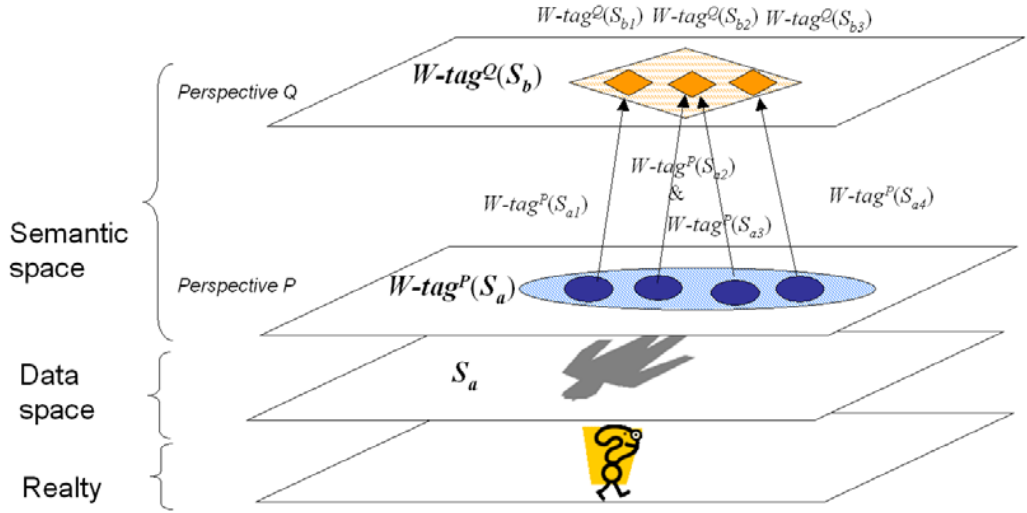

**Figure 25.** A graphical illustration for project operation. Areas in semantic space within one decomposition structure is converted to areas with different decomposition structure in a different perspective. The changes may happen not only in semantic space, but also in data space that the associated shadows may go through some kind of transformation or aggregation into different values, for example, unit change from inch to cm.

The results of a select operation include shadows with proper W-tags in the form of the specified decomposition structure. Note that the decomposition structure $D^P(s)$ is an optional parameter in select operation, unlike table (relation) is required in Relational Algebra. That is, if not specified, the select operation can be applied for any kind of decomposition structures (from any perspective) in semantic space. The select operation may also be provided with a range of different decomposition structures and perspectives. If no or more than one decomposition structure is provided, then the return results will be in the form of a combination of different decomposition structures.

The goal of Shadow Theory is to support data operations based on meanings in semantic space without constraints of logical representations from different sources. Therefore, the select operation is designed to be as generic as possible, such that input parameters can be data values for shadows in data space or relations between W-tags for their semantic relations in semantic space. If users do need to specify certain kind of logic representations for their specific applications, we will need to use the project operation to satisfy this need. We will now move to next section to introduce this operation.

### 5.3.3 Project shadows into a different decomposition structure

The project operation is to convert areas in semantic space within one decomposition structure (due to a specific perspective) to areas in a different decomposition (due to a different perspective). The changes may happen not only in semantic space, but also in data space that the associated shadows may go through some kind of transformation or aggregation into different values, for example, unit change from inch to cm.

Figure 25 provides a graphical illustration that the area of $W\text{-}tag^P(S_a)$ in perspective $P$ with decomposition structure $D^P(S_a) =$ { $W\text{-}tag^P(S_{a1})$, $W\text{-}tag^P(S_{a2})$, $W\text{-}tag^P(S_{a3})$, $W\text{-}tag^P(S_{a4})$} is projected to the areas in perspective $Q$ with decomposition structure $D^Q(S_b) = $ { $W\text{-}tag^Q(S_{b1})$, $W\text{-}tag^Q(S_{b2})$, $W\text{-}tag^Q(S_{b3})$}.

Similar to the project operation in Relational Algebra, we use the Greek letter pi($\pi$) to denote project as the following:

$$\pi_{f(D^P >> D^Q)}(D^P(s))$$

where $f(D^P >> D^Q)$ represents the description for how the elements in decomposition structure $D^P$ is projected into $D^Q$. Compared with project operation in Relational Algebra where the criteria is limited to select attributes (columns) of a relation (relation), we provide the project operation more flexibilities like to merge multiple W-tags into one or to derive extra W-tags in semantic space, as well as corresponding value transformation for shadows in data space.

For example, if we need to convert the P2:location-based ECID in Figure 23 into a different decomposition structure in perspective *P5*,

$$\pi_{P2:Customer >> P5:CustomerID}(D^{P2}(s))$$

$P2:Line1 >> P5:CustomerName$
$P2:Line2||Line3||Line4||ZIP(Line2||Line3||Line4) >> P5:Address$
$P5:CountryCode = USA$

The results is like the following:

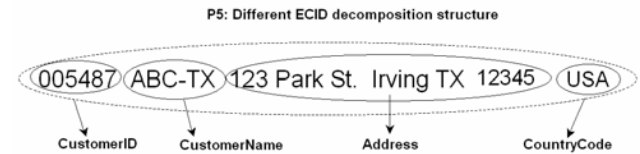

The descriptions for how the elements in decomposition structure $D^{P2}$ is projected into $D^{P5}$ includes: (1) the sub-shadow with W-tag

*P2:Customer* is copied without any transformation into $D^{P5}$ and attached with a new W-tag *P5:CustomerID*, (2) the sub-shadow with W-tag *P2:Line1* is copied without any transformation into $D^{P5}$ and attached with a new W-tag *P5:CustomerName*, (3) the sub-shadows of *P2:Line2*, *P2:Line3*, and *P2:Line4* are combined together with an extra information (provided by a function ZIP to retrieve the associated zip code) and attached with W-tag *P5:Address*, (4) a new sub-shadow is created with value "USA" and attached with W-tag *P5:CountryCode* (since all of the enterprise customer data from perspective *P2* are limited to location in USA).

Readers can observe from this simple example that, not only the logical structure can change in project operations (e.g. filtered, merged, or aggregated), but also that extra data and meanings can be added based on information outside of what the original data model represents (e.g. zip code and country code). Although this is actually a common practice in ETL process, especially happens frequently in data flow from upstream systems into downstream systems, the project operation in Relational Algebra cannot support such practical usage since it is constrained by the principle that the results must be within the information existing in the original table (relation).

The reasons we do not follow the same principle is due to the basic philosophy of Shadow Theory that meanings are mental entities in viewers' cognitive structure, hence they evolve gradually or significantly when data go through projection process from one perspective (i.e. in upstream systems) into another (i.e. in downstream systems). Since the viewers are from different background and in different business operations, the meanings are switched from one set of mental entities in one perspective to a different set of mental entities in another perspective.

The extra information added in this process may or may not be consistent with the original data model. It is simply due to the fact that correctness is evaluated not against the original perspective, but to the new converted one. In this example, the *P5:CountryCode* is totally inserted based on some kind of understanding for the system functionalities and business operation, and this is an example that common knowledge may be excluded in the a data model since the original users have the same consensus about the implied meaning. For zip code, it is associated with address following some commonly accepted standard for the specific country, so we may think it is functionally dependent of the aggregated W-tag *P5:Address* (not to the primary key *P5:CustomerID*), and the original data model simply did not include it into its scope.

For the project operation in Relational Algebra, although even it is limited to the definition of a table (relation), the meanings of the data may be lost when the remaining attributes are not proper chosen due to the removal of duplicate rows. This is especially a problem when applied under semantic heterogeneity environments: the duplicate rows are judged by data values without considering the semantic heterogeneous meanings hidden from the surface. Our approach to represent meanings explicitly by W-tags with the extra expression power of our project operations can better manage such issues, and it comes with the cost of potential issue of consistency and correctness, which is a natural result for supporting multiple version of the truth.

We can now provide the following definition for project operation as the following.

**OP 2. Project shadows by their meanings into a different decomposition structure.**

The project operation is defined as the following:

$$\pi_{f(D^P \gg D^Q)}(D^P(s)) = \{D^Q(s) \mid W\text{-}tag^P(s) \text{ and } f(D^P \gg D^Q)\ \}$$

where $D^Q(s)$ is a given decomposition structure, represented by a set of semantic relations through the W-tags of the sub-shadows in perspective *P*. The formula $f$ is a set of descriptions for which sub-W-tags in $D^P$ are selected and/or converted into sub-W-tags in $D^Q$. The results of a project operation include shadows with proper W-tags in the form of the target decomposition structure $D^Q$ in perspective *Q*.

### 5.3.4  Union, Intersection, and Difference

Next, we will discuss the operation for union, intersection, and difference. In section 2.2 we have explained that we have difficulties to apply Set Theory due to the following reasons:

(1) A set is a collection of **distinct *things*** considered as a whole. However, under semantic heterogeneous environments, different primitive units can be used from different perspectives or at different levels of abstraction; hence, the same *things* can be represented differently due to chosen ontology. Therefore, we do not have a common ground to identify distinct *things* for union, intersection, and difference operations in semantic space. On the contrary, the notion of set is the center of Relational Algebra for operation in data space, since a relation is defined as a set with distinct tuples as its elements with distinct attributes. When encountering semantic heterogeneity, the difficulties are due to that the meanings represented by these distinct *things* in data space cannot be identified as distinct *things* in semantic space.

(2) **Equivalence of two sets** requires that each one of the two sets must have exactly the same distinct *things*. When two meanings as mental entities are treated as **the same** in semantic space, there are unlimited different ways to further decompose the area they occupied in semantic space. That is, for the same *thing* in semantic space, there are different decomposition structures in data space. This includes different logical and physical data types (e.g. INTEGER, STRING), different data values representing the same *thing* from different domains, as well as different logical structure (e.g. flat, hierarchical, or graph).

(3) The root can be traced back to the required **subjective decision** in semantic space for what is defined to be the same or to be different. It is a subjective judgment made according to individual viewer's cognitive structure and chosen perspective(s). Applying the notion of set for operation in semantic space requires this subjective decision made and accepted from every perspective. Further, equivalence of sets has no flexibility to represent rich varieties of similarity that can be treated as the same under specific context or criteria of data integration. In other words, we need to support multiple versions of truth, while sets are not for such purpose.

Due to these issues, we propose to model meanings as regions and use this as the foundation to support operations for in semantic space. In section 5.2 we summarized the basic properties we can use based on point-free geometry. Since we use W-tags to anchor with unique mental entities in viewers' cognitive structures, we can think intuitively that union operations for areas in semantic space are to aggregate W-tags, and the overlapped areas represents common meanings or similar characteristics which can be reduced only if we know how to identify them.

Figure 26 illustrates two simple scenarios for union operation. The diagram (a) shows that union of two different W-tags for the same shadow from different perspectives. The associated areas can be treated as with strong semantic equivalence and are completely overlapped. However, since there are different decomposition structures for the same area, and we cannot further identify same sub-areas with common meanings acceptable from these two different perspectives, the results are the union of the two different decomposition structures (with descriptions about their strong semantic equivalence attached if users need full interpretations).

Diagram (b) shows the case of union for three different areas within the same perspective for three different shadows. Since we do not know if they overlap or not, we simply return the results as union of these three different W-tags without any reduction for overlapped areas. It is possible that the different shadows may be from the same *thing* or things with overlap (or similarity) in reality, we need to know the semantic equivalence relations between them or their sub-shadows in their decomposition structures. If further evidences are provided and if the perspectives can recognize the overlapped areas, then we can reduce duplicates like union operations in Relational Algebra.

Readers can see the difficulties for such union operations include

(1) how we can represent the results of union operations, and

(2) how the duplicates/overlapped areas can be identified and reduced from the results.

For issue (1), our solution is to return the aggregation of W-tags with their decomposition structures based on their perspectives. If users need the results to be expressed in a specific perspective with uniform decomposition structure, then mapping between perspectives must be provided such that the union operations can first translate all W-tags into the chosen perspective, and return the aggregation of W-tags and their decomposition structures (with descriptions about the semantic equivalence relations attached if users need full interpretations). When there are difficulties to identify such mappings, then the results are union of W-tags with different decomposition structures.

For issue (2), it depends on the modeling capabilities of the involved perspectives for how they can recognize the same sub-areas in semantic space, which represents common sub-meanings or similar characteristics. It would be easier to use an example to illustrate this challenge. For example, if the only concepts available in perspective *P* are {*tomato*, *broccoli*}, the only concepts in perspective *Q* are {*fruit*, *vegetable*, *flower*}, and viewers observe the same shadow from their individual perspective *P* and *Q*. Let's assume the same shadow $S_a$ is attached with W-tag *broccoli* from perspective *P*, and is attached with W-tag *flower* from perspective *Q*.

Obviously these two perspectives are not at the same level of abstraction, and we can establish mapping between them as weak semantic relations such that a *tomato* is a *fruit* or a *vegetable*, and a *broccoli* is a *vegetable* or a *flower*. The union of *broccoli* and *flower* comes with the result of *flower* since we recognize *broccoli* is a kind of *flower*, so we can reduce the duplicates with the result expressed in terms of perspective *Q*, but not expressible in terms of perspective *P*. In the same way, if we union *broccoli* and *vegetable*, the result is *vegetable*, which can be expressed in terms of perspective *Q*, but not expressible in terms of perspective *P*.

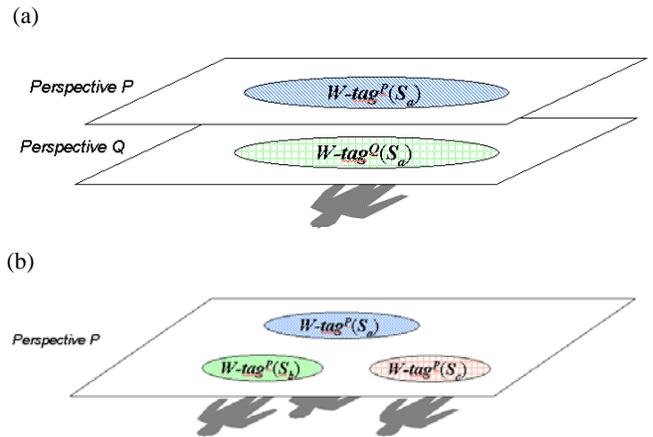

Figure 26. Union operations in semantic space can be thought as aggregating areas marked by W-tags, and the overlapped areas can be reduced only if we know how to identify them.

From this example, readers can see that the deeper challenge is about how we can model common sub-meanings or similarity among meanings attached to shadows. Specifically, how we can represent the common characteristics between mental entities from different viewers' cognitive structures. It is due to this reason that human data integrators are required in data integration projects. They rely on the interactions with those people who hold the mental entities from different perspectives to recognize what are the common sub-meanings or similar characteristics, and what are not. In our observations or practical data integration projects, the results of union operations in semantic space (and similarly for intersection and difference operations) often need something not available in existing perspectives. This is another reason why we assume we do not have full knowledge of the truth in Shadow Theory, and there are always needs for new concepts (mental entities) in order to answer such operations efficiently in semantic space.

In essences, reduction of overlapped areas for union operation in semantic space and the representation for the results are limited by the chosen perspective(s). We return the aggregation of different W-tags from their original perspectives to serve as the uniform representation, and reduction of duplicates is optional since it is possible only when users choose a specific perspective that has the capability to recognize the overlapped areas. Although the aggregated W-tags may hold different decomposition structures in data space, in semantic space the associated representation is a set of semantic relations between W-tags.

**OP 3. Selection of shadows based on union of their meanings.**

The union operation in semantic space is defined as the following:

$$D^P(s) \cup D^Q(s) = \{\ D(s)\ |\ \textit{W-tag}^P(s)\ \text{or}\ \textit{W-tag}^Q(s)\ \}$$

where $\textit{W-tag}^P(s)$ is in perspective *P* with decomposition structure $D^P(s)$, and $\textit{W-tag}^Q(s)$ is in perspective *Q* with decomposition structure $D^Q(s)$. The results $D(s)$ are aggregation of decomposition structures $D^P(s)$ or $D^Q(s)$ where shadows attached with either one of the specified W-tags.

Compared with union operation in Relational Algebra, we do not need the shadows with the same logic structure (i.e. same attributes, same arity) since the union operation is based on W-tags, which can be with different decomposition structure. In addition, unlike the requirement in Relational Algebra that attribute domains should be compatible, we do not need to request the shadows and their attached meanings to be compatible. Since any shadow with any meanings can be union together based on the users' needs, the advantage of our approach is to relax the constraints due to logical structures in data space, and allow user to perform their data operations by meanings in semantic space.

Similarly, intersection operations are to find the overlapped areas, which represent common meanings or similar characteristics; difference operations are to find not overlapped areas, which represent meanings or characteristics that are not common. We can have the following definitions:

**OP 4. <u>Selection of shadows based on difference of their meanings.</u>**

The difference operation in semantic space is defined as the following:

$$D^P(s) — D^Q(s) = \{ \ D(s) \ | \ W\text{-}tag^P(s) \text{ and not } W\text{-}tag^Q(s) \ \}$$

where $W\text{-}tag^P(s)$ is in perspective $P$ with decomposition structure $D^P(s)$, and $W\text{-}tag^Q(s)$ is in perspective $Q$ with decomposition structure $D^Q(s)$. The result $D(s)$ represents the decomposition structures for the unique sub-meanings or characteristics for $W\text{-}tag^P(s)$, and it may not be the same as $D^P(s)$ or $D^Q(s)$.

**OP 5. <u>Selection of shadows based on intersection of their meanings.</u>**

The intersection operation in semantic space is defined as the following:

$$D^P(s) \cap D^Q(s) = \{ \ D(s) \ | \ W\text{-}tag^P(s) \text{ and } W\text{-}tag^Q(s) \ \}$$

where $W\text{-}tag^P(s)$ is in perspective $P$ with decomposition structure $D^P(s)$, and $W\text{-}tag^Q(s)$ is in perspective $Q$ with decomposition structure $D^Q(s)$. The result $D(s)$ represents the decomposition structures for the common sub-meanings or characteristics, and it may not be the same as $D^P(s)$ or $D^Q(s)$.

*5.3.5 Alternative way for meaningful Cartesian-product operation in semantic space*

By Cartesian-product operations in Relational Algebra, information from any two relations can be combined together no matter whether it is meaningful or not to do so. When Cartesian-product operators and other operators (especially select, join, and project), users can filter the combinations and keep only those with the desired semantics.

The question here for supporting operations based on Shadow Theory is that, do we need a special Cartesian-product operator to achieve similar functionality in semantic space? In other words, do we need a special operator in order to combine meanings into new meaning?

We can try to directly translate this operation from data space into semantic space such that the results are the all of the possible combinations between two sets of shadows. In section 3.4.3, we have discussed that any decomposition shadows are still shadows, and the associated meanings in semantic space for such decomposition in data space can be either homogeneous or heterogeneous (as illustrated by the example in Figure 14). Hence, in the opposite direction, any shadows combined together are still shadows, and Cartesian-product operations are like performing heterogeneous aggregation (opposite to heterogeneous decomposition) since the decomposition structure of the results are always the combinations of the decomposition structures of the original two sets of shadows.

Following this direction, several questions may arise. First, whether any combinations of shadows should be allowed if they are not meaningful. We rely on semantic relations between W-tags to describe the decomposition structures. When shadows can be generated through Cartesian-product operations, how we should represent the decomposition structure by the semantic relations between W-tags?

In Shadow Theory, we request any shadow be tagged with their meaning(s). Obviously, there exist shadows that people may not understand their associated meanings, and we should allow some kind of temporary W-tags as placeholders to be attached to these shadows. Since we do not know their meaning, we cannot expect such placeholder W-tags to be with homogeneous meanings. Therefore, we simply cannot apply our rules for modeling the usage of W-tags for such placeholders, and we cannot use semantic relations to describe the decomposition structures associated with Cartesian-product operations.

Based on such consideration, we propose to not support Cartesian-product operations in semantic space, and we need to require the usage of placeholder W-tags must be homogeneous, with some basic meanings. For example, specific data records received from different sources should not be tagged with the same kind of placeholder W-tags, since the existence of such shadows are in different places.

In this approach, we also avoid the difficult issue for how to combine the decomposition structures of two shadows into a new one. In section 5.1 when we defined decomposition structures, we do not limit a decomposition structure must be only one level deep, like how a tuple is decomposed to a list of (atomic) attribute values in Relational Model. When the decomposition structures have multiple levels, it is difficult to generate the new decomposition structure for the results of Cartesian-product operations.

Then, the next question is how we can support the operations in semantic space without Cartesian-product operations. In other words, the issue is about the mechanism for how information can be combined together in semantic space.

Since any shadows combine together are again shadows (i.e. we use shadows as the uniform representation for modeling both entities and relationships in terms of Entity-Relationship Model), we actually do not need Cartesian-product operations to combine shadows. Users can always define a new shadow and it decomposition structure to include the existing shadows and their associated decomposition structures. Such decomposition can be either heterogeneous or homogeneous, compared with only heterogeneous one if we support Cartesian-product operation.

As a result, when users need to combine information, they can always create new shadows with associated decomposition structures to include what existing shadows. It is different than the traditional way performed in Relational Algebra that users can dynamically combine information by Cartesian-product operations. Since there is no schema concept in Shadow Theory, our approach is also dynamic such that users can create new shadows for their query purpose, not only for their insert-update operations.

In summary, the steps for creating new shadows by combing existing shadows through W-tags are:

(1) Define a new W-tag but without any shadow attached yet, and define the decomposition structure by semantic relations between this new W-tag and existing W-tags for the sub-shadows to serve as observable for the new shadow within this decomposition structure.

(2) Specify the criteria for which existing sub-shadows can be selected into the decomposition structure to serve as which observable role for which instance of the new shadows.

(3) Perform the selection for each of the sub-shadows to serve as observable roles, and combine them according to the decomposition structure to create new shadows. Each instance of the new shadows should be attached with the new W-tag, and this completes the operation.

We can define the operation in a declarative way as the following:

OP 6. **Create meaningful shadows by combining existing shadows into newly defined decomposition structure.**

The alternative way for meaningful Cartesian-product operation in semantic space is defined as the following:

$$X_{f(D^P D^Q)}(D^W(S)) = \{D^W(S) \mid \sigma_{f1}(D^P(s_a)) \subseteq D^W(S)$$
$$\text{and } \sigma_{f2}(D^Q(s_b)) \subseteq D^W(S)\}$$

$D^W(S)$ is the newly defined decomposition structure for the desired target shadow $S$ which do not exist yet. It is represented by a set of semantic relations between the desired new $W\text{-}tags^W(S)$ and existing $W\text{-}tag^P(s_a)$ and $W\text{-}tag^Q(s_b)$ for the shadows $s_a$ in perspective $P$ and $s_b$ in perspective $Q$ which will serve as observable for this new shadow. The formula $f1$ and $f2$ are the selection predicates for $s_a$ and $s_b$ individually, and how the different select results combined together depends on the decomposition structure.

In this way, we rely on the newly defined decomposition structure to regulate our alternative way for supporting Cartesian-product operations. Possible scenarios include: (1) results like Cartesian-product operations if the product results of selection predicates for $s_a$ and $s_b$ are meaningful, (2) filtered results if there are semantic relations in the decomposition structure that correlate between selection predicates for $s_a$ and $s_b$. Join operation between shadows is an example that can generate filtered results, and we will discuss the details in next section.

### 5.3.6  Join Shadows by Their Meanings in Semantic Space

Now, we can move on to the most different operations in semantic space: join shadows by their meanings. It is based on the notion of semantic equivalence in Shadow Theory that two areas in semantic space can be treated as the same (i.e. overlapped) subjectively, based on supporting evidences and the chosen perspective. The difference between their associated data values in data space can be ignored, so does the difference between their decomposition structures. The join operation is used to bridge shadows together to achieve specific functional objectives.

To illustrate this operation, Figure 27 shows an example that different meanings are attached to the same shadow $S_x$ of the same subject matter due to different perspectives, represented as $W\text{-}tag^P(S_x)$ and $W\text{-}tag^Q(S_x)$. The shadow $S_x$ is in the role of observable for shadow $S_a$ and $S_b$ with different decomposition structures, represented as circles and diamonds in the graph. Assuming this is a case of synchronization point, there exists a strong semantic equivalence, represented as $W\text{-}tag^P(S_x) \supseteq\subseteq W\text{-}tag^Q(S_x)$.

Similar to the natural join operation in Relational Algebra, we use the symbol $\bowtie$ to denote join operation as the following:

$$\bowtie_{f(D^P D^Q)}(D^W(S)) = \{D^W(S) \mid \sigma_{f1}(D^P(s_x)) \subseteq D^W(S)$$
$$\text{and } \sigma_{f2}(D^Q(s_x)) \subseteq D^W(S)\}$$

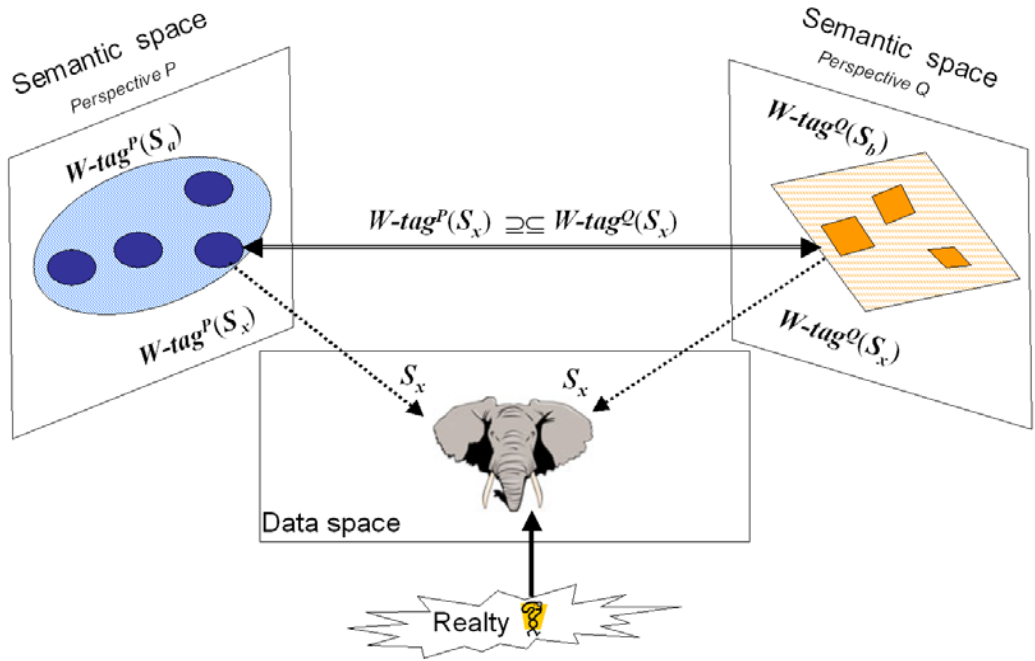

**Figure 27. Illustration for join operation:** different W-tags attached to the same shadows for the same meaning.

and $W\text{-}tag^P(S_x) \supseteq \subseteq W\text{-}tag^Q(S_x)$ )

$D^W(S)$ is the newly defined decomposition structure for the desired target shadow $S$ which do not exist yet. It is represented by a set of semantic relations between the desired new $W\text{-}tags^W(S)$ and existing $W\text{-}tag^P(s_x)$ and $W\text{-}tag^Q(s_x)$ for the shadows $s_a$ in perspective $P$ and in perspective $Q$. $s_x$ will serve as observable for this new shadow, so do sub-shadows in the results of select predicates $f1$ and $f2$ which select shadows from perspective $P$ and $Q$ independently but joined with the strong semantic equivalence $W\text{-}tag^P(S_x) \supseteq \subseteq W\text{-}tag^Q(S_x)$.

For example, in Figure 23, for W-tag *P3:EC_Num* of *P3:Contract-based ECID,* and *P3:Parent_Num* of *P3:Contract_map* are attached to the same shadow value 87936-965042. In semantic space, this value uniquely represents an enterprise customer from contract-based ECID perspective, and it meaning as a contract-based ECID can be decomposed into a set of attributes (*P3:Contract-based ECID*) or into a unit in the organizational structure (*P3:Contract_map*). With the assumption of semantic homogeneous data from perspective *P3*, this example can be simplified properly into primary key and foreign key relation. The join operation can be established due to the meaning for the same shadow value of *P3:EC_Num* and *P3:Parent_Num* to associate *P3:Contract-based ECID* and *P3:Contract_map*. The result is like the following:

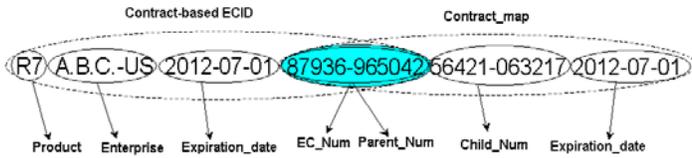

Note that in section 3.4.3, we have requested that there exists only one decomposition structure for a W-tag in a perspective to avoid another identifier to guarantee the uniqueness of meaning for W-tag. In this example, we have two different W-tags for the same shadow due to the original data model from *P3*, which can be an alternative way to reach the same objective. Hence we do not need to create a different perspective as required by W-tag Rule # 8.

Figure 28 illustrates another example where different W-tags are attached to different shadows for the same subject matter. $S_x$ and $S_y$ represent the different shadows, and $W\text{-}tag^P(S_x)$ and $W\text{-}tag^Q(S_y)$ represent their W-tags due to different perspectives. The shadow $S_x$ is in the role of observable for shadow $S_a$, and $S_y$ is in the role of observable for $S_b$. Let's assume this exist evidences to support the E-tag for their strong semantic equivalence, we can have $W\text{-}tag^P(S_x) \supseteq \subseteq W\text{-}tag^Q(S_y)$. Join operations can be used to combine shadow $S_a$ with $S_b$ to form a new shadow, or to associate sub-shadows in $S_a$ with sub-shadows in $S_b$. We can describe the join as the following:

$$\bowtie_{f(D^P D^Q)}(D^W(S)) = \{D^W(S) \mid \sigma_{f1}(D^P(s_x)) \subseteq D^W(S)$$
$$\text{and } \sigma_{f2}(D^Q(s_y)) \subseteq D^W(S)$$
$$\text{and } W\text{-}tag^P(S_x) \supseteq \subseteq W\text{-}tag^Q(S_y) \text{ )}$$

$D^W(S)$ is the newly defined decomposition structure for the desired target shadow $S$ which do not exist yet. It is represented by a set of semantic relations between the desired new $W\text{-}tags^W(S)$ and existing $W\text{-}tag^P(s_x)$ and $W\text{-}tag^Q(s_y)$ for the shadows $s_a$ in perspective $P$ and shadows $s_b$ in perspective $Q$. $s_x$ and $s_y$ will serve as the same observable for this new shadow, so do the shadows in the results of select predicates $f1$ and $f2$ from perspectives $P$ and $Q$, joined by the criteria of the strong semantic

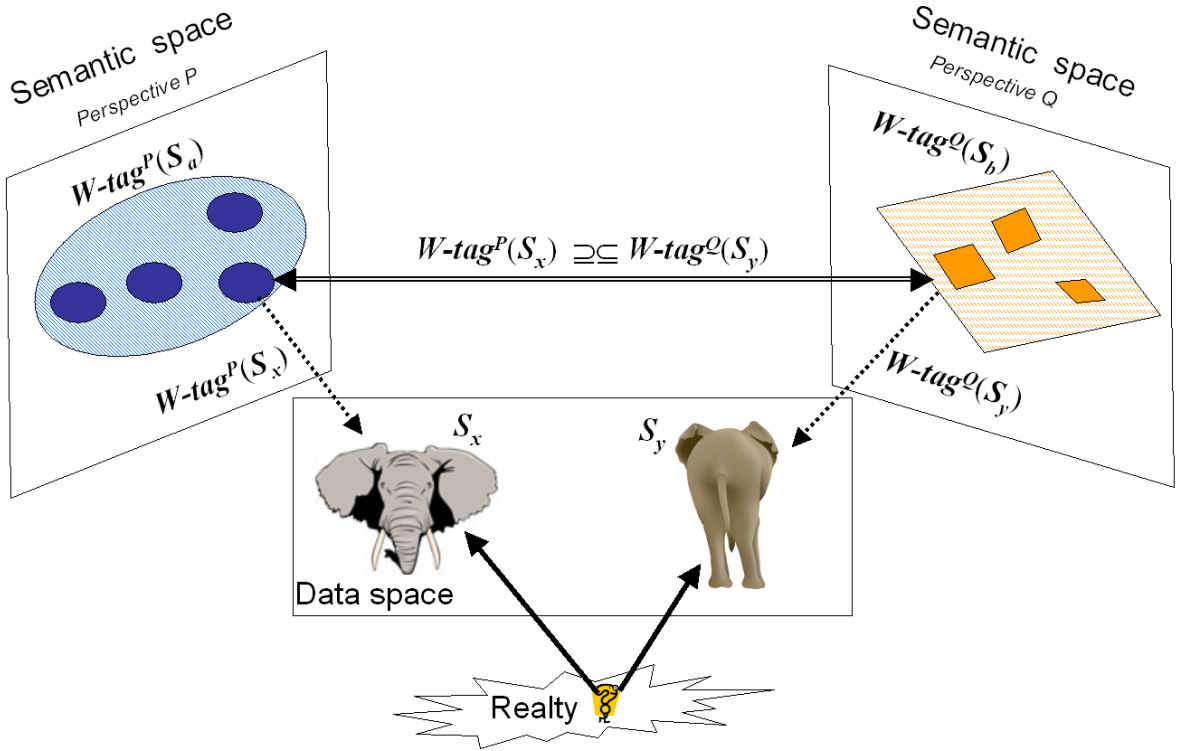

**Figure 28. Illustration for join operation:** different W-tags attached to different shadows for the same subject matter.

equivalence $W\text{-}tag^P(S_x) \supseteq \subseteq W\text{-}tag^Q(S_y)$ .

For example, in Figure 23, let's assume we have evidences to support strong semantic equivalence between *P3:EC_Num* of *P3:Contract-based ECID,* and *P2:Customer* of *P2:location-based ECID*, which are W-tags attached to different shadow values 87936-965042 and 005487 from different sources. In semantic space, these two different values can uniquely represent the same enterprise customer from their specific perspective (contract-based ECID versus location-based ECID). The result of the join operation is like the following:

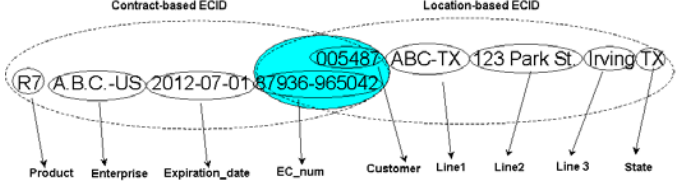

This is an example of mapping which is usually implemented as a tuple/record with two of the attributes representing the unique keys from original tuples/records. Since Relational Model relies on the data values in the same domain to determine equivalence, the two attributes representing different entities in the mapping table cannot be treated as the same in any of the operations of Relational Algebra. Hence, this example of strong semantic equivalence (between shadow values 87936-965042 and 005487) cannot be operated within the join operator; it must rely on interpretation through the relations of the two attributes.

Our objective is to include such semantic equivalence in the algebra operations. That is, due to certain common characteristics in semantic space, a mental entity represented by a W-tag can be joined with another mental entity represented by a different W-tag. Although the common characteristics may be modeled as sub-shadows with different values in data space due to different perspective, we can treat them as the same in the algebra operations. The corresponding operation in data space is like two shadows joining together by some common sub-shadow(s). The commonality of sub-shadow(s) is defined not due to their data values, but due to their meanings (be treated as equivalent) in semantic space, representing the same subject mater under the conditions of established E-tags with supporting evidence(s).

We can now define join operation based on strong semantic equivalence.

**OP 7. Bi-directional join operation in semantic space.**

Bi-directional join operation is based on the strong semantic equivalence, defined as:

$$\bowtie_{f(D^P D^Q)} (D^W(S)) = \{D^W(S) \mid \sigma_{f1}(D^P(s_x)) \subseteq D^W(S)$$
$$\text{and } \sigma_{f2}(D^Q(s_y)) \subseteq D^W(S)$$
$$\text{and } W\text{-}tag^P(S_x) \supseteq \subseteq W\text{-}tag^Q(S_y) \text{ )}$$

$D^W(S)$ is the newly defined decomposition structure for the desired target shadow $S$ which do not exist yet. It is represented by a set of semantic relations between the desired new $W\text{-}tags^W(S)$ and existing $W\text{-}tag^P(s_x)$ and $W\text{-}tag^Q(s_y)$ for the shadows $s_a$ in perspective $P$ and shadows $s_b$ in perspective $Q$. $s_x$ and $s_y$ will serve as the same observable for this new shadow, so do the shadows in the results of select predicates *f1* and *f2* from perspectives $P$ and $Q$, joined by the criteria of the strong semantic equivalence $W\text{-}tag^P(S_x) \supseteq \subseteq W\text{-}tag^Q(S_y)$ .

For weak semantic equivalence, there are some differences due to its limited uni-direction for applying semantic equivalence. The definition is as the following:

**OP 8. Uni-directional join operation in semantic space**

Uni-directional join operation is based on the weak semantic equivalence, defined as:

$$\bowtie_{f(D^P D^Q)} (D^W(S)) = \{D^W(S) \mid \sigma_{f1}(D^P(s_x)) \subseteq D^W(S)$$
$$\text{and } \sigma_{f2}(D^Q(s_y)) \subseteq D^W(S)$$
$$\text{and } W\text{-}tag^P(S_x) \subseteq W\text{-}tag^Q(S_y) \text{ )}$$

$D^W(S)$ is the newly defined decomposition structure for the desired target shadow $S$ which do not exist yet. It is represented by a set of semantic relations between the desired new $W\text{-}tags^W(S)$ and existing $W\text{-}tag^P(s_x)$ and $W\text{-}tag^Q(s_y)$ for the shadows $s_a$ in perspective $P$ and shadows $s_b$ in perspective $Q$. $s_x$ and $s_y$ will serve as the same observable for this new shadow, so do the shadows in the results of select predicates *f1* and *f2* from perspectives $P$ and $Q$, joined by the criteria of the weak semantic equivalence $W\text{-}tag^P(S_x) \subseteq W\text{-}tag^Q(S_y)$ .

The only direction of join is limited by the direction of weak semantic equivalence. For example, since an enterprise customer may have multiple contracts for different products, we can establish a weak semantic equivalence between legal entity-based ECID and contract-based ECID. That is, a contract-based ECID is the same as a legal entity-based in terms of representing the enterprise customer, but not in the reverse direction since a legal entity-based ECID may represents the same subject matter as the combination of several contract-based ECID. For the example in Figure 23, let's assume we have evidences to support weak semantic equivalence between *P3:EC_Num* of *P3:Contract-based ECID,* and *P1:Customer_ID* of *P1:Legal entity-based ECID*, which are W-tags attached to different shadow values 87936-965042 and 763810 from different sources. In semantic space, these two different values can uniquely represent the same enterprise customer from their specific perspective (contract-based ECID versus legal entity-based ECID). The result of the join operation is like the following:

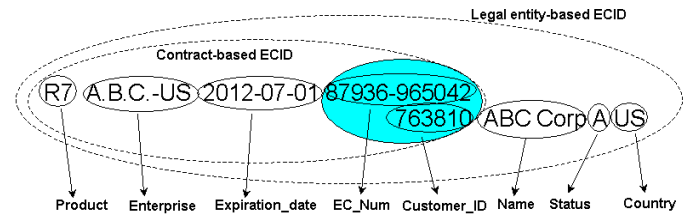

## 5.4 Simulating relational schema by templates in semantic space

For practical application for data integration in information ecosystems (which are dominated by Relational Databases), we need to be backward compatible with Relational Algebra such that we can simulate relational schema if users need to mimic the data model design in the original sources for business reasons.

Since our objective is for data model design in semantic heterogeneous environments, to support compatibility with Relational Model can be treated as simplification for semantic homogeneous environments. W-tags can be simplified to table names or column names, assuming there exist only homogeneous data values with unique meanings. P-tags are simplified to data formats and integrity constraints. E-tags are simplified to equivalence by data values in the same domains, since the meaning of data values is homogeneous such that equivalence of data values implies equivalence of the meanings they represent.

The extra advantage is that we can manage data by their patterns in semantic space if we can simulate relational schema. In section 3.4.2, when we introduced W-tag rules, we mentioned the idea to simulate relational schema by templates of W-tags in semantic space. In section 4.3.1, we use an example to show that we can use semantic equivalence as a different way to model the relationship of primary key and foreign key when we discussed about decomposition structures represented by semantic relations between W-tags. These semantic relations can function like a template, not only we can use them to understand patterns of shadows in data space, we can also manage data in the reverse direction by filling shadows into templates.

That is, we can pre-define W-tags and their semantic relations without attaching to specific shadows. When some shadows are attached with involved W-tags, they actually fit into the template in semantic space and we can perform meaningful inference even without all shadows being attached with every involved W-tag. This can be viewed as an enhancement for how Relational Model can model about unknown or missing information (which now relies on the notion of null value).

A **template** $T^P(W\text{-}tag^P)$ is a decomposition structure represented by a set of semantic relations between $W\text{-}tag^P$ and its sub-components (W-tags) with or without being attached to shadows. Just like that there is only one decomposition structure for a shadow in a specific perspective, we expect only one template for the involved W-tags within a single perspective. That is, all of the shadows attached to the same kind of W-tags have the same decomposition structure even they may not have all of the sub-shadows identified.

Note that the difference between decomposition structures (represented by a set of semantic relations) and templates is that, the former is for individual shadows and different shadows attached with the same kind of W-tags are not required to be decomposed in the same way; the later is for the consistent pattern for the meanings such that the involved W-tags should behave consistently in semantic space.

To simulate relational schema, we can use a template combined with some logical properties in data space. A **simulated schema** $M^P(W\text{-}tag^P)$ is a template satisfying the following conditions:

(1) A simulate schema is a one level structure, containing a set of sub-components $W\text{-}tag^P_x$ like attribute names in Relational Model.

(2) Shadows that can be attached with $W\text{-}tag^P_x$ are assumed to be (locally) atomic within this perspective $P$.

(3) For each $W\text{-}tag^P_x$, there is an associated $P\text{-}tag_x$ specifying the required properties for shadows which can be attached with, functioning like a domain in Relational Model.

(4) A shadow $S_a$ that satisfy these criteria can be attached with $W\text{-}tag^P$ like a row in a table, which includes a set of sub-shadows $(s_1, s_2, …, s_n)$ like attribute values in Relational Model.

(5) If there is no shadow $s_x$ attached with $W\text{-}tag^P_x$, then it is the situation where null value is used in Relational Model.

In a simple way, we can think $M^P(W\text{-}tag^P)$ is like the relational schema and $W\text{-}tag^P$ is the table name. Each sub-component $W\text{-}tag^P_x$ in the template is like a column name. When shadows are attached with $W\text{-}tag^P_x$, they are like attribute values in a tuple, and $P\text{-}tag_x$ are their associated domains.

The basic operators we have defined in previous section can be applied for simulated schema, or they can be simplified into the operators for data values in Relational Algebra. The relations of primary key and foreign key are performed by strong semantic equivalence, and the referential integrity is performed by extra constraints for the existence of W-tag instance (indicated by WID assignment), not by shadow values.

In this way, we can overcome the issues of ghost problem in information ecosystems (as discussed in section 2.1) that when upstream systems delete the unique keys, downstream systems can move the shadow values into history archive of the W-tag instances, but keep the WID and their connections with other W-tags. For example, when upstream system physically delete a unique key for *P1:legal entity-based ECID* for some reason, downstream systems which still hold the associated ordering records or billing invoices just need to move the key values into archive of the W-tag instance, and keep all of the W-tags still connected together.

This also resolves the unique key reuse problem that when the same unique key is used for representing something else in the upstream systems, the downstream systems still hold the same WID for the W-tag instance, but with the new shadow values from upstream systems. We also resolve the rollback problems, as all we need to do is just move shadow values from archive to where they were before.

One may wonder when we should really delete the W-tags and their relations in the integrated database of downstream systems. It depends on the business requirements or legal constraints, just as it is now in practical applications. Compared with current practice to design extra historical tables without referential integrity, our approach provides a simpler solution at data model level to support the gap between systems requirements (e.g. upstream systems hold historical data for 2 years, while downstream systems may be required to hold for 7 years).

User may notice that in our definition, we enforce simulated schema to hold only semantic homogeneous data due to the definition of W-tags: a column ($W\text{-}tag^P_x$) can hold only one kind of data with homogeneous meaning, and a table ($W\text{-}tag^P$) can hold only one kind of tuples with the same meanings. If users need to continue existing design (for compatibility with existing applications) that a single table is overloaded with semantic heterogeneous data, we can simulate such usage by union of multiple simulated schemas.

The assumption of atomic attribute values can also be relaxed like what current Relational Database can support. If a shadow that serves as an attribute values in a simulated schema can be further decomposed, we just need to establish an E-tag with strong semantic equivalence between the attribute in simulated schema and its decomposition structure.

Again, the goal of simulated schema is to be backward compatible with schema in Relational Model such that users can have the flexibility to continue their existing design features. We do not have the notion of schema for rigid logical structure in data space for Shadow Theory. The notion of template is for having consistent decomposition structure for the same kind of shadow within a perspective; however, templates are not required, they are only for satisfying users needs when they do not want the W-tag instances of the same kind W-tag be decomposed in different ways.

## 6. Discussion and future work

Next, we will briefly discuss related issues and make comparisons with existing data models and data integration approaches.

## 6.1 The challenges to support multiple versions of the truth

In this paper, we have used the example of enterprise customer data integration in information ecosystems to show readers the problem of semantic heterogeneity, how it makes Relational Algebra less efficient. Since Relational Calculus and Relational Algebra have the same expression power, readers may wonder how semantic heterogeneity impacts Relational Calculus from logic perspective. In essence, the challenge is about the requirement to support multiple versions of the truth, when there are inconsistencies or even conflicts between meanings as mental entities in different viewers' cognitive structures.

We will start the discussion with reviewing C. J. Date's claim that database must be relational (which is valid if under the assumption of semantic homogeneity) [6]. Then we will introduce the factors of semantic heterogeneity and characteristics of information ecosystems to show where issues may arise. Then we will use one of the diagrams in S. Russell and P. Norvig's AI textbook [81] to highlight the fundamental logical difference for how we need to make judgment about the correctness for data integration in information ecosystems.

### 6.1.1 Date's claim that database must be relational

In Date's book, *SQL and Relational Theory*, he has a interesting claim that "*database must be relational*" (see p.287 in [6]). The argument is summarized as the following:

(1) A database isn't just a collection of data, but rather a collection of true propositions, e.g. Joe's salary is 50k.

(2) Propositions can be encoded as ordered pairs, e.g. (Joe, 50k).

(3) Specifically, we want to record all propositions that happen to be true instantiations of certain predicates, e.g. the predicate x's salary is y.

(4) We can use a set of ordered pairs to record the extension of the predicate x's salary is y.

(5) In mathematical sense, a set of ordered pairs is a binary relation.

(6) A binary relation can be depicted as a table.

(7) We need to deal with n-ary relations, not just binary ones, and n-tuples, not just ordered pairs.

(8) Hence, we replace the ordering concept by the concept of attributes identified by names.

(9) Since a relation is both a logical construct (the extension of a predicate) and a mathematical one (a special kind of set), Codd was able to define a relational calculus and a relational algebra to support data operations.

### 6.1.2 Impacts of semantic heterogeneity

Now let's consider the impacts of semantic heterogeneity and characteristics of information ecosystem to Date's claim. The following potential issues may happen:

### *I. Different predicates for the same subject matter due to different perspectives.*

The first issue may happen between step (1)~(3). If the concept of the subject matter is accepted by every one, for example, Joe is a person and we use US dollars as the units of salary, then every one will reach the same predicate in his example. That is, in a semantic homogeneous environment, every one think in the same perspective (and choose the same ontology implicitly), then, data in information ecosystem can be easily integrated together.

Unfortunately, even we assume every one accepts the same primitive ontological units for Joe, we may have different predicates to describe what Joe is, for example, a person, a people, an employee, a salve, a workforce, a contractor, a prisoner, and so on. What is even worse is that, the different ontology can make very different meanings about the same subject matter. For example, the predicates for Joe's salary can have many variations with meanings that is not even explicitly described in the data model, for example, paid hourly, daily, weekly, bi-week, monthly, quarterly, yearly, and son on.

Therefore, different versions of the predicates for the same subject matter demand a mechanism to identify what conceptual notions should be treated as the same (in a subject way). Shadow Theory answers the needs to allow users establishing E-tags to recognize different predicates (i.e. W-tags) for the same subject matter due to different perspectives. Relational Model simply misses the mechanism to model the same predicates since it is designed to manage dynamic data, not dynamic meanings.

Since information ecosystems are naturally in the reality of semantic heterogeneity with multiple version of the truth, if data integrators use Relational Model, then equivalence of different predicates is forced to be modeled as different attributes in a relational schema, not be treated as the same naturally by the operators of Relational Algebra. This is why we believe that the current difficulties of data integration are the natural results due to the weakness of existing data models.

### *II. Different meanings for the same predicate due to overloaded with semantic heterogeneity.*

The second possible issue is due to the hidden assumption in step (3) that the name for the terms in predicates should contain clear meanings, not causing confusions, such that we can understand precisely the meanings of data values. Unfortunately, if overloaded with semantic heterogeneity, this assumption becomes invalid. Different perspectives or different ontologies make the issue even more complicated as the same predicate can actually have very different meanings in different systems. Further, we may not even determine the precise meaning of a single attribute value by only its table name and column name. For example, the meaning of the status column that we have illustrated in the ECID example of section 2.2 depends on the type of the ECID; status 'A' for legal entity-based ECID means active legal status, status 'A' for contract-based ECID indicates an active contract, and status 'A' for location-based ECID represents that services at that location are active.

For Date's example, if we use the same predicate for (Joe, 50k) to describe dog Cooky and its salary for serving in army, the meanings of the predicates are overloaded. We may say that we have abstract the meaning of the predicate x's salary is y, salary(Joe, 50k) is for a human and his salary, while salary(Cooky,

100) is for a dog and its rewards. There is commonality between the different semantics, and there are differences that prevent us to apply Relational Algebra efficiently.

We believe that the root of the issue is due to the fact that semantics is not explicitly represented in Relational Model, it is only partially and implicitly indicated by table names, column names, domain names, and data values in the specific domains. Resolving semantic heterogeneity demands explicit meanings, but not necessary represented in a formal and complete way. Shadow Theory address this need by recognizing that meanings are really just mental entities existing in viewers' cognitive structure, and we can use W-tags to uniquely anchor these mental to make them explicit. In this way, we can distinguish different meanings explicitly, and recognize what meanings are treated as the same based on data integrators' subjective decisions. In this way, we can avoid the burden of formal and complete representation: when extra details are needed, we can simply include them, since it is not possible and not necessary to have a complete and formal representation that is designed to answer any kind of questions.

### III. Criteria for evaluation

The most important and fundamental issue is in step (1) about "true" propositions. The tradition approach for such evaluation is based on the truth value; propositions (or predicates if with variables) are expressed according to the syntax of the chosen representation languages, and the semantics of the languages defines the truth of each proposition with respective to each possible world (i.e. the term *model* used in Artificial Intelligence). The required assumption is that all of these true propositions or predicates should consistent within the same model.

For data integration in information ecosystems, if we assume each individual data source is a model with consistent propositions/predicates, then the challenge is really about how we can integrate these models established by different perspectives. For the ECID example, legal entity-based ECID are true propositions in its original data model, and contract-based ECID are true propositions in a different one. When we need to integrate them together in a downstream system, we actually bring them out of their birth places and force them to co-exist in a new integrated model. The semantic heterogeneity simply forces inconsistencies or even conflicts to co-exist in this integrated model, and the approach of truth values cannot help us to evaluate the correctness (since these propositions are not designed for the same possible world).

Based on Figure 7.6 in S. Russell and P. Norvig's agent-oriented AI textbook [81], we can use Figure 29 and Figure 30 to illustrate the difference. Diagram (a) in Figure 29 shows that sentences (propositions/predicates) are expressed according to the syntax of the chosen representation languages, and the semantics of the languages defines the truth of each sentence with respective to each model. Diagram (b) in Figure 29 shows our proposal that all kinds of data we can observe about the subject matter are just shadows, including the representation created by different people based on their specific perspectives. The meanings of these shadows are based on mental entities existing in the viewers' cognitive structures, not necessarily due to something in reality. Therefore, the chosen perspective or ontology limits what the meanings can be, and we propose to use W-tags to uniquely anchor these mental entities with perspectives as their explicit boundaries.

The fundamental challenge for supporting multiple version of the truth is how we can perform reasoning within and across the boundaries of perspectives or ontology. Figure 30(a) shows that traditional logical reasoning should ensure new sentences generated based on existing ones should represents the aspects of real world follow the aspects of real world represented by the existing ones in a consistent way. This figure is from Figure 7.6 in

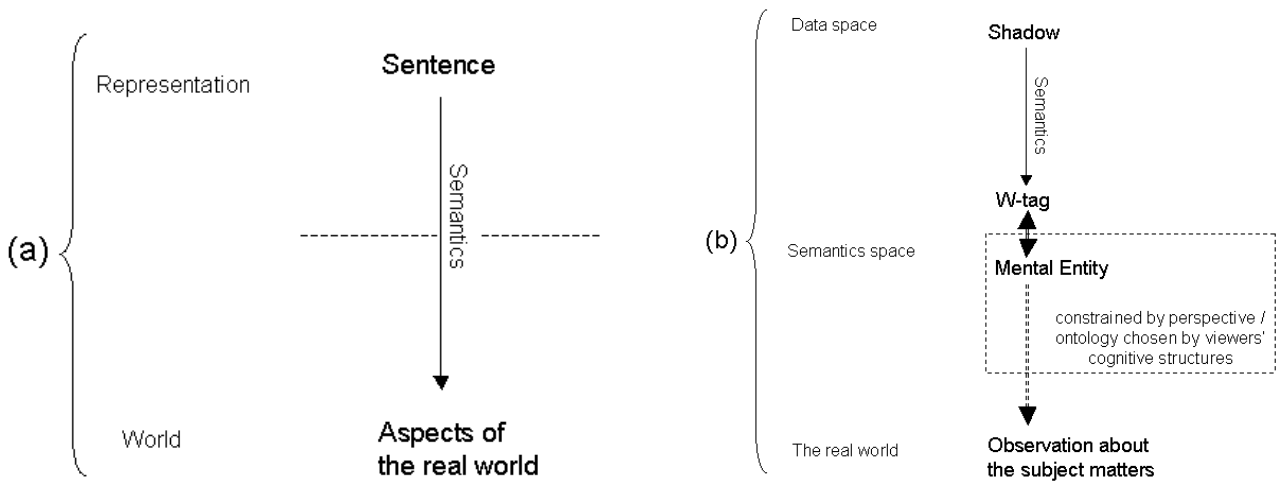

**Figure 29.** Comparison of different approaches to evaluate correctness. Figure (a) shows the traditional logic representation that sentences (propositions/predicates) are expressed according to the syntax of the chosen representation languages, and the semantics of the languages defines the truth of each sentence with respect to each model. Figure (b) shows our proposal that all kinds of data we can observe about subject matter are just shadows, including representation created by other people based on their specific perspectives. The meanings of these shadows are based on mental entities existing in the viewers' cognitive structures, not necessarily due to something in reality. Therefore, the chosen perspective or ontology limits what the meanings can be, and we propose to use W-tags to uniquely anchor these mental entities with their perspectives as the explicit boundaries.

S. Russell and P. Norvig's agent-oriented AI textbook [81] (p.200), we include here for readers convenience to make comparison with our proposal in Figure 30(b), which illustrates that not only semantics is based on the existence of mental entities constrained by viewers' cognitive structures, there are also limitations for consistency for how new sentences generated based on the old ones. That is, consistency of reasoning is only valid within a single perspective or ontology, or within different perspectives/ontologies that are consistent.

Based on Shadow Theory, data integration should not only bring different shadows and their associated meanings into the integrated model, but also bring their mental entities with associated boundaries of their cognitive structures into consideration. That is why we propose to use E-tags to represent semantic equivalence between W-tags with supporting evidences, and these W-tags are bridges (with directions) across the boundaries of different perspectives. And the most important part is that the evaluation is performed against the mental entities in viewers' cognitive structure to support multiple versions of the truth.

There are still issues need further investigation in the future. Unavoidably, we have the issue of grounding that the connection between reasoning process and the real environments of the information ecosystems. For the agent example in Russell and Norvig's book, such connection can be created through agent's sensors (see p.204 in [81]). Similarly, we can think the data collected and modeled by the original data models are like the data collected through their "sensors". However, the challenge is that the meanings of data evolve when data flow from upstream systems into downstream systems (as we described in section 2.1). We have difficulties to know whether the second hand (or third hand, forth hand…) data still truly matches with the original data model, and distinguish the added meanings from the original meanings. This is the place we resort to the philosophical foundation of Shadow Theory that anything we can store in database are just shadows, which can be direct observation of the subject matter in reality from chosen perspectives, or indirect information received through process with unknown factors (including physical data transportation and semantic data interpretations). Further research is need to understand related impacts.

Next, the issue of soundness: under what conditions the reasoning is sound across boundaries? i.e. the inference algorithm will derive only entailed sentences (e.g. new W-tags or E-tags) based on existing E-tags. Further, the issue of of completeness: is it possible to design inference algorithm to derive any sentence that is entailed by existing ones? i.e. an inference algorithm to derive any sentence (e.g. new W-tags or E-tags) that is entailed by existing W-tags and E-tags. Figure 31 illustrates a simple scenario that if we can establish semantic equivalence between two W-tags from different perspectives, what are the semantic relations between other W-tags that are derived based on these two W-tags independently in their individual perspectives?

These are critical challenges for data integration to overcome semantic heterogeneity in information ecosystems, since the basic mapping process involves make inferences across boundaries of perspectives of logical consistency (i.e. the consistency due to functional dependency). In our observation for industrial practices, these are traps that many data integrators fall into.

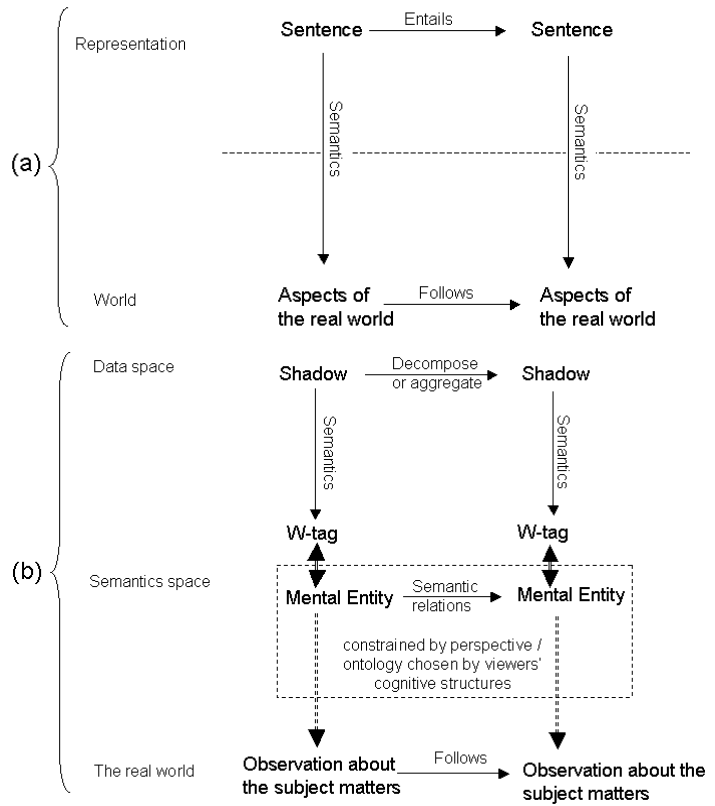

**Figure 30.** Figure (a) shows that traditional logical reasoning should ensure new sentences generated based on existing ones should represents the aspects of real world follow the aspects of real world represented by the existing ones in a consistent way. Figure (b) illustrates that not only semantics is based on the existence of mental entities constrained by viewers' cognitive structures, there are also limitations for consistency for how new sentences can be generated based on old ones.

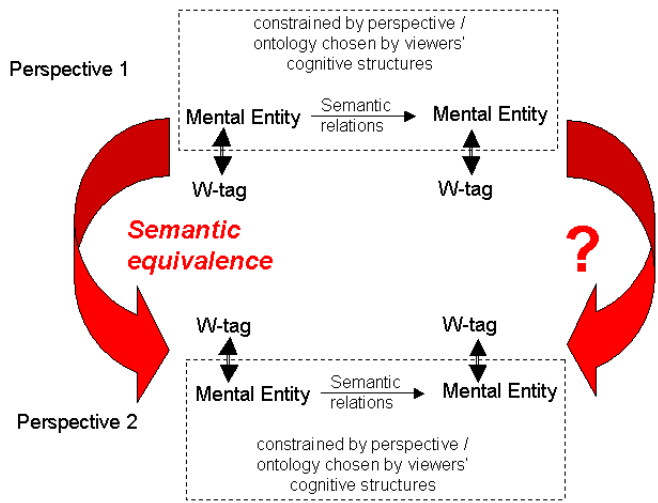

**Figure 31** If we can establish semantic equivalence between two W-tags from different perspectives, what are the semantic relations between other W-tags that are derived based on these two W-tags independently in their individual perspectives?

For example, if *A* maps to *B*, *B* maps to *C*, can we expect that *A* maps to *C*? In the ECID example, if we find a mapping from *legal entity-based ECID A* to *location-based ECID B*, and we also have a mapping from *location-based ECID B* to *contract-based ECID C*, can we infer the mapping from *legal entity-based ECID A* to *contract-based ECID C*?

If *A*, *B*, *C* are semantic homogeneous, the answer is obvious correct. But they are not in this example, for the legal entity *A* modeled by service provider 1 can be an international enterprise, while the identified customer of location *B* modeled by service provider 2 can be just the US domestic sub-company that use the specific services. Further, the mapping from *B* to customer *C*, which are identified by the contract for the products of provider 3 available only in Europe areas. In addition to the different scopes and focuses (e.g. service areas and available products) for each different ECID model, the individual mappings of A-B and B-C are for designed for different purposes.

Since the available operators of Relational Algebra cannot support equivalence of heterogeneous data of *A, B,* and *C* to be treated as the same (by their meanings), they can only be modeled as normal attributes in tables, i.e. modeling *A* and *B* as attributes in a mapping table *X* and modeling *B* and *C* as attributes in another table *Y*. In this way, there are no warning or any protection mechanism for users to join them together to get relations between *A* and *C*. By doing so, they simply just pass the boundaries of logical consistency based on the design principle of functional dependency.

This is the reason why in section 4.3.3 we mentioned that different semantic relations associated to W-tags from different perspective may not be transitive, and in section 5.2, Property 6, we defined extra criteria for helping to determine when we can do so. Our approach based on Shadow Theory needs to carefully identify the boundaries of logical consistency and represent them as different perspectives explicitly. Inconsistencies or conflicts between different perspectives may not be resolved in the real world, and we bring them into the data model world in order to help users to recognize the issues of data integration. By providing such modeling, our goal is for users to identify their ways for how they can map shadows due to the desired characteristics of their meanings such that they can reach their functional objectives.

## 6.2    Comparison with Relational Model

Next, we will make further comparison with Relational Model. In Codd's own description (p.400 in [22]), the Relational Model consists of three parts:

(1) A collection of time-varying tabular relations with related properties, especially keys and domains.

(2) The insert-update-delete rules.

(3) The Relational Algebra.

In addition, there are some closely associated decomposition concepts that are semantic in nature. Specifically, Codd mentioned the concepts of nonloss (natural) joins and functional dependencies, multivalued dependencies, and normal forms.

In contrast, we propose Shadow Theory to serve for the philosophical foundation to provide guideline for decomposition concepts, specifically for the semantic heterogeneous environments in information ecosystems under the following situations:

(i) Semantic heterogeneity creates serious issues for unique keys to functional properly, for example, uniqueness due to logical representations instead due to their meanings, multiple versions of the truth due to different perspectives/ontologies; or the nature of information ecosystems does not fit with unique key concept, for example, ghost problem, unique keys re-use problem, and rollback problems.

(ii) Functional dependencies are no longer valid when there exist multiple versions of the truth, i.e. when data integrators need to subjectively decide mapping between semantic heterogeneous data across the boundaries of different perspectives or ontologies.

(iii) Join is made by semantic equivalence (of mental entities) instead of by the same data values (of shadows). Since semantic equivalence is based on subjective design decisions, we sure may lose some meanings that are unique in specific perspectives.

We proposed six principles to guide the overall data integration process. The goal is to help users to recognize & manage inconsistencies or conflicts that exist in the real world by bring them into the model world, such that users can work on the real solutions if possible. Therefore, the functionalities of a data model is extended to include helping users to explore unknown semantic space incrementally, not just limited to the assumption that data modelers should design based on full knowledge of the data. This also correlates to the current business needs of agile software development process which forces data model designed be made quickly (and also make adjustment quickly) without fully understand everything about their dependencies.

Corresponding to the three parts of the Relational Model, data model designed based on Shadow Theory have the following parts:

(1) A collection of W-tags for anchoring mental entities in semantic space, a collections of E-tags to recognize which mental entities are treated as the same subjectively, and a collection of optional P-tags when users need to specify the logical formats of shadows. WID is proposed to replace the role of unique keys such that the specific instance of the mental entity can continue to exist even if the unique key disappeared or changed in upstream systems.

(2) A set of rules for insert-update-delete shadows in data space, and assign-change assignment for W-tags, E-tags, and P-tags, briefly summarized in the beginning of section 5.3.

(3) The basic operators we need to perform operations in semantic space, which can be simplified and backward compatible to those operators in Relational Algebra. These operators are designed based on point-free geometry to treat any meaning anchored by a W-tag as a region, which can be decomposed in different ways concurrently without limitation like the notion of a atomic point. Semantic equivalence is modeled as overlapped areas, which can range from complete overlap to various kinds of partial overlap (i.e. the ratio of overlapped area versus non-overlapped area satisfies certain criteria such that data integrators treat them as *approximately* the same).

Data independence is an important concept for Relational Model, and meaning independence we proposed is built on top of the data independence. In next section, we will discuss how these two can work together, and make comparisons with the logical data independence proposed for the Universal Relation model.

## 6.3 Meaning independence versus logical data independence

When we proposed the notion of meaning independence in section 3.5.1, we discussed the differences compared with data independence that meaning independence is to allow data integration be performed without the constraints of the underlying logical structures, just like data independences allows data to be modeled & managed without knowledge of their physical structures [65]. The notion is established based on the philosophical foundation of Shadow Theory that shadows themselves do not hold meanings, and meanings of shadows only exist in viewers' cognitive structures. Therefore, we can have separated semantic space for all kinds of meanings and data space for all kinds of data with their logical structures.

Note that meaning independence is one level higher in abstraction than data independence, and it relies on data independence to hide the details of physical storage. We use W-tags and E-tags to model & manage meanings in semantic space, and eventually we have to go to their associated shadows in data space for databases, not a single database system, but heterogeneous databases in the information ecosystems.

There are some major differences compared with logical data independence proposed in the Universal Relation model proposed in 80s [82, 83]. The basic idea is that users do not need to specify the logical navigation path to compute a query. However, we need users to explicitly identify their desired meanings in semantic space for specific perspectives, i.e. the specific W-tag instances that anchor uniquely with the mental entities in the particular viewers' cognitive structures. Like Relational Model, schema is the kernel representation mechanism for Universal relational model, while we try to avoid the constraints of schema in data space in order to allow user to subjectively decide what meanings can be treated as the same in semantic space.

In general, Vardi's approach is not for the purpose of modeling semantic heterogeneity across different perspectives, but rather for semantic homogeneous environments (like Relational Model is intended for). If applied for information ecosystems with semantic heterogeneous data, it will suffer the same problems, for example, the issues for unique keys, different meanings not explicit represented, and structural heterogeneity.

There were some attempts to apply the notion of logical data independence to help data integration (schema integration). Universal relation can be viewed as a special case for the global schema [84]. Following LAV approach (Local-as-view), that each relation in a local database is defined by a view in the global schema [1], the universal relation can be further classified pure universal relation assumption (that global instances yield exactly the given local relations) versus weak universal relation assumption (that global instances yields supersets of the given local relations) based on whether closed-world assumption or open world assumption is used.

However, we cannot not agree this approach due to the following reasons:

(1) Global schema is not practical for many fields, simple due to the fact that involved people may have difficulties to reach consensus [15, 31], or too slow/too costly to reach agreement due to scale of the information ecosystems.

(2) Even there is a global schema, it is more likely a compromised solution such that individual local model must re-interpret the global schema from its specific perspective.

(3) As a result, this encourages semantic heterogeneity (that there are different meanings for the same representation) to happen to happen in information ecosystems.

(4) The more heterogeneous meanings overloaded to the global schema, the harder for data to be integrated semantically for efficient use.

On the other side, both meaning independence and logical data independence are efforts to raise level of abstraction for data modeling. We do believe this is the direction to go for a complete solution to overcome difficulties encountered during data integration, as described by Hass: that "*Experience with a variety of integration projects suggests that we need a broader framework, perhaps even a theory, which explicitly takes into account requirements on the result of the integration, and considers the entire end-to-end integration process*" [11].

From the viewpoint of Shadow Theory, our ultimate wish is this: since meanings as mental entities only exist in viewers' cognitive structure, the objective of database design should to help users to manage data by meanings, not by manipulating complicated logical representations in the model which requires layers of layers interpretations. Under such wish, then the job for data integration can really focus on the real issue: differences between business semantics and procedures, on top of which business operations are established. The data models can help users to identify or make subject decisions about what can be treated as the same meaning without worrying about logical structures from different sources. There is a long way to go for this wish, and many issues are not resolved yet. In next section, we will make further comparisons with semantic data models, especially Entity-Relationship Model and Entity Set model.

## 6.4 Comparisons with Entity-Relationship Model

There are many semantic data models proposed in past several decades. Here we will briefly compare the main features with Entity-Relationship Model [54]. When we provide definition for shadows in section 3.3, we explained the issue of **semantic relativism** [55], which motivates us to choose a different design principle than commonly accept Entity-Relationship Model. Semantic relativism concerns about the ability to view and manipulate data in the way most appropriate for the viewers semantically, not forced by the chosen data model. Take marriage as an example, it can be modeled as an entity, a relationship, an attribute of an entity, or an attribute of a relationship. If a data model makes a rigid choice for its users, than there will have difficulties for data integration later as such choice has nothing to do with the subject matter in reality, but a subjective decision made during modeling process.

In terms of Shadow Theory, this is about the difficulties of data integration triggered during projection process from subject matter to meanings as mental entities then represented by shadows in databases, or from a shadow in one system into another system, or from one mental entity existing in a viewer's cognitive structure into another one. Since our goal is to reduce such difficulties, we cannot follow Entity-Relationship Model to (subjectively) classify the conceptual entities into different kinds of categories [54].

Further, there are some differences about the hidden assumptions like the existence of entities or relationships in reality, versus mental entities that are only in viewers' cognitive structures. The generic notion for an entity is for *something* out there in the real world, and relationships are for the interactions/associations between these *things*. This is classified as the realist semantics approach in Gärdenfors's survey (ch5 in [56]) as the meanings of the entities or relationships are due to *something* out there in the real world (extensional) or possible worlds (intensional).

In addition to the philosophical foundation of Shadow Theory that we only assume the existence of shadows, we do not choose this direction due the semantic heterogeneity difficulties that occurs when we compare different entities represented in different models about the same subject matter. For example, the legal entity-based ECID versus location-based ECID or contract-based ECID. If their meanings are due to *something* in reality, then we sure cannot integrate them together as the *something* are so different. We can only integrate them if they are mental entities such that their differences are due to different perspectives chosen by individual viewers' cognitive structure.

Our objective to support for multiple version of the truth is not in the scope for ER model, neither does the need to support incremental design for agile development process. For we do not assume the data modelers have full knowledge of everything when they make design decisions, and they may need to make frequent revisions as their understandings evolve. On the contrast, in the ER model, the very first decision requires designers to choose among entity, relationship, and attributes such that it creates difficulties if there is a need later to revise this decision.

## 6.5 Business push on Customer Data Integration (CDI)

Since we use enterprise customer data integration as the example in this paper, we need to discuss recent business push in this specific application area. Customer Data Integration (CDI) is promoted in industry since 2004, and summarized in Gartner report as "*the combination of the technology, processes and services needed to create and maintain an accurate, timely and complete view of the customer across multiple channels, business lines, and potentially enterprises, where there are multiple sources of customer data in multiple application systems and databases*"[49].

The basic idea is from business perspectives to expect (or to wish) that there can be a single version of the truth about customer data [85] to satisfy different kinds of business needs. It is another wave of business push after people recognized the failure rate of CRM (Customer Relationship Management) could be 65% (it is debatable, and can be between 50% - 70% depends on different sources) [86-89].

Further, Radcliffe summarized the four approaches to pursue the single truth [49]: (i) External reference to absolutely identify a customer (complementary to the other three styles), (ii) Central registry of global identities to link to master data in source systems with transformation rules. At runtime, the CDI hub accesses the source master data and assembles a point-in-time single customer view. (iii) Coexistence and harmonizes the master data across these heterogeneous systems for greater consistency and data quality, (iv) CDI transaction hub as the primary repository of customer reference information.

The rationale of such expectation can be recognized as being based on the commonsense example of a library: different kinds of publications are collected and systematically stored without asking users how they want to use the information. Why customer data cannot be collected and systematically stored in the same way without asking how applications or end users want to use them? In another way, we can think data like water: just like water can be "integrated" in reservoir and provide users to use as they wish, why can't we "integrate" customer data such that any one can use it later for any kind of purpose?

We understand the business needs, but we cannot agree to over simplify data integration without considering the factor of semantic heterogeneity and the nature of information ecosystems. If the data need to be integrated are semantic homogeneous, and if the information ecosystems are totally within the enterprise' control, then yes, it is possible to accomplish data integration with a single version of the truth. That is, the challenge of data integration is to force every business process (e.g. billing, ordering, repair and so on) to choose the same perspective (e.g. legal entity-based ECID). If these conditions cannot be met, then it is not wise to promote and sell the unrealistic expectations as it will hurt the industry after the bubbles broke in real world (like what happened in history for exaggerated promises of new technologies).

There are also chicken-and-egg problems between data integration driven by individual departments versus those for overall enterprise wide data integration. For example, it is commonly accepted that customer data integration need to be done first since other applications like integrated ordering need to refer to such data. However, during the data model design for the integrated customer data, if the requirements of the integrated ordering are not available, such integration is like shooting in the air without precise targets, the reverse system dependency problem we mentioned in section 2.1. It relies on data integrators' prediction about how the integrated data will be used in the future (by the integrated ordering systems as one of the downstream systems). It often ends up with change requests for the integrated customer data model, when the integrated ordering systems are in the design process. That is why we have to assume data integrators do not have full knowledge about the data when their perform data model design, and we need easy ways to adjust data model (to avoid the know schema evolution difficulties), since there are always revisions needed later when downstream systems start using the integrated data.

Under such conditions, applying the criteria of completeness to evaluate data integration is not appropriate. If it is evaluated from perspectives chosen by upstream system (i.e. the integrated customer data), completeness is not possible due to the inconsistencies or conflicts of the original model perspectives. If it is evaluated from downstream systems by their local requirements (i.e. integrated ordering), completeness may not be possible as the available perspectives are limited by what the upstream systems can provide.

The efforts in this paper are based on practical application experiences to pursue such over-simplified business wishes and difficulties encountered during such process. It is not only due to technical problems, but also due to the expectation of management. For example, in the example of Scenario 1, can a high level management understand why there are inconsistent answers from different systems for the answers of the simplest questions like: "*how many enterprise customers do we have in total across service provider B, G, and W*?" The reality is that, there is no single perspective from legal entity-based ECID, location-based ECID, or contract-based ECID holds data for all involved enterprise customers due to different kind of services and geological areas. With different mappings based on local systems' subjective criteria, there is no way to come out a single and consistent number if business cannot provide a single perspective to integrate their business semantics and operations. Of course, the scale of the information ecosystems also play a big role here, the more individual databases involved, the more complexities the data integration is.

Hence, the Shadow Theory proposed here can be viewed as a way to justify why certain data integration approaches will work or will not work in practical applications. In addition, it helps to explain the role for involved human, that the uniqueness of unique keys is due to the unique existence of mental entities in specific viewers' cognitive structure (i.e. the chosen perspectives), not necessarily due to the *things* in reality. It also explains about the subjective decisions made by data integrators, for how and why they may or may not be combined together to infer extra mappings.

The most important point we want to raise here is a simple common sense that if there are inconsistencies or even conflicts that people cannot resolve in the real world, don't expect data integrators can magically resolve them in model world. Business management needs to understand that only if they can integrate business operations (e.g. choose the same or coordinated perspectives about the subject matter), then data models can help them to fulfill their wish to increase efficiencies of business operations.

Without such understanding, and under the conditions that inconsistencies or conflicts cannot be resolved, what we proposed is to extend the responsibilities of data models to help users to understand the issues in order to manage semantic heterogeneity, not to hide them under a uniform logical structure. This is why we proposed in section 4.3 to add the functionality of explanations by information collected with W-tags and E-tags, with extra business logic description provided by application programs during basic insert-update-delete operations. In this way, users can interact with the database systems to ask questions and receive answers about questions like

(A) **Who** performed the changes in either data space or semantic space? and **why** they make such changes (i.e. following which business operation rules) ?

(B) **What** happen to shadows in data space? and **how** does it happened? What does it means in terms of meanings as mental entities in the specific viewers' cognitive structures?

(C) **When** and **where** (i.e. which upstream systems) did changes happen?

## 6.6 Related data integration approaches

Next, we will briefly review progress made in different approaches for data integrations. Due to the huge amount of related literatures, we will only highlight related ones and interested readers can find more details in the following survey or analysis reports: [3-5, 9, 11, 15, 90-92].

The fundamental difference in our proposal is to first address the weakness of existing data models (to model semantic heterogeneity and multiple versions of truth), since we believe the difficulties encountered during data integration are the natural results due to such weakness. Therefore, we proposed Shadow Theory in order to use explicit W-tags to uniquely anchor meanings as mental entities in viewers' cognitive structures.

Next, we propose to use E-tags to model the rich varieties of semantic equivalence, and push the operations of semantic equivalence to be included in the algebra at data model level. For example, in the survey of [4], Rahm and Bernstein summarized the concept of similarity and treated a mapping as a similarity relation, which can be directional or no directional over scalars (=, <), functions (addition), semantic relations (is-a, part-of), or set oriented operations. Batini classified equivalence as three kinds: behavioral, mapping, transformational [9]. The work in [45] classified mapping into three major categories: equivalence, set theory, and generic (semantic) relation. Sheth introduced the concept of semantic similarity with the following four levels, semantic equivalences, semantic relationship, semantic relevance, and semantic resemblance, to be used with abstraction mechanism like aggregation, generalization, mapping (1:1, n:1), and functional dependency in related context representation for describing relations between objects [46] [47].

With generic W-tags and E-tags for operations in semantic space, we can model these different kinds of mapping and control their mapping directions by strong versus weak semantic equivalence. We can use IS-A / HAS-A to control the direction of semantic inheritance, for how we model meanings can be propagated in semantic space as complex nested semantic relations. Further, in order to resolve the constraints of subjective ontological primitive units, we choose point-free geometry as the logical foundation instead of Set Theory and First Order logic. Still, we can support set oriented operations like join, intersection, and difference.

Our goal is to be generic as possible such that existing data integration algorithms or techniques can be migrated into this data model with minor revisions. For example, semantic similarity can be viewed as combination of semantic equivalence and level shifting that across different perspectives, and different measurement like confidence levels, semantic distance, or probabilistic can be viewed as a measurement based on the supporting evidences for E-tags to support semantic equivalence.

Larson's theory for attributed equivalence for schema integration is a very interesting approach that we need to mention here [42]. In general, it is based on the Entity-Relationship Model, and the foundation of equivalence by attribute is established on uniqueness, cardinality, domain, static and dynamic integrity constraints, allowable operators, and scale to determine semantic equivalence. We can re-interpret these criteria in terms of the three categories of shadow properties: uniqueness, cardinality, domain, allowable operators, and scale are due to projection process (mixed with characteristics due to subject matter), while static and dynamic integrity constraints are due to the wall-like system requirements. Whether these should be considered as criteria to make decisions about strong/weak semantic equivalence between meanings (as mental entities of different viewers' cognitive structures) depend on the scope of data integration projects. For those intended to integrate every related properties, these should be included as supporting evidence for E-tags; but for those only focus on integrating properties due to the subject matter, these criteria can be subjectively ignored.

Although we do not have schema concept in semantic space, we can simulate such structure by templates (with the cost of less modeling flexibilities) in order to support those applications or algorithms depending on schema notions. On the opposite direction, we also can model unstructured or semi-structure data and apply semantic equivalence between their meanings. Overall, Shadow Theory is an effort to mimic how we human think and integrate data with meaning(s) from other people, with or without rigid logical structures like schema is not the most important factors, but rather something we can adapt with.

## 6.7   Model management in semantic space

Next, we need to discuss Model Management, which is an important development for meta-data management that can be used for schema integration [50] [93]. Its advantage (against object-at-a-time programming) is to treat models and mappings as abstractions and manipulate them by model-at-a-time or mapping-at-a-time operators. Three meta levels are defined as: (1) model instance, (2) meta- model that consists of the type definition for the objects of the model, and (3) meta-meta-model which is the representation language that models and meta models are expressed in [93 ].

A set of model management operators is defined and they are generic by treating models and mappings as graph structures. However, there is a semantic gap between such operators and applications under specific business semantics. To fulfill such gap, three approaches are proposed in [93]: (1) to make meta-meta-model and behaviors of the operators more expressive, (2) to extend operators to produce expressions for any generated mapping objects, (3) to design a special design tool to adding semantics to mappings.

Shadow Theory can contribute to manage this semantic gap, since our explicit representation of meanings are based on the philosophy to model meanings as mental entities that exist in different viewers' cognitive structures. Instead of the other choices like modeling meanings as functions of the communication, or as something in real world or in possible worlds, we can better model the nature of mapping as subjective decisions made by data integrators based on supporting evidence for their desired semantic equivalence with direction control. That is, we can explicit represent different meanings from different perspectives, and which can be subjectively treated as the same as which.

The success of model management depends on how much and how precise meanings of the underlying schema can represent. In the situation where schema is overloaded with semantic heterogeneity, and thus it can only provide a common logical structure with confusing meanings, model management is hard to apply as such model is actually a combination of heterogeneous models mixed together under a common format.

For example, very common we see data integrators convert customer data from different sources into a standard format and loaded into a data warehouse. A logical unique key is assigned for each record, but the semantics of a single data instance cannot be determined by the schema or by the data values. In the ECID example of Scenario **4**, when legal entity-based ECID is mixed with location-based ECID and contract-based ECID, but without clear mappings (with explicit meanings and directions) between them, it is difficult to apply operators of Model Management to further integrate this model with other models.

Therefore, we feel that Shadow Theory can work with Model Management together to fully take the advantage of model-at-a-time. That is, for a polluted data model that holds semantic heterogeneous data, we need to use Shadow Theory to first clarify every involved perspective, and identify the meanings for each data instances by W-tags and each mapping by E-tags. Then we can think each perspective is a semantic homogeneous model that we can manage by operators proposed in Model Management.

In this way, the constraints due to rigid logical representation like schema can be bypassed, and Model Management can handle the differences due to different decomposition structure (or templates) in semantic space. Further work is needed to investigate the details and valid related theoretic characteristics. For example, consistency is a difficult challenge when mapping across the boundaries of individual perspective, in other words, when meanings from evidences collide due to multiple versions of the truth. In precise engineering mapping described in [50], any semantic equivalence is actually an engineered design for specific data integration projects; therefore, it is limited by the specific business semantics and chosen perspectives. This will impact model operators like composition, which combines successive schema mappings into a single schema mapping, i.e. model *A* to *B*, *B* to *C*, versus model *A* to *C*. Fagin pointed out that it is critical to provide semantics to the composition operator for precisely what equivalence means (between the successive mapping and direct mapping) [32]. One approach is to define equivalence due to the same query results against the original data sources versus the mapped one [35]. Another way is defined by space of instances of schema mappings (the binary relations between source instances and target instances) [32].

In terms of Shadow Theory, data instances are just shadows, and schema is just logical structures in data instance. Mapping among them cannot avoid the semantic heterogeneity difficulties as each of them can represent multiple different meanings, and the specific meanings used by mapping cannot not explicitly represented (in Relational Model). In addition, equivalence due to the same query results is like to define equivalence based on data space, the rich varieties of semantic equivalence due to similarity cannot be included. As we have discussed in section 4.3.3, (and section 5.2 Property 6 for transitive property) for ECID example, combining subjective mapping from legal entity-based ECID to location-based ECID and from location-based ECID to contract-based ECID may not generate the same mapping from legal entity-based ECID to contract-based ECID.

Several issues remain and need future work. For example, Fagin explained [32] that the composition of certain kinds of first-order mappings might not be expressible in any first-order language; therefore they introduced a second-order mapping language that can quantify over function and relation symbols. In Shadow Theory, we use the notion of region as the only ontological primitive unit to model meanings as mental entities (that can represent anything, including function and relations), and regions like sets are Monadic Second Order (MSO) objects as any region contains sub-regions. Although we simplify the overall complexities by operations in semantic space (without constraints by data space), the challenge is about how to represent the complexities of mapping due to rich varieties of similarities. E-tags with supporting evidence are compromised solutions for simplicity, since they relies on human to interpret or evaluate the meanings behind supporting evidences.

## 6.8 Ontology, XML, and Semantic Web

In this section we will briefly discuss related issues about ontology, XML, and Semantic web. First is about ontology versus epistemology, the fact that reality can be studied at different levels in various forms of levellism. For example, ontological, methodological, and epistemological. In the recent debate among different forms, Floridi and Sanders proposed that refined version of epistemological levellism should be the approach for conceptual analysis, since ontological levellism may be untenable [94, 95]. Here we will not further discuss the details and refer interested readers to [96] for the debates and arguments.

The reason we need to mention this is due to the common phenomena happened during data integration in information ecosystems that different people have different understanding about the subject matter due to their limitations to interact with reality. It is like different people can only see a portion of the elephant[12], while no one can see every aspect to get the overall picture. However, to support business operations, data integration must be performed and we need to manage the situation that we do not even know what really exist in the reality. That is, the difficulty is more **epistemological** about what different people know about the subject matter, combined with the difference due to their chosen ontology.

Therefore, we have no choice but to assume that the only *things* exist out there are shadows, and we must rely on meanings as mental entities to perform data integration. This is the fundamental difference between Shadow Theory and semantic data models (e.g. Entity-Relationship), Object-Oriented modeling, and various ontology approaches. We simply cannot assume or classify our representations as entity, relationship, object, class, attribute, and so on. Instead, we focus on users' interpretations of the data and their (intentional) usage of the data, not data itself either at data instance level or schema level.

Next, due to its popularity in enterprise applications, especially for interface between systems in information ecosystems, XML is another big area that we need to mention. For the perspective of data modeling, the basic difference is about how we use tags: we do not use tags for logical structures, instead we use tags (i.e. W-tags) to anchor meanings of data as mental entities viewers' cognitive structures. In addition, the organizations of W-tags are not limited to hierarchy; it can be in graph, flat, or any structure.

---

[12] In terms of the ECID example in Scenario **4**, different kinds of ECID are designed to model the behaviors of enterprise customers only in specific products, geological areas, and within specific service provider. They are just some of the unlimited portions of the elephant with inconsistent patterns, and the worst of all is that the elephant can even merge with another one or split into multiple ones in never ending M&A activities.

The notion of equivalence in XML is limited to by the same data values, while we need E-tags to support rich varieties of semantic equivalence with subjective criteria. E-tags are also for the purpose to support operations for the algebra, not intended for a classification for different kinds of tags. As for P-tags, they are for denoting properties of data space like format and data type, which are mainly for being backward compatible with existing data model.

The fundamental objectives are also different. As self-description documents, XML expect readers can understand the meaning of the documents by themselves, while our goal is to design data model for database systems that can explain the meanings interactively with users. Specifically, the goal is to manage semantic heterogeneity and to help users to understand the differences for what different data represent.

We borrow some results from the development of XML, especially the use of Monadic Second Order (MSO) logic, which is applied for XML queries through automata theory [97, 98]. MSO is an extension of First Order (FO) logic by allowing quantified variables denoting set of elements. We are interested to use MSO since we use the notion of regions as the primitive units, which like sets are MSO objects such as any region contains sub-regions. This allows us to model the same W-tags that can be further decomposed in different ways in semantic space for heterogeneous logical structures and non-atomic data elements.

However, more precisely, we can only say the predicates are MSO-like since we miss the atomic data elements and the contents of any set cannot be explicitly represented (i.e. a region can be viewed like a set of points, however, we cannot explicitly represent all of the points included or all of the possible ways of decomposition for this region). We adapt Thomas's logical representation of MSO proposed in [80] with some revision, which has the same expressive power as traditional MSO, that all second order quantifiers are shifted in front of first order quantifiers, and first order variables are cancelled by simulating elements with singletons. That is, if lowercase variable $x$, $y$ represent atomic data elements, and uppercase variable $X$, $Y$ to represent sets (monadic second order objects), Thomas uses the atomic formula $\{x\} \subseteq X$ to replace $X(x)$ to indicate that $x$ is an atomic element in $X$, uses Sing($X$) and Sing($X$) to indicate that $X$ and $Y$ are singleton, and uses $X \subseteq Y$ to indicate an ordering relation $R(X,Y)$ that $X$, $Y$ are singletons $\{x\}$, $\{y\}$ such that $R(x, y)$ exists. The revision we need is to drop the atomic formula Sing($X$) and any first order variables since any W-tag for its associated shadow is modeled as a region in semantic space, such that we will never have any W-tag that is singular.

MSO is also used to in Modular Decomposition, which can establish structure properties describing the partial order of decomposition graphs [99, 100]. The notion of a module is similar to the notion of a region since both can be further decomposed in different ways concurrently. However, there is a major difference due to our representation constraint that we need to use a single type of ontological primitive unit (to reduce the complexities of data integration), while there are two types of primitive units in the work of Modular Decomposition, vertex and edge, following the traditional definition of a graph. If we ignore the atomicity assumption of a vertex and view a vertex as a region, and treat an edge as also a region that includes its two vertexes as sub-regions, we can convert the Modular Decomposition into the decomposition structure based on Shadow Theory. In this way, we may be able to use the axioms identified in this filed to help our representation in semantic space.

Next, we will make a brief comparison with Semantic Web, specifically RDF (Resource Description Framework)[101] and OWL (Web Ontology Language)[102]. There are many similarities of the challenges to resolve, for example, Open World Assumption, no single truth, inconsistencies and conflicts must be managed, and no perfect one single data model or one single view for everything [103-106].

The fundamental difference is the assumption about the existence of the entity / the subject matter. The vision of Semantic web is on real world entities with implicit assumption about their existence, consistent with the ontology and description logic, their logical foundation. As we explained earlier, Shadow theory can only assume the existence of shadows, not the existence of subject matter as entities. Therefore, our approach has the flexibility to model semantic heterogeneity and inconsistencies due to different perspectives or different levels of abstraction.

Second, the objectives are different. Semantic web is targeting to allow machines to understand semantics, while Shadow Theory is to serve human as a conceptual modeling methodology in order to understand the complexities encountered during large-scale data integration in information ecosystems. Hence, we need to explicit represent meanings and we choose meanings as mental entities to be the foundation for such representation purpose. Explicitness simply indicates that some one holds the meanings in his specific cognitive structure. For Semantic Web, we are not sure if this approach can be used, further investigation is needed to see if there is potential to allow machine to understand semantics in this way.

Next, the hierarchy concept in Semantic Web is based on the principle of commonality versus variability: the nodes in the higher levels indicate the common properties that the lower levels share. However, the hierarchy notion we have for decomposition structure is not limited by this principle, since we try to use a uniform representation to include not only IS-A relations, but also HAS-A relations. In general, we use hierarchy to describe the partial order relations among different meanings as mental entities, the nodes in higher level may not have commonality with the nodes in lower levels.

Next, the fundamental representation mechanism in Semantic web is based on statement, i.e. triples of (subject, predicate, object). Therefore, the mechanism forces distinction between subjects, objects, and predicates. For Shadow theory, we try to avoid such distinction in order to overcome semantic relativism problem that creates more issues for data integration. Therefore, the graphical structures are also different: in Semantic web and the graph theory it established on, a node represents a real world entity and an edge represents relations between nodes. In Shadow theory, we only use regions to represent any kinds of partial order relations in semantic space.

For the notion of equivalence, Semantic web treats two statements as the same when the subjects, objects, and predicates are the same. Although RDF does provide "equal" operation, but it is limited to express the rich varieties of semantic equivalence, not enough to allow human to fully recognize the semantic differences while treat them as the same for certain purpose. Since the nature of semantic heterogeneity is that there exist different representations for the same meaning, and there exist different meanings for the same representation, our main focus is about the underlying model perspectives, levels of abstraction, and chosen ontology (explicitly or implicitly) such that we need an explicit way to represent these characteristics and express the subjective decisions of viewers to treat different meanings as the same. For example, we can convert the example of *Ownership* in Figure 14 (2a) into the triple (*Enterprise Customer, owns, Account*). Assuming we identify a legal entity-based ECID *X1*, a location-based ECID *X2*, and a contract-based ECID *X3* for "the same" enterprise customer *X*, given the fact that *X1*, *X2*, and *X3* are modeled from different products, different areas, and different service providers *B*, *G*, and *W*. If we know (*X1*, owns, *A*), i.e. *X1* owns account *A*, can we infer the triples (*X2*, owns, *A*) and (*X3*, owns, *A*)? If an application needs to treat them as the same for one purpose, and also treat them as different for another purpose, in addition, use the two results together in a report, we may have difficulties to describe the complexities in terms of triples as there are different meanings we intend to use for the same subjects, objects or predicates.

Shadow Theory may be used in Semantic web in the future since the same difficulties of semantic heterogeneity also exist on internet. For example, different news reports about the same subject matter or events: since we cannot control the exact meanings people use in terms of the same vocabulary, and we cannot limit people to use the same vocabulary to describe the same meaning, data integration (in order to search news by semantics, one example application) requires some explicit way to model semantic heterogeneity, not just by the definitions provided in dictionaries, but also allows users to make subjective judgments about what can and cannot be treated as the same.

## 6.9    Other related issues

Top-down inheritance of class hierarchy in Object-Oriented (OO) programming is used to extend existing definition into new class such that we can re-use existing class and code. In knowledge representation systems, approaches like LOOM can support not only inheritance of concepts (entities/objects), but also inheritance of relations, where entities and relations are two kinds of primitive units in the representation [107] [108].

Since the hierarchies of decomposition structures in semantic space are not based on commonality versus variability, the two-way **Semantic Inheritance** we described in section 3.4.3 works fundamentally differently, We intend to address a different need: how meanings of the model subjects can be shifted among different levels of abstraction and cross the boundaries of different perspectives in semantic spaces. In other words, a meaning as a mental entity in a viewer's cognitive structure can inherit meanings from higher or lower levels of abstraction, and such process can happen across different perspectives when certain E-tags are established to support semantic equivalence.

The concept of two-way inheritance is not new, as researchers have already reported them in 80s [109] where "less than" was used as an example to explain upward inheritance that the parent node is less than child nodes. In terms of Shadow Theory, we can view the notion of "less than" as an implementation of inclusion relations, where regions in semantic space are translated into numbers, and the area includes sub-areas is assigned a smaller value than its sub-areas.

Here we also need to refer readers to the concept of basic-level effects in **prototype theory** [71, 72], that psychologically the most basic level is in the middle of the taxonomic hierarchies, which are most commonly used by people. Remember that data modelers are human and they use "their appropriate basic levels" in their data model designs. Inevitably, during integration happened later, data integrators have the natural tendencies to use their existing models as the foundation to understand other perspectives by making comparisons.

This is the reason we mentioned middle-out approach [76, 77] in section 4.2.2, such that users just need to model the concept they need without complete representation for all top levels. In addition to the saving of extra work for completeness, the more important feature is to support the need of data integration, since the results of comparison between different representations of the same subject matter may need higher (or lower) levels of abstraction to describe their semantic differences. That is, we have the need to allow users to represent their concepts at the most appropriate levels they need (without the requirements to start from the top level), and they can add higher levels or lower levels later during data integration.

## 7. CONCLUSION

In this paper, we have proposed Shadow Theory to serve as the philosophical foundation for data model design. The objective is to support the modeling needs of semantic heterogeneity, which is the most difficult issue in current data integration due the weakness of existing data models. Specifically, we consider the requirements of information ecosystems where data flow from upstream systems into downstream systems, and hence used in the downstream data models to satisfy local business operations.

In section 2, we raised six basic questions, and in section 3 and section 4 we answer these questions with six basic principles to guide the overall data integration process as the following:

Question 1: What does a unique key actually represent?

→Principle 1. What we can observe and store in database are only shadows.

Question 2: What are meanings? How can we explicitly represent meanings in order to manage semantic heterogeneity?

→Principle 2. The meanings of shadows exist as mental entities in viewers' cognitive structures, and we can use W(hat)-tags, short as W-tag, to anchor such mental entities uniquely.

Question 0: What is semantic heterogeneity?

→Principle 3. Semantic Heterogeneity is the overall aggregated result due to differences among meanings as mental entities in viewers' cognitive structure, and differences of how shadows are projected onto wall-like surface of system requirements about the same subject matter.

Question 3: What is the nature of mapping? In what sense can mapped data be treated as equivalence? Does such equivalence uni-directional or bi-directional?

→Principle 4. Equivalence between meanings (as mental entities from different viewers' cognitive structure) is a subjectively decision, and we can model such equivalence by E(quivalence)-tag, short as E-tag, with supporting evidences, just like a bridge to cross the boundaries of different perspectives.

Question 4: What kinds of characteristics of the data should be integrated and re-use in the local model? How data integrators should perform their design activities such the integrated data can be meaningful for users and re-useable again later for different needs? And how we should handle the inconsistencies or conflicts?

→Principle 5. Meaningful data integration should be performed only with required shadow properties, and the scope of the subjective equivalence decisions should be explicitly represented with meanings of the data.

Question 5: How should data models help users to understand the meanings of data? For data integration, can data models help users to recognize the problems due to semantic heterogeneity such that users can manage inconsistencies or even resolve conflicts in their business operations first?

→Principle 6. To helps users to understand and use integrated data properly, data models need some features to explain the meanings of data, including modeling perspectives, business logic rules, and the criteria for decision decisions made for semantic equivalence.

In section 5 we proposed algebra based on point-free geometry to support operations in semantic space. We explained the constraints due to Set Theory and First Order logic, especially those prevent data integrators from treating different representations of the same meaning to be equivalent, even such semantic equivalence is a subjective decision only valid from specific perspective and criteria.

There are still many issues need further investigation. In section 6, we briefly discuss them as we make comparisons with existing data models and data integrations approaches. Due to the nature of the difficulties encountered during data integration, there are needs of support from many different fields. Shadow Theory we proposed here is an attempt to use simple common sense notions to understand these issues based on practical experiences, and hence the efforts here are more like to find ways to justify why such practices can work. The ultimate wish is a new generation of data models implemented in database systems that can interactively explain to users about the meanings of data, the differences from different perspectives, and help people involved with information ecosystems to know what happened, why changes happened in the way, when changes occur, such that people can recognize the root difficulties, manage semantic heterogeneity, even to resolve the inconsistencies/conflicts in the real world.

## ACKNOWLEDGMENTS

The author would like to thank Michael Brodie for his encouragement and support for the theory development. The author also needs to thank the practical data integration experiences, techniques, and arts developed by professional data integrators in the field, especially Robert Long, Julie Lin, Yunyan Wang, Fang Xie, Eric Greene, Dong Zhong, Wei Miao, and Thomas Moroney.